\documentclass[aps,pre,superscriptaddress,groupedaddress,nofootinbib]{revtex4}

\usepackage{graphicx} 
\usepackage{bm}        % for math
\usepackage{amssymb}   % for math
\usepackage{amsmath}
\usepackage{enumerate}
\usepackage{color}
\usepackage{pgfplots}
\usepackage{tikz}
\usepackage{tikz, adjustbox}
\usepackage[most]{tcolorbox}
\usetikzlibrary{calc,decorations.markings}
\usepackage{nomencl}
\makenomenclature
\usepackage[toc,page]{appendix}

\begin{document}

\title{Fluctuations of dynamical observables in linear diffusions with time delay: a Riccati-based approach
}\author{ M.L. Rosinberg}
\email{mlr@lptmc.jussieu.fr}
\affiliation{Laboratoire de Physique Th\'eorique de la Mati\`ere Condens\'ee, CNRS-UMR 7600, Sorbonne Universit\'e, 4 place Jussieu, 75252 Paris Cedex 05, France}
\author{G. Tarjus}
\affiliation{Laboratoire de Physique Th\'eorique de la Mati\`ere Condens\'ee, CNRS-UMR 7600, Sorbonne Universit\'e, 4 place Jussieu, 75252 Paris Cedex 05, France}
%\email{tarjus@lptmc.jussieu.fr}
\author{T. Munakata}
\affiliation{Department of Applied Mathematics and Physics, 
Graduate School of Informatics, Kyoto University, Kyoto 606-8501, Japan
} 
%\email{tmmm3rtk@hb.tp1.jp}

\begin{abstract} 

Our current understanding of  fluctuations of dynamical (time-integrated) observables in non-Markovian processes is still very limited. A major obstacle is the lack of an appropriate theoretical framework to evaluate the associated large deviation functions.  In this paper we bypass this difficulty in the case of linear diffusions with time delay by using a Markovian embedding procedure that introduces an infinite set of coupled differential equations. We then show that the generating functions of current-type observables can be computed at arbitrary finite time by solving  matrix Riccati differential equations (RDEs) somewhat similar to those encountered in optimal control and filtering problems. By exploring in detail the properties of these RDEs and of the corresponding  continuous-time algebraic Riccati equations (CAREs), we identify the generic fixed point towards which the solutions converge in the long-time limit. This allows us to derive the explicit expressions of the scaled cumulant generating function (SCGF), of the pre-exponential factors, and  of the effective (or driven) process that describes how fluctuations are created dynamically. Finally, we describe the special behavior  occurring at the limits of the domain of existence of the SCGF, in connection with fluctuation relations for the heat and the entropy production.
\end{abstract}
%\date{\today}
 
\maketitle

\tableofcontents

 \section{Introduction}

A common problem in nonequilibrium statistical mechanics consists in estimating the statistics of a time-integrated observable,  such as the   heat or the entropy production for a system in contact with a heat reservoir and driven by an external force.  Under quite general conditions, the probability density of such an observable  obeys a large deviation principle in the limit of large integration times and  the fluctuations  are then characterized by  the so-called rate or large deviation function (LDF). In the case of Markovian stochastic dynamics, a powerful mathematical framework has been developed to compute this quantity and the closely related scaled cumulant generating function (SCGF)  (for reviews, see, e.g., Refs. \cite{T2009,J2020}).

However, many stochastic systems in biology, physics and technology are affected by memory effects and long-range temporal correlations. These occur in particular in the presence of feedback loops when the time lag between the signal detection and the system response (or the control operation) makes the dynamics inherently non-Markovian~\cite{AM2008,A2010,JPSS2010,M2015,KK2019,B2021}. 
The theoretical description of the  fluctuations of a dynamical observable then becomes problematic.  The major difficulty is that the time evolution of the  generating function can no longer be formulated as a linear partial differential equation~\cite{T2018}. Accordingly, the SCGF can no longer be obtained as the largest eigenvalue of the corresponding ``tilted" generator~\cite{T2018,CT2015}.  

The main goal of this paper is to bypass this obstacle in the case of linear diffusions with time delay.  In this respect, the paper is a sequel of previous work in which we studied  dynamical fluctuations for an underdamped particle trapped in a harmonic potential and submitted to a position-dependent, time-delayed feedback force~\cite{RTM2017}. Such a model  describes  the motion of  feedback-cooled mechanical resonators, e.g., a microcantilever in the vicinity of its fundamental mode resonance~\cite{P2007,M2012,KK2016}. The most significant outcome of Ref. \cite{RTM2017}, recently tested experimentally~\cite{DRLK2022}, was that the delay strongly affects the regime of large deviations and that the fluctuations of  heat,  work, and  entropy production in the nonequilibrium steady state  are quite different (whereas their expectation values are identical). This feature cannot be rationalized from the sole knowledge of the LDF: one needs to determine the {\it complete} asymptotic behavior of the probability distributions and generating functions, including the pre-exponential factors. This nontrivial task (which  could not be achieved in Ref. \cite{RTM2017}) is fulfilled in the present paper.  

To this end, and to make the problem mathematically tractable, we  use a procedure known in the literature as the ``linear chain trick", which consists in replacing the discrete delay by the larger class of gamma-distributed delays~\cite{McD1978}. It is indeed a well-known fact  in the theory of  delay integro-differential  equations that an equivalent system of ordinary differential equations is  obtained whenever the delay is gamma-distributed~\cite{S2011}.  This procedure, which requires one to introduce auxiliary variables, is widely used  in the context of biological modeling, population dynamics or evolutionary systems~\cite{C1979,PW1986,MD1986,McD1989,HK2019}.  The discrete delay is  recovered when the number of auxiliary variables goes to  infinity~\cite{LK2019}. Besides  the fact that a distributed delay is often more likely to capture reality  than a discrete one (which  justifies studying the  properties of such a system per se~\cite{LHK2021}), the bonus is that  the dynamics of the augmented system is  Markovian. This allows us to work within the standard  framework of Markov processes. 

More generally, we propose a method  to calculate the generating functions {\it beyond} the large-deviation regime, i.e., for stochastic trajectories of arbitrary duration. Owing to the linearity of the dynamics and  of the current-type observables, the calculation boils down to solving  continuous matrix Riccati  differential equations  (RDEs) similar to those encountered in optimal control and filtering problems. We can thus benefit from the extensive mathematical literature devoted to the analysis and the numerical solution of such equations (see, e.g., Refs. \cite{AFIJ2003, K2010} and references therein), including for large-size systems~\cite{BM2004}. However, there is a crucial difference with the standard situation treated in linear quadratic (LQ) optimal control which significantly complicates the theoretical analysis. In particular, the solutions of the RDEs (and in turn the generating functions) may diverge in a finite time or  converge to different fixed points. This  latter feature can be missed when only focusing on the spectral problem for the dominant eigenvalue of the tilted generator.  Our study thus requires a detailed (and occasionally rather involved) exploration of the properties of the RDEs and of the corresponding  continuous algebraic Riccati equations (CAREs)  in order to anticipate the various possible scenarios. The reward is that  many of the results presented in this work  are applicable beyond the specific case of time-delayed  Langevin equations and can be used to compute the fluctuations of any  linear current-type observables in multi-dimensional  linear diffusions\footnote{While completing the writing of this paper- which took much longer than  expected-  we learned of Refs. \cite{DB2023, DBT2023} that also use Riccati differential equations to study dynamical large deviations of linear diffusions. Similarities and differences with our approach will be discussed in the text (see in particular Sec. \ref{SubsecIVB5}).  A preliminary account of the present work was presented orally at a conference in June 2022~\cite{ConfNice}.}. 
We therefore hope that the present  analysis will  not only provide a better understanding of the influence of memory effects on dynamical fluctuations, but   more generally will  be useful for the  application of  large deviation theory to nonequilibrium stochastic systems.

The content of the paper is the following:

 In Sec. II.A  we  present the  stochastic underdamped model with a discrete time delay and we introduce the linear chain trick that makes it possible to replace the original non-Markovian dynamics by an infinite set of coupled linear equations without delay.  We then define in Sec. II.B the three linear currents (work, heat, and entropy production) whose fluctuations are commonly studied  in the framework of stochastic thermodynamics in connection with fluctuation theorems (see, e.g., Refs. \cite{S2012,PP2021} and references therein).  We also briefly discuss in Sec. II.C how the delay affects the stability of the nonequilibrium steady state (NESS). 

Section \ref{SecRic} is mainly devoted to the study of the matrix Riccati equations that play a major role in this work. In Sec. \ref{SubsecRic1} we first derive the exact form of the moment generating functions of the fluctuating observables in terms of real symmetric matrices that are solutions of Riccati differential equations (RDEs).  A specific feature of our treatment is that, for a given observable, we study together the generating function conditioned on the initial state (solution of a ``backward"  PDE) and  the generating function conditioned on the final state  (solution of a ``forward"  PDE).  Although this modus operandi may appear redundant at first sight,  it will turn out to be very useful for analyzing the long-time behavior. In Sec. \ref{SubsecRic2} we then investigate in detail the properties of the RDEs, making heavy use of the concepts and  methods available in the mathematical literature. We  first discuss the global existence of the solutions  (Sec. \ref{SubsecRic2a}) and then  provide a closed-form representation of these solutions in terms of the eigenvalues and eigenvectors of the associated Hamiltonian matrices (Sec. \ref{SubsecRic2b}). Next, in Sec. \ref{SubsecRic3}, we use this  method to construct explicitly  all fixed points of the RDEs, which are solutions of the corresponding  CAREs.  The so-called ``maximal" solution is singled out as it is generically the fixed point  towards which the solutions of the RDEs converge asymptotically. 

Section \ref{SecIV} is the central and longest part of the paper in which we study the time evolution of the generating functions. To make it more concrete, the theoretical analysis is illustrated by numerical results obtained for the gamma-distributed delay. In Sec. \ref{SubsecIVA}, we  first investigate the domain of existence of the generating functions at finite time. In Sec. \ref{SubsecIVB}, we then  focus on the long-time behavior. We first determine the domain of existence of the SCGF (Sec. \ref{SubsecIVB1}) and  then derive the  explicit expressions of the SCGF and of the pre-exponential factors  in terms of the maximal solutions of the CAREs (Sec. \ref{SubsecIVB2}).  We also give a representation of  the SCGF in terms of an integral of the spectral density of the process, which shows the connection between the  Riccati-based approach and  general results in the mathematical literature for quadratic observables of stationary Gaussian processes~\cite{BD1997,BGR1997,GRZ1999,ZS2023}. In Sec. \ref{SubsecIVB3}, we make contact with the standard spectral problem for the dominant eigenvalue of the tilted generators, which allows us in Sec. \ref{SubsecIVB4} to characterize  the so-called effective or driven process that  describes how fluctuations  are created dynamically in the long-time limit.  In Sec. \ref{SubsecIVB5}, we then discuss the role of temporal boundary terms in relation with  the recent  work of De Buisson and Touchette~\cite{DBT2023}. Finally, in Sec. \ref{SubsecIVC}, we discuss the nontrivial behavior occurring  at the limits of the  domain of existence of the SCGF. We show  that  the solution of the RDEs may be attracted to a non-maximal solution of the CARE or may oscillate between two fixed points.  In both cases the SCGF  displays a positive jump discontinuity. 

Finally,  in Sec. \ref{SecV}, we focus on the special case where $\lambda$, the conjugate field to the dynamical observable, is equal to $1$,   in connection with the fluctuation relations for the heat and the entropy production.  In particular, we provide  the analytical proof of the conjecture relating the fluctuations of the entropy production at large times to the ``Jacobian" contribution induced by the breaking of causality in the backward process~\cite{RTM2017,MR2014,RMT2015}.

We end the paper  in Section \ref{SecVI} with a brief summary of the main results. Several technical derivations  are given in the Appendices as well as an application of the Riccati formalism to an active particle model.  A summary of the main notations used in this work is also provided at the end of the paper.

\section{Model and observables}

\subsection{Langevin equation and linear chain trick}

\label{Sec:trick}

As in previous works~\cite{RTM2017,MR2014,RMT2015}, we consider a Brownian particle of mass $m$ trapped in a harmonic potential and immersed in a thermal environment with viscous damping $\gamma$ and temperature $T$. The dynamical  evolution is governed  by the one-dimensional  underdamped Langevin equation 
\begin{align}
\label{EqLinit}
m\dot  v_t=-\gamma v_t-kx_t+F_{fb}(t)+\sqrt{2\gamma T}\xi_t\ ,
\end{align}
where $k$  is the spring constant and $\xi_t$ is a zero-mean Gaussian white noise with unit variance (throughout the paper Boltzmann's constant is set to unity). $F_{fb}(t)$ is a feedback control force which  is originally taken proportional to the position of the particle at the time $t-\tau$,
\begin{align}
F_{fb}(t)= k'x_{t-\tau} \ ,
\end{align}
where $\tau>0$ is the time delay. We  generally assume that the  feedback is positive ($k'>0$).  Eq. (\ref{EqLinit}) accurately describes the motion of the levitated nanoparticle studied in the experiments of Refs.~\cite{DRLK2022,DGALK2020}.  
We stress that the non-Markovian character of the dynamics results from the feedback  and not from the interaction with the environment.  By choosing the inverse angular resonance frequency $\omega_0^{-1}= \sqrt{m/k}$ as the unit of time and $x_c=\sqrt{2\gamma T\omega_0}/k$ as the unit of length, the Langevin equation takes the dimensionless form~\cite{RMT2015} 
 \begin{align}
\label{EqLlinred}
\dot v_t=-\frac{1}{Q_0} v_t -x_t+\frac{g}{Q_0}x_{t-\tau}+\xi_t\ ,
\end{align}
where $Q_0=\omega_0 \tau_0$ is the  quality factor of the oscillator ($\tau_0=m/\gamma$ is the viscous relaxation time) and $g=k'/(\gamma \omega_0)=(k'/k)Q_0$ is the gain of the feedback loop. The  dynamics  is thus fully characterized by the three  dimensionless parameters $Q_0$, $g$ and $\tau$. 

In order to apply the linear chain trick, we need to smooth the discrete  delay kernel $\delta(t-\tau)$ and replace Eq. (\ref{EqLlinred}) by
\begin{align}
\label{EqlinearTrick1}
\dot  v_t&=-\frac{1}{Q_0} v_t-x_t+\frac{g}{Q_0} \int_{-\infty}^t ds\:g_n(t-s,n/\tau)x_s+\xi_t\ ,
\end{align}
where 
\begin{align}
\label{Eqgamma1}
g_j(t,a)=\frac{a^j }{(j-1)!}t^{j-1}e^{-at}\ , \ \ t\ge 0 \ , 
\end{align}
is the probability density function of the gamma distribution (more exactly, the Erlang  distribution~\cite{I2013}). Note that the lower limit  of the integral in Eq.~(\ref{EqlinearTrick1}) is sent to $-\infty$ since  we will only be interested in the steady-state regime.
At the lowest order, the  kernel $g_1(t,1/\tau)$ describes an exponentially fading memory with a decay rate $\tau^{-1}$ (or a low pass filter with bandwidth $\tau^{-1}$ in another language). For $n>1$,  the kernel has a maximum around $t=\tau$ and the peak becomes sharper as $n$ increases (see e.g. Fig. 7.1 in \cite{S2011} or Fig. 2 in \cite{LHK2021}).  The discrete delay is recovered in the limit $n\to \infty$. In the frequency domain (i.e., in Fourier space\footnote{We here define the Fourier transform of a function $f(t)$ by $f(\omega)=\int_{-\infty}^{+\infty} dt\: f(t) e^{i\omega t}$.}) this simply amounts to approximating the delay function $e^{i\omega \tau}$ as~\cite{note1}
\begin{align}
\label{Eqexp}
e^{i\omega \tau}\approx \frac{1}{(1-i\omega \tau/n)^n}\ .
\end{align}

The Erlang density functions $g_j(t,a)$ satisfy the recursion relation $dg_j(t,a)/dt=a[g_{j-1}(t,a)-g_j(t,a)]$ for $j\ge 1$ which is the basis of the linear chain trick. Eq. (\ref{EqlinearTrick1})  is then equivalent  to the set of $n+1$  differential equations
\begin{align}
\label{EqL}
\dot  v(t)&=-\frac{1}{Q_0} v(t)-x(t)+\frac{g}{Q_0} x_n(t)+\xi(t),\nonumber\\
\dot x_j(t)&=\frac{n}{\tau}[x_{j-1}(t)-x_j(t)] \ ,\ \ j=1\cdots n\ ,
\end{align}
where  the  auxiliary dynamical variables  $x_j(t)$ are defined by
 \begin{align}
\label{EqlinearTrick2}
x_j(t)&=\int_{-\infty}^t ds\: g_j(t-s,n/\tau)x(s)\ ,
\end{align}
with $x_0(t)\equiv x(t)$.  

Thanks to this alternative representation of the dynamics, we are now dealing with a  Markov process in  the enlarged space $\{v(t),x(t), x_1(t),...x_n(t)\}$, whereas the marginal dynamics of  $x_t$ of course remains non-Markovian.  In general, the auxiliary variables do not represent actual physical degrees of freedom but it may be the case for $n$ small, in particular $n=1$ (see, e.g., Ref. \cite{CLSV2021}). Note also that  $x_j(t)\approx x(t-j\tau/n)$ as $n\to \infty$~\cite{LK2019}, so that the linear chain trick  in the large-$n$ limit may be interpreted  as a discretization of the trajectory of the particle in the time interval $[t-\tau,t]$. This  illustrates the well-known fact that a delay-differential equation such as Eq. (\ref{EqLlinred}) defines an infinite-dimensional dynamical system (since an infinite number of initial conditions -actually, a function - is needed to uniquely specify the time evolution)~\cite{L2010}.

To simplify the forthcoming analysis, it is convenient to recast the set of Eqs. (\ref{EqL}) into a matrix form by introducing  the $n+2$-dimensional vector ${\bf u}$ with components $u_1=v,u_2=x,u_3=x_1,...,u_{n+2}=x_n$. Defining the drift matrix 
\begin{align}
\label{EqDrift}
A=
\begin{bmatrix}
 -1/Q_0&-1 & 0 &0 &...&g/Q_0\\
 1&0& 0 &0&...& 0 \\
  0 & n/\tau & -n/\tau & 0 &...& 0\\
  0&0&n/\tau & -n/\tau &...&0&\\
   \vdots & \vdots & \vdots & \vdots & \vdots& \vdots \\
   0&...&0&n/\tau&-n/\tau&0&\\
   0&...&0&0&n/\tau&-n/\tau&
\end{bmatrix}\ ,
\end{align}
the equations in~(\ref{EqL})  become 
 \begin{align}
\label{EqLmatrix}
\dot {\bf u}_t=A {\bf u}_t+{\boldsymbol \xi}_t\ ,
\end{align} 
where ${\boldsymbol \xi}_t=(\xi_t,0,0...0)^T$.

\subsection{Dynamical observables}
 
Assuming that the system has reached a NESS, we are interested in studying the fluctuations of three time-integrated  stochastic currents. These are (in reduced units):
 
a) the work done by the feedback force during the time window $[0,t]$,
\begin{align}
\label{EqW}
\beta {\cal W}_t=\frac{2g}{Q_0^2}\int_{0}^tx_n(t')\circ dx(t')\ ,
\end{align}
where $\beta=(k_BT)^{-1}$ and $\circ$ denotes the Stratonovich product, 

b) the corresponding heat dissipated into the environment~\cite{S2010}
\begin{align}
\label{EqQ}
\beta {\cal Q}_t&=\frac{2}{Q_0}\int_{0}^t[\frac{1}{Q_0}v(t')-\xi(t')]\circ dx(t')\nonumber\\
&=\beta {\cal W}_t-\frac{2}{Q_0}\int_{0}^t\left[x(t')\circ dx(t')+v(t')\circ dv(t')\right]\nonumber\\
&=\beta {\cal W}_t-\Delta E \ ,
\end{align}
where $\Delta E=(1/Q_0)[x^2_t-x^2_0+v^2_t-v^2_0]$ is the change in the internal energy of the system, 

c) the entropy production (EP)
\begin{align}
\label{EqEP1}
\Sigma_t&=\beta {\cal Q}_t+\ln \frac{p(x_0,v_0)}{p(x_t,v_t)}\ ,
\end{align}
where $\beta {\cal Q}_t$ is the entropy change in the medium and $p(x,v)$ is the  stationary PDF (see Eq. (\ref{Eqp}) below).  Specifically, 
\begin{align}
\label{EqEP2}
\Sigma_t&=\beta {\cal W}_t+\frac{1}{Q_0}\big[(\frac{T}{T_x}-1)(x^2_t-x^2_0)+(\frac{T}{T_v}-1)(v^2_t-v^2_0)\big] \ ,
\end{align}
where $T_x=(2T/Q_0) \langle x_t^2\rangle$ and  $T_v= (2T/Q_0) \langle v_t^2\rangle$  are the configurational and kinetic temperatures of the system, respectively~\cite{RMT2015} (angle brackets indicate a steady-state average). 

If the dynamics were Markovian, the second term in Eq. (\ref{EqEP1}) would correspond to the change in the Shannon entropy of the system~\cite{C1999,S2005} and  $\Sigma_t$ would be the total stochastic EP in the time interval $[0,t]$. Here, $\Sigma_t$  defines the ``apparent"  EP that an observer unaware of the existence of the non-Markovian feedback would regard as the total  EP. This quantity  can be extracted from the stochastic trajectories collected in experiments, as done in Ref. \cite{DRLK2022}.  Note in passing that  the linearity of the Langevin equation (\ref{EqLlinred}) implies that the probabilities of a  trajectory and its  time reversal are Gaussian and equal  in the NESS. In consequence,  the log ratio of these two probabilities, which is usually taken as the  definition of  EP  in stochastic thermodynamics  (see Ref. \cite{PP2021} and references therein), is zero and does not properly accounts for the irreversible character of the feedback process~\cite{RMT2015}.

Since the three observables only differ by temporal boundary terms, they share the same average rate in the NESS,
\begin{align} 
\langle \beta \dot {\cal W}_t\rangle=\langle \beta \dot{\cal Q}_t\rangle=\langle \dot \Sigma_t\rangle=  \frac{2}{Q_0}[ \frac{1}{Q_0}\langle v_t^2\rangle -\langle \xi_t v_t\rangle]=\frac{1}{Q_0}(\frac{T_v}{T}-1) \ .
\end{align}
This expression illustrates the fact that the exchange of heat with the thermal environment in an underdamped system proceeds through the kinetic energy of the system independently of the form of the potential function~\cite{S2010}. On the other hand, the fluctuations of the observables may differ, as shown experimentally  in Ref. \cite{DRLK2022} in the case of the discrete delay. In particular, one has 
\begin{align}
\label{EqIFTheat} 
\langle e^{-\beta {\cal Q}_t}\rangle=e^{t/Q_0} 
\end{align}
at all times~\cite{RTM2016}, whereas it is conjectured  that 
\begin{align} 
\label{EqIFTentropy}
\langle e^{-\Sigma_t}\rangle\sim e^{\dot S_{\cal J} t}
\end{align}
as $t\to \infty$, where $\dot S_{\cal J}$ is a nontrivial quantity related to the Jacobian originating from time reversal~\cite{MR2014,RTM2017}.

 In the following it will be convenient to collectively define the three stochastic currents by
\begin{align} 
{\cal A}_t= \int_0^t {\bf g}_{\rm o}({\bf u}_{t'})\circ d{\bf u}_{t'}\ ,
\end{align}
where\footnote{With the present definition (see also Ref. \cite{RTM2017}), ${\cal A}_t/t $ is  intensive in time and converges in  probability to a constant  as $t\to \infty$. This differs from the definition adopted in Refs. \cite{CT2015, T2018} where  ${\cal A}_t$ itself is the intensive quantity.}  
\begin{align} 
\label{EqgaU}
{\bf g}_{\rm o}({\bf u})=B_{\rm o}  {\bf u} \ ,
\end{align}
 and $B_{\rm o}$ is a matrix of dimension $(n+2)\times (n+2)$. (Henceforth, the subscript $\rm o$ refers to the observable, with ${\rm o}=w,q$ and $\sigma$ for the work, the heat, and the entropy production, respectively.)  
From Eqs. (\ref{EqW})-(\ref{EqEP2}),  we  have
\begin{align}  
\label{EqBa}
B_{\rm o}&=B_w+S_{\rm o}\ ,
\end{align}
where 
\begin{align}
\label{EqBw}
B_w= \frac{2g}{Q_0^2} 
\begin{bmatrix}
 0&0&0 &...&0\\
 0&0&0&...& 1 \\
  0&0&0 &...& 0\\
  \vdots & \vdots  & \vdots & \vdots& \vdots \\
0&0&0 &...& 0&
\end{bmatrix}\ ,
\end{align}
and
\begin{align}
\label{EqSq}
S_q=- \frac{2}{Q_0}
\begin{bmatrix}
1&0 & 0&...&0\\
 0&1& 0&...&0 \\
  0 & 0 & 0 &...& 0\\
   \vdots & \vdots & \vdots & \vdots & \vdots\\
  0 & 0 & 0 &...& 0&
\end{bmatrix}\ ,
\end{align}

\begin{align}
\label{EqSsigma}
S_{\sigma}=\frac{2}{Q_0}
\begin{bmatrix}
\frac{T}{T_v}-1&0 & 0&...&0\\
 0&\frac{T}{T_x}-1& 0&...&0 \\
  0 & 0 & 0 &...& 0\\
   \vdots & \vdots & \vdots & \vdots & \vdots\\
  0 & 0 & 0 &...& 0&
\end{bmatrix}\ .
\end{align}

For convenience, we also define the matrix $S_w$ as the null matrix. Note that the matrices $B_{\rm o}$  characterizing the observables have the same anti-symmetric part. This will have an important consequence in the following.

\subsection{Stability of the non-equilibrium steady state}

In the present study, we assume that the system has reached a stable NESS. For given values of the parameters $(Q_0,g,n)$, this requires to choose the delay $\tau$ appropriately.  Indeed, as is well known, a time delay induces a complex  dynamical behavior~\cite{N2001}.  In particular,  there may be a series of stability switches as $\tau$ increases, corresponding to destabilizing/stabilizing Hopf bifurcations~\cite{DVGW1995}. 
As usual with linear stochastic systems,  the boundaries of the domain of stability in the parameter space can be determined by computing the roots of the characteristic polynomial $p_A(s)$ of the drift matrix $A$. The  loss of stability is then associated with the occurrence of a root with  a positive real part. From Eq. (\ref{EqDrift}), a straightforward calculation yields   
\begin{align} 
\label{Eqchar}
p_{A}(s)\equiv \det(sI_{n+2}-A)=(\frac{n}{\tau})^n[(s^2+\frac{s}{Q_0}+1)(1+\frac{s\tau}{n})^n-\frac{g}{Q_0}] \ .
\end{align}
We thus have to deal with a polynomial of degree $n+2$, for which there exist powerful root-finding algorithms. As  $\lim_{n\to \infty}(1+\frac{s\tau}{n})^n=e^{s\tau}$ and 
\begin{align} 
\label{Eqchar1}
\lim_{n\to \infty}(\frac{n}{\tau})^{-n}p_{A}(s)=(s^2+\frac{s}{Q_0}+1)e^{s\tau}-\frac{g}{Q_0} \ ,
\end{align} 
this may be compared with the task of solving a transcendental  equation  in the case of the discrete delay (see Appendix  C of Ref. \cite{RMT2015}). 

However, the calculation of  the stability diagram in the whole parameter space $(Q_0,g,\tau,n)$  is a formidable task which is beyond the scope of the present work. We will content ourselves with the  example shown in Fig. 1 that  illustrates the influence of $n$. This corresponds to a case for which only two stability domains exist for $n$ finite. In this figure,  the kinetic temperature $T_v= (2T/Q_0) \langle v^2\rangle$ (with $\langle v^2\rangle$ given by the element $(1,1)$ of  the covariance matrix $\Sigma$,  see Eq. (\ref{EqLya}) below), is plotted as a function of $\tau$, and  the loss of stability of the stationary state is signaled by the divergence of $T_v~$\cite{RMT2015}. It can be  seen that the width of the  unstable region increases with $n$ and that only the first stability domain survives for $n=\infty$\footnote{Moreover,  the critical delay $\tau_{c,n}^{(1)}$ corresponding to  the first destabilizing Hopf bifurcation  is numerically found to decreases to its limit $\tau^{(1)}_{c,\infty}$ as  $1/n$.}. This is in line with the general lore that a discrete delay is more destabilizing than a distributed delay~\cite{S2011,B1989,CJ2009,C2010,ABG2020} (see also Ref.~\cite{LHK2021}).
 \begin{figure}[hbt]
\begin{center}
\label{FigStability}
\includegraphics[trim={0cm 0cm 0cm 0cm},clip,width=10cm]{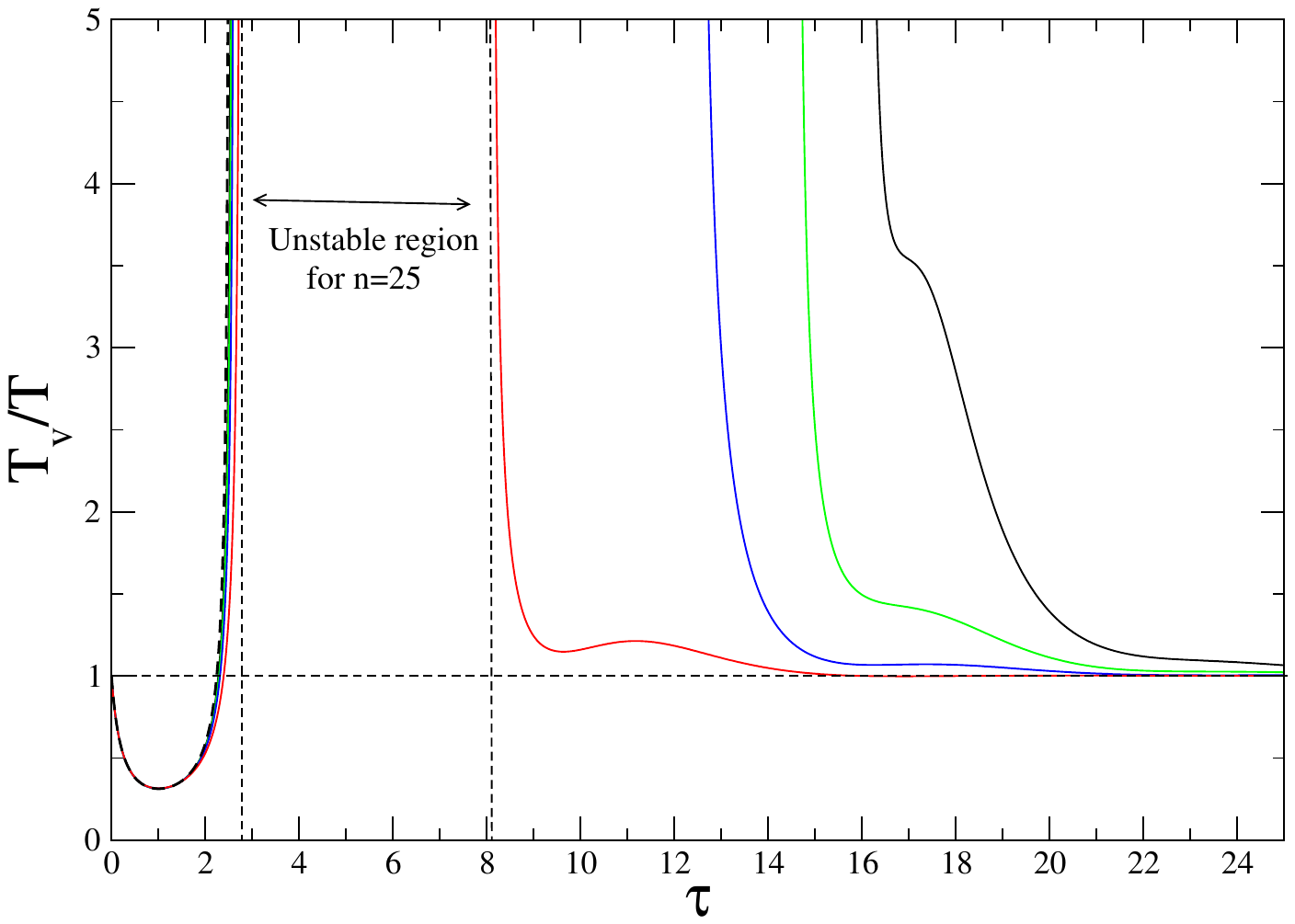}
\caption{\label{Fig1} (Color on line)  Kinetic temperature $T_v$ in the NESS as a function of $\tau$ for $Q_0=4$, $g=3.9$, and different values of $n$: $n=25$ (red solid line), $n=50$ (blue), $n=75$ (green), $n=100$ (black), and $n=\infty$ (black dashed line).  For $n=25$, the stationary state is stable for $0\le\tau <2.79$ and $\tau > 8.07$. The second stability region is displaced to larger $\tau$ as $n$ increases and no longer exists for $n=\infty$ (discrete delay). In the latter case,  the system is stable for $0\le \tau<\tau_{c,\infty}^{(1)}\simeq 2.53$ only.}
\end{center}
\end{figure}

Due to the linearity of  Eq.~(\ref{EqLmatrix}), the  probability distribution function (pdf) in the NESS  is  a  multivariate  Gaussian distribution,
\begin{align} 
\label{Eqp}
p({\bf u})=\frac{1}{\sqrt{(2\pi)^{n+2}\det  \Sigma }} e^{-\frac{1}{2}{\bf u}^T. \Sigma^{-1}.{\bf u}}\ ,
\end{align}
where $\Sigma$, the $(n+2)\times (n+2)$ covariance matrix (not to be confused with the entropy production defined above), is solution of the Lyapunov equation~\cite{G2004} 
\begin{align} 
\label{EqLya}
A \Sigma+ \Sigma A^T=- D\ ,
\end{align}
with  $D={\bf d}.{\bf d}^T$ has only one non-zero element with ${\bf d}^T=(1,0,0...0)$. To solve this equation  for large $n$,  one can use the property that the eigenvalues of $ \Sigma$, ordered as $\lambda_1\ge \lambda_2\ge...\ge\lambda_{n+2}$,  decay very fast as their order increases. Therefore, although  $\Sigma$ is  full-rank in the stability region, the numerical rank is very low.   This feature is commonly encountered in large-scale Lyapunov equations when the matrix in the right-hand side of the equation has a low rank~\cite{P2000,ASZ2002}.  Such equations typically arise in the study of the controllability and observability of linear time-invariant (LTI) dynamical systems, with  the matrix in the right-hand side having a rank equal to the number of inputs or outputs in the system~\cite{B2021}. One can then use accurate low-rank approximations to the solution and consider  large values of $n$~\cite{BS2013,S2016}.  When extremely high precision is required, one can also use the closed-form expression of $ \Sigma$ in terms of the Cauchy matrix built from the eigenvalues of $A$~\cite{P2000,ASZ2002,S2016}.

\section{Matrix Riccati equations and their properties}

\label{SecRic}

\subsection{Basic equations}

\label{SubsecRic1}

We now come to the heart of the matter and  derive the Riccati differential equations (RDEs) that will allow us to investigate the  time evolution of the moment generating functions $G_{\rm o,\lambda}(t)\equiv \langle e^{-\lambda {\cal A}_t}\rangle$ in  the steady state and to extract  the complete asymptotic form  at large time:
\begin{align}
G_{\rm o,\lambda}(t)\sim g_{\rm o}(\lambda)e^{\mu_{\rm o}(\lambda)t}\ ,
\end{align}
where  
\begin{align}
\label{EqSCGF}
\mu_{\rm o}(\lambda)\equiv \lim_{t\to \infty}\frac{1}{t}\ln G_{\rm o,\lambda}(t)
\end{align}
 is the SCGF and $g_{\rm o}(\lambda)$ is the pre-exponential factor. 

As  will be discussed in detail in the following (see Sec. \ref{SubsecIVB1}), the domain of definition of the SCGF (i.e., the values of $\lambda$ for which $\mu_{\rm o}(\lambda)<\infty$) may depend on the observable. This is due to rare but large fluctuations of the initial or final points of the stochastic trajectories that induce singularities in the pre-exponential factors. Therefore,  the knowledge of both  $\mu_{\rm o}(\lambda)$ and $g_{\rm o}(\lambda)$ is required to obtain the large-time behavior of the pdf  $P_{\rm o}(a,t)\equiv\langle \delta ({\cal A}_t-at)\rangle$ characterized by the  rate function
\begin{align}
I_{\rm o}(a)\equiv -\lim_{t\to \infty}\frac{1}{t}\ln P_{\rm o}(a,t)\ .
\end{align}
Large deviation theory tells us that if $\mu_{\rm o}(\lambda)$ exists and is differentiable, then $I_{\rm o}(a)$ is given by the Legendre-Fenchel transform~\cite{T2009,J2020} 
\begin{align}
\label{EqLF}
I_{\rm o}(a)=\max_{\lambda} [-\lambda a-\mu_{\rm o}(\lambda]\ .
\end{align}

In order to compute  $G_{\rm o,\lambda}(t)$, it is convenient to start from the restricted generating function  $G_{\rm o,\lambda}({\bf u},t\vert {\bf u}_0)=\langle e^{-\lambda {\cal A}_t}\rangle_{{\bf u}_0,{\bf u}}$, where  the initial and final configurations of the trajectories of duration $t$ are  fixed at ${\bf u}_0$ and ${\bf u}$, respectively. By definition, $G_{\rm o,\lambda}({\bf u},0\vert {\bf u}_0)=\delta ({\bf u}-{\bf u}_0)$ and 
\begin{align}
G_{\rm o,\lambda}(t)&= \int d{\bf u}_0 \: p({\bf u}_0) \int d{\bf u}\: G_{\rm o,\lambda}({\bf u},t\vert {\bf u}_0)
\end{align}
provided the integrals over ${\bf u}_0$ and ${\bf u}$ converge. This suggests considering the time evolution of 
\begin{align}
\label{EqGr} 
G^r_{\rm o,\lambda}({\bf u}_0,t)&=\int d{\bf u} \:G_{\rm o,\lambda}({\bf u},t\vert {\bf u}_0)
\end{align}
and
\begin{align} 
\label{EqGl} 
G^l_{\rm o,\lambda}({\bf u},t)&=\int d{\bf u}_0\: p({\bf u}_0)G_{\rm o,\lambda}({\bf u},t\vert {\bf u}_0)
\end{align}
separately. They satisfy the initial conditions
\begin{align}
\label{Eqinit1}
G^r_{\rm o,\lambda}({\bf u}_0,0)=1
\end{align}
and
\begin{align}
\label{Eqinit2}
G^l_{\rm o,\lambda}({\bf u},0)=p({\bf u})\ .
\end{align}
(The superscript  r and l denote ``right" and ``left", respectively.  This notation will be justified later when considering the long-time limit.)

Standard application of the Feynman-Kac formula (see, e.g., Ref.~\cite{M2005} for a pedagogical review) shows that $G^r_{\rm o,\lambda}({\bf u}_0,t)$  evolves in time according to the backward Fokker-Planck equation
\begin{align}
\label{EqFP1}
\partial_t G^r_{\rm o,\lambda}({\bf u}_0,t)={\cal L}_{\rm o,\lambda} G^r_{\rm o,\lambda}({\bf u}_0,t)\ ,
\end{align}
 where ${\cal L}_{\rm o,\lambda}$ is the so-called tilted (or biased) generator  given  by~\cite{CT2015,T2018}
\begin{align} 
\label{Eqtiltedgen}
{\cal L}_{\rm o,\lambda}={\bf F}\cdot ({\bf \nabla} -\lambda {\bf g}_{\rm o})+\frac{1}{2}(\nabla -\lambda{\bf  g}_{\rm o})\cdot D(\nabla -\lambda {\bf g}_{\rm o})\ ,
\end{align}
with ${\bf F}=A{\bf u}$ in the case at hand.  Explicit expressions of ${\cal L}_{\rm o,\lambda}$ for ${\rm o}=w,q,\sigma$ are given in Appendix A.   
 Likewise, $G^{l}_{\rm o,\lambda}({\bf u},t)$ satisfies the forward Fokker-Planck equation
\begin{align}
\label{EqFP2}
\partial_t G^{l}_{\rm o,\lambda}({\bf u},t)={\cal L}_{\rm o,\lambda}^{\dag} G^{l}_{\rm o,\lambda}({\bf u},t)\ ,
\end{align}
where ${\cal L}_{\rm o,\lambda}^\dag$ is the adjoint of ${\cal L}_{\rm o,\lambda}$. 

Solving such linear partial differential equations beyond the long-time limit requires one to determine the whole spectrum of the operators ${\cal L}_{\rm o,\lambda}$ and ${\cal L}_{\rm o,\lambda}^\dag$ and the associated eigenfunctions, which is a daunting or even  impossible task. In the present case, however,  both the drift ${\bf F}$ and the vector function ${\bf g}_{\rm o}$ depend  linearly  on  ${\bf u}$ so that we can anticipate that the solutions of Eqs.~(\ref{EqFP1}) and (\ref{EqFP2}) with  initial conditions  (\ref{Eqinit1}) and (\ref{Eqinit2}), respectively, are just multivariate Gaussians. Specifically, we show in Appendix \ref{SecB1} that  
\begin{subequations}
\label{EqSol1:subeqns}
\begin{align}
G^{r}_{\rm o,\lambda}({\bf u}_0,t)&=\exp \left(-\frac{1}{2}{\bf u}_0^TC^r_{\rm o}(\lambda,t){\bf u}_0+\int_0^t  f^{r}_{\rm o}(\lambda,t') dt' \right)\label{EqSol1:subeq1}\\
G^{l}_{\rm o,\lambda}({\bf u},t)&=\left[(2\pi)^{n+2}\det\Sigma\right]^{-1/2}\exp \left(-\frac{1}{2}{\bf u}^TC^l_{\rm o}(\lambda,t){\bf u}+\int_0^t  f^{l}_{\rm o}(\lambda,t') dt' \right) \label{EqSol1:subeq2}\ ,
\end{align}
\end{subequations}
with 
\begin{subequations}
\label{Eqmua1:subeqns}
\begin{align} 
f^{r}_{\rm o}(\lambda,t)&=-\frac{1}{2}\mbox{Tr}[D(C_{\rm o}^{r}(\lambda,t)+\lambda B_{\rm o})]\label{Eqmua1:subeq1}\\
f^{l}_{\rm o}(\lambda,t)&=-\frac{1}{2}\mbox{Tr}[D(C_{\rm o}^{l}(\lambda,t)-\lambda B_{\rm o})]-\mbox{Tr}(A)\label{Eqmua1:subeq2}\ .
\end{align}
\end{subequations}
$C^r_{\rm o}(\lambda,t)$ and $C^l_{\rm o}(\lambda,t)$ are symmetric matrices of dimension $(n+2)\times (n+2)$  which are  solutions of the matrix differential equations  
\begin{subequations}
\label{EqRic:subeqns}
\begin{align}
\dot C_{\rm o}^r(\lambda,t)&={\cal R}_{\rm o,\lambda}\big[C_{\rm o}^r(\lambda,t)\big]\label{EqRic:subeq1}\\
\dot C_{\rm o}^l(\lambda,t)&={\cal R}_{\rm o,\lambda}\big [-C_{\rm o}^l(\lambda,t)\big ]\label{EqRic:subeq2}
\end{align}
\end{subequations}
 with initial conditions
\begin{subequations}
\label{Eqinit:subeqns}
\begin{align}
C_{\rm o}^{r}(\lambda,0)&=0\label{Eqinit:subeq1}\\
C_{\rm o}^{l}(\lambda,0)&=C\label{Eqinit:subeq2}\ ,
\end{align}
\end{subequations}
respectively, where $C$ is the inverse of the covariance matrix $ \Sigma$.  
In these equations, ${\cal R}_{\rm o,\lambda}$ is a  quadratic Riccati operator acting on  a general $(n+2)\times (n+2)$ matrix $X$ as
\begin{align} 
\label{EqRa1}
{\cal R}_{\rm o,\lambda}[X]= A_{\rm o}(\lambda)^TX+XA_{\rm o}(\lambda)- X D X+K_{\rm o}(\lambda) \ ,
\end{align}
with 
 \begin{align} 
 \label{EqtildeA}
A_{\rm o}(\lambda)=A-\lambda D B_{\rm o}\ ,
\end{align}
and 
 \begin{align} 
 \label{EqKa}
K_{\rm o}(\lambda)= \lambda(A^T B_{\rm o}+B_{\rm o}^T A)-\lambda^2 B^T_{\rm o}  D  B_{\rm o}\ .
\end{align}
The  explicit expressions of the symmetric matrices $K_{\rm o}(\lambda)$  are given in Appendix~\ref{SecB2}. Furthermore,  since $D_{ij}=\delta_{i1} \delta_{j1}$ and $\mbox{Tr}(A)=-(1/Q_0+n^2/\tau)$ in the present model, Eqs.~(\ref{Eqmua1:subeqns}) become
\begin{subequations}
\label{Eqmua2:subeqns}
\begin{align} 
f^r_{\rm o}(\lambda,t)&=-\frac{1}{2}(C_{\rm o,11}^r(\lambda,t)+\lambda B_{\rm o,11})\label{Eqmua2:subeq1}\\
f^l_{\rm o}(\lambda,t)&=-\frac{1}{2}(C_{\rm o,11}^l(\lambda,t)-\lambda B_{\rm o,11})+\frac{1}{Q_0}+\frac{n^2}{\tau}\label{Eqmua2:subeq2}\ .
\end{align}
\end{subequations}
By integrating  $G^{r}_{\rm o,\lambda}({\bf u}_0,t)p({\bf u}_0)$ over ${\bf u}_0$ and $G^{l}_{\rm o,\lambda}({\bf u},t)$ over ${\bf u}$, we then obtain  two  expressions of the generating function $G_{\rm o,\lambda}(t)$,
\begin{subequations}
\label{EqDynZ:subeqns}
\begin{align} 
G_{\rm o,\lambda}(t)&=\Big[\frac{\det\big(C^r_{\rm o}(\lambda,t)+C\big)}{\det C}\Big]^{-1/2}\exp \big(\int_0^t f^{r}_{\rm o}(\lambda,t') dt' \big)\label{EqDynZ:subeq1}\\
&=\Big[\frac{\det C^l_{\rm o}(\lambda,t)}{\det C}\Big]^{-1/2}\: \exp \big(\int_0^tf^l_{\rm o}(\lambda, t')dt'\big)
\label{EqDynZ:subeq2}\ . 
\end{align}
\end{subequations}
By construction, these two expressions\footnote{Eq. (\ref{EqDynZ:subeq2}) is similar to the expression of the generating function derived  by C. Kwon {\it et al.}~\cite{KNP2011} via a path-integral method for the non-equilibrium work in linear diffusion systems. The role of the matrix $C^l(\lambda,t)$ is played by  a matrix $\tilde A(\tau,\lambda)$ that  obeys a non-linear matrix equation similar to  our Eq. (\ref{EqRic:subeq2}) (with  the matrix $\Lambda$ playing the role of the matrix $K_w(\lambda)$). Likewise, Eq. (\ref{EqDynZ:subeq1}) corresponds to Eq. (43) in the recent paper of J. du Buisson and H. Touchette~\cite{DBT2023}, with  $-(1/2)C_{\rm o}^r(\lambda,t)$  replaced by the matrix $B_k(t)$. From the dictionary $\lambda \to -k$, $A\to -M$, $B_{\rm o}\to \Gamma$, one can easily check that the RDE  (\ref{EqRic:subeq1})  corresponds to Eq.~(60) in Ref. \cite{DBT2023} (with $\Gamma^T=-\Gamma$ in Eq. (60) because $\Gamma$ is assumed to be antisymmetric).} give the same result as long as  the solutions of Eqs. (\ref{EqRic:subeqns}) exist (see the discussion in the next section).  But  for the generating function $G_{\rm o,\lambda}(t)$ to be finite  it is mandatory the matrices $C+C_{\rm o}^r(\lambda,t)$ and $C_{\rm o}^l(\lambda,t)$ be positive definite. As we shall see later in Sec. \ref{SubsecIVA}, this crucial condition is not always satisfied.   Moreover, the equivalence between Eqs.~(\ref{EqDynZ:subeq1}) and (\ref{EqDynZ:subeq2}) may result from a nontrivial mathematical mechanism.

Finally, let us note that an alternative expression  of $G_{\rm o,\lambda}(t)$ is available when the matrix $C_{\rm o}^l(\lambda,t')$ is invertible for all $t'\in [0,t]$. If so, a few manipulations detailed in Appendix \ref{SecB3}  lead to\footnote{A similar, but slightly more complicated expression of the generating function can be obtained in terms of the inverse of the matrix $C+C^r_{\rm o}(\lambda,t)$, when the latter exists.}
\begin{align} 
\label{EqZAnew}
G_{\rm o,\lambda}(t)&=\exp \left(-\frac{1}{2}\int_0^t \mbox{Tr}\big(K_{\rm o}(\lambda)\Sigma^l_{\rm o}(\lambda,t')+\lambda DB_{\rm o}\big)dt'\right)\ ,
\end{align}
where $\Sigma^l_{\rm o}(\lambda,t)$, the inverse matrix of $C_{\rm o}^l(\lambda,t)$, is solution of the complementary RDE 
\begin{align} 
\label{EqRicSigma}
\frac{\partial}{\partial t} \Sigma^l_{\rm o}(\lambda,t)= A_{\rm o}(\lambda) \Sigma^l_{\rm o}(\lambda,t)+ \Sigma^l_{\rm o}(\lambda,t)A^T_{\rm o}(\lambda)- \Sigma^l_{\rm o}(\lambda,t) K_{\rm o}(\lambda)\Sigma^l_{\rm o}(\lambda,t)+ D\ ,
\end{align}
 with initial condition
 \begin{align} 
 \Sigma^l_{\rm o}(\lambda,0)= \Sigma\ .
\end{align}
In particular, using the fact that $B_{q,11}=-2/Q_0$ and $K_q(1)=0$ [Eq. (\ref{EqKq})],  Eq. (\ref{EqZAnew}) readily yields
\begin{align} 
\label{IFTheat}
G_{q,\lambda=1}(t)&=e^{t/Q_0}\ ,
\end{align}
which is the universal IFT for the fluctuating heat in underdamped Langevin processes~\cite{RTM2016}. This result is also directly obtained  from Eq.~(\ref{EqRic:subeq1}) since the unique solution of the initial value problem is $C^{r}_q(\lambda=1,t) = 0$,  which implies  that $f^{r}_q(\lambda=1,t) = 1/Q_0$ from Eq.~(\ref{Eqmua2:subeq1}).

\subsection{Properties of the  Riccati differential equations (RDEs)}

\label{SubsecRic2}

RDEs similar to Eqs. (\ref{EqRic:subeqns}), as well as the corresponding continuous algebraic Riccati equations  (CAREs) whose solutions are stationary solutions of the RDEs (see Sec. \ref{SubsecRic3} below), appear in many  branches of applied mathematics, the  most prominent application being linear optimal control and filtering problems~\cite{B2021}. Within this framework, several important issues have been extensively discussed in the literature such as the global existence of the solutions as one varies the coefficients of the differential equation or the initial data, the convergence toward a particular solution of the  corresponding CARE as $t\to \infty$, and the mechanism of attraction (see Ref. \cite{AFIJ2003} and references therein).  In particular, it is a standard result that the solution of the RDE  (which is unique for a given initial condition) exists for $t\in [0,\infty)$ and is symmetric, positive semidefinite if the source term (in the present case, the matrix $K_{\rm o}(\lambda)$ in Eq.~(\ref{EqRa1})), the quadratic term  (i.e., the term $XDX$), and the initial condition are positive semidefinite. With additional conditions on the coefficients,  it is also proven that the solution converges monotonically to the maximal solution $X^+$ of the CARE  (to be defined later) which turns out to be the unique symmetric positive semidefinite solution\footnote{From  the viewpoint of  control theory, the important feature is that the maximal solution $X^+$ is  ``stabilizing", i.e.,  all the eigenvalues of the ``closed-loop" matrix $A_{\rm o}(\lambda)-DX^+$ have negative real parts (see e.g. Ref. \cite{K2010}).}. 

Unfortunately, it is readily seen from Eqs. (\ref{EqKw})-(\ref{EqKsigma}) that neither  $K_w(\lambda)$ nor $K_{\sigma}(\lambda)$ are positive semidefinite ($n$  eigenvalues are equal to  $0$, one is positive, and one is negative). Only $K_q(\lambda)\ge 0$ for $0\le \lambda\le 1$.  This is an essential difference with the standard situation treated in  linear quadratic (LQ) optimal control, and it  significantly complicates the present study~\cite{note3}.
 The consequences that will be explored in the rest of this paper and illustrated numerically are the following: 
 
 1) the solutions of Eqs.~(\ref{EqRic:subeqns})  may exhibit a finite-time escape phenomenon, which means that they may blow up in a finite time, 
 
  2) they may fail to converge  (i.e., the solution may oscillate), 
  
  3) they may converge to  a solution of the CARE which is not  the maximal solution, 
  
  4) the maximal solution  is not automatically positive semidefinite.  
 
 \vspace{0.2cm}
On the positive side, the Riccati operator defined by Eq. (\ref{EqRa1})  has a remarkable  property that holds  for arbitrary matrices $A$, $B_{\rm o}$, and $D$ symmetric. Indeed, if  $S$ is  a symmetric matrix, then 
\begin{align} 
\label{Eqinvar}
{\cal R}_{\rm o,\lambda}[X]={\cal R}'_{\rm o,\lambda}[X-\lambda S]\ ,
\end{align}
where   ${\cal R}'_{\rm o,\lambda}$ is the modified operator obtained by changing $B_{\rm o}$ into $B'_{\rm o}=B_{\rm o}+S$.  Therefore, when the  matrices $B_{\rm o}$ characterizing the various observables only differ  by their symmetric part $S_{\rm o}$, which is the case for the three observables ${\cal W}_t,{\cal Q}_t,\Sigma_t$ under consideration, one has
\begin{align} 
\label{Eqinvar1}
{\cal R}_{\rm o,\lambda}[X]={\cal R}_{\rm o',\lambda}[X+\lambda (S_{\rm o}-S_{\rm o'})]\ .
\end{align}
This ``invariance" property allows one to  compute all matrices $C^r_{\rm o'}(\lambda,t)$ and $C^l_{\rm o'}(\lambda,t)$ by solving  the RDEs  corresponding to a single operator ${\cal R}_{\rm o,\lambda}$ but with different initial conditions. Specifically, by letting $X_{\rm o'}(\lambda,t)$ be the solution of the RDE $\dot X_{\rm o'}={\cal R}_{\rm o,\lambda}[X_{\rm o'}]$ with initial condition 
\begin{equation}
\label{EqRqinit1}
X_{\rm o'}(\lambda,0)=\lambda (S_{\rm o'}-S_{\rm o})
\end{equation}
we obtain that
\begin{equation}
\label{EqRq1}
C^r_{\rm o'}(\lambda,t)= X_{\rm o'}(\lambda,t)-\lambda (S_{\rm o'}-S_{\rm o})
\end{equation}
and, similarly, if $X_{\rm o'}(\lambda,t)$ is the solution of the RDE $\dot X_{\rm o'}={\cal R}_{\rm o,\lambda}[-X_{\rm o'}]$ with initial condition 
\begin{equation}
\label{EqRqinit2}
X_{\rm o'}(\lambda,0)=C-\lambda (S_{\rm o'}-S_{\rm o})
\end{equation}
we have that
\begin{equation}
\label{EqRq2}
C^l_{\rm o'}(\lambda,t)= X_{\rm o'}(\lambda,t)+\lambda(S_{\rm o'}-S_{\rm o}).
\end{equation}

\subsubsection{Global existence of the solutions}

\label{SubsecRic2a}

We begin our study of the solutions of the RDEs (\ref{EqRic:subeqns}) by briefly discussing the issue of their global existence. 
First of all, we note that  the non-negativity of the diffusion matrix $D$ implies that  the solutions are bounded from above  by the  solutions of  the corresponding  Lyapunov differential equations (i.e., Eqs.~(\ref{EqRic:subeqns}) with $D=0$)~\cite{note4}. Since these upper bounds do not blow up in finite time, the  finite-time escape phenomenon, when it occurs, is  due to the absence of a lower bound and  manifests itself  by the divergence of the smallest eigenvalue of $C^r_{\rm o}(\lambda,t)$ or  $C^l_{\rm o}(\lambda,t)$ toward $-\infty$.  This is a crucial  observation because it means that this eigenvalue, which is initially positive (as $C+ C^r_{\rm o}(\lambda,0)=C^l_{\rm o}(\lambda,0)=C$), vanishes before diverging to $-\infty$. Therefore,  this  finite-time escape phenomenon is {\it always} preceded by the divergence of the generating function $G_{\rm o,\lambda}(t)$\footnote{Note that the consistency between the two expressions of the generating function [Eqs.~(\ref{EqDynZ:subeq1}) and (\ref{EqDynZ:subeq2})] imposes that the  determinants of $C^r_{\rm o}(\lambda,t)+C$ and $C^l_{\rm o}(\lambda,t)$ vanish simultaneously, as will be observed later in the numerical examples (see, e.g., Fig.~\ref{Fig9} in Appendix \ref{AppendF}).}.  
 
It is in general impossible to predict from the outset the global existence of the solutions of the RDEs and of the generating function. However, if $\lambda\in[0,1]$, we can take advantage of the semi-positiveness of the source term $K_q(\lambda)$  to obtain some  valuable  results.
 To this end, it suffices  to  consider the ``right" matrices $C^r_{\rm o}(\lambda,t)$ since  the condition $C+ C^r_{\rm o}(\lambda,t)>0$ for all $t\in[0,\infty)$ implies  that $C^l_{\rm o}(\lambda,t)>0$ (otherwise, the two  expressions   (\ref{EqDynZ:subeq1}) and (\ref{EqDynZ:subeq2}) of $G_{\rm o,\lambda}(t)$ would not be consistent). 

The case of  the generating function $G_{q,\lambda}(t)$ is straightforward. Indeed, since $C^r_q(\lambda,0)=0$ and the RDE  $\dot X ={\cal R}_{q,\lambda}[X]$  has all the properties of the RDEs  encountered in LQ optimal control~\cite{AFIJ2003}, we know that the  matrix  $C^r_q(\lambda,t)$ exists and is positive semidefinite on $[0,\infty)$. Moreover,  $\dot C^r_q(\lambda,0)=K_q(\lambda)\ge 0$   and thus $C^r_q(\lambda,t)$ is  monotonically non-decreasing~\cite{note80}.
 Hence $C^r_q(\lambda,t)+C>0$ and we conclude that  $G_{q,\lambda}(t)$ is always finite.
 
Reaching a conclusion about the existence of $G_{w,\lambda}(t)$ and $G_{\sigma,\lambda}(t)$ is less trivial  and, as shown in Appendix \ref{AppendC},  only partial results can be obtained.
 
 \subsubsection{Solutions of the RDEs and asymptotic behavior}

\label{SubsecRic2b}

Various  methods for solving RDEs are available in the literature, including for large-scale problems~\cite{BM2004}. Here, we will use the classical approach that consists in transforming each  quadratic differential equation into a linear system of first-order Hamiltonian differential equations of double size~\cite{AFIJ2003}.  We can then obtain a closed-form representation of the solution which is suitable for analyzing the asymptotic behavior  and the dependence on the initial condition. Although this procedure is standard, it is worthwhile to replicate the derivation in the case at hand. 

Consider first Eq. (\ref{EqRic:subeq1}) with initial condition  (\ref{Eqinit:subeq1}). It can be checked by direct substitution that the solution  can be expressed as
 \begin{align}
\label{Eqsol}
C^{r}_{{\rm o}}(\lambda,t)&=V^{r}_{\rm o}(\lambda,t)U^{r}_{\rm o}(\lambda,t)^{-1}\ ,
\end{align}
where the matrices $U^{r}_{\rm o}(\lambda,t)$ and  $V^{r}_{\rm o}(\lambda,t)$ are solutions of the linear  system
\begin{align}
\label{EqSys}
\begin{bmatrix}
\dot U^{r}_{\rm o}(\lambda,t)\\
\dot V^{r}_{\rm o}(\lambda,t)
\end{bmatrix}
=H_{\rm o}^r(\lambda)
\begin{bmatrix}
U^{r}_{\rm o}(\lambda,t)\\
V^{r}_{\rm o}(\lambda,t)
\end{bmatrix}\ ,
\end{align}
with 
\begin{align}
\label{EqHr}
H_{\rm o}^r(\lambda)=
\begin{bmatrix}
-A_{\rm o}(\lambda)&D\\
K_{\rm o}(\lambda)& A_{\rm o}^T(\lambda)&
\end{bmatrix}\ ,
\end{align}
and  initial condition 
\begin{align}
\begin{bmatrix}
 U^r_{\rm o}(\lambda,0)\\
 V^r_{\rm o}(\lambda,0)
\end{bmatrix}
&=\begin{bmatrix}
I_{n+2}\\
0
\end{bmatrix}\ .
\end{align}
As a consequence,
\begin{align}
\label{Eqsol1}
\begin{bmatrix}
 U^{r}_{\rm o}(\lambda,t)\\
V^{r}_{\rm o}(\lambda,t)
\end{bmatrix}
=e^{H^r_{\rm o}(\lambda)t}
\begin{bmatrix}
I_{n+2}\\
0
\end{bmatrix}\ .
\end{align}
The existence of the solution of  Eq. (\ref{EqRic:subeq1}) for all $t'\in [0,t]$ ensures that the corresponding matrix $ U^{r}_{\rm o}(\lambda,t')$  is invertible in this interval.  Conversely, if $ U^{r}_{\rm o}(\lambda,t')$ is nonsingular for all  $t'\in[0,t]$, then $C^r_{\rm o}(\lambda,t')$  exists in the same interval and is given by Eq. (\ref{Eqsol}).

 A similar transformation holds for the solution $C^l_{\rm o}(\lambda,t)$ of  Eq. (\ref{EqRic:subeq2}), with $ U^r_{\rm o}(\lambda,t),  V^r_{\rm o}(\lambda,t)$ replaced by $ U^l_{\rm o}(\lambda,t),  V^l_{\rm o}(\lambda,t)$ and Eqs. (\ref{Eqsol1}) replaced by
 \begin{align}
\begin{bmatrix}
 U^{l}_{\rm o}(\lambda,t)\\
V^{l}_{\rm o}(\lambda,t)
\end{bmatrix}
=e^{H^l_{\rm o}(\lambda)t}
\begin{bmatrix}
I_{n+2}\\
C
\end{bmatrix}\ .
\end{align}
with 
\begin{align}
\label{EqHl}
H^l_{\rm o}(\lambda)=
\begin{bmatrix}
 A_{\rm o}(\lambda)&D\\
  K_{\rm o}(\lambda)&- A_{\rm o}^T(\lambda)&
\end{bmatrix}\ .
\end{align}
 The  $2(n+2)\times 2(n+2)$ matrices $H^r_{\rm o}(\lambda)$ and $H^l_{\rm o}(\lambda)$ are Hamiltonian matrices which play a central role in the forthcoming analysis\footnote{We recall that a real Hamiltonian matrix $H$ of dimension $2(n+2)\times 2(n+2)$ satisfies the equation $\left[\begin{smallmatrix}
0&-I_{n+2} \\
  I_{n+2}&0
\end{smallmatrix}\right] H=-H^T \left[\begin{smallmatrix}
0&-I_{n+2} \\
  I_{n+2}&0
\end{smallmatrix}\right]$,
where $I_{n+2}$ is the $(n+2)\times (n+2)$ identity matrix.  The eigenvalues of $H$ then come in quadruples: if $s\in \mathbb{C}$ is an eigenvalue,  so are $\bar s,-s$ and $-\bar s$, where an overline indicates complex conjugation.}. Their spectral properties are investigated in Appendix \ref{AppendD}. Observe in particular that $H^r_{\rm o}(\lambda)$ and $H^l_{\rm o}(\lambda)$ are invertible and that  the eigenvalue spectrum only depends  on the anti-symmetric part of the matrices $B_{\rm o}$ (a property that is  shared  by all linear current-type observables). In  the present case, the anti-symmetric part is the same for the three observables [Eq.~(\ref{EqBa})], and owing to the fact that $D_{ij}=\delta_{1i}\delta_{1j}$, the characteristic polynomial of the Hamiltonian matrices is shown in Appendix~\ref{AppendD1} to be given by
\begin{align}
\label{EqpH}
 (-1)^np_H(\lambda,s)=p_A(s)p_A(-s) 
-\frac{2\lambda g}{Q_0^2}(\frac{n}{\tau})^{2n}\: s [(1-\frac{s\tau}{n})^n-(1+\frac{s\tau}{n})^n]\ ,
\end{align}
where $p_A(s)$, the characteristic polynomial of $A$, is given by Eq. (\ref{Eqchar}).   

For the sake of simplicity,  we will assume in the following that  the matrices $H^r_{\rm o}(\lambda)$ and $H^l_{\rm o}(\lambda)$  are  diagonalizable.  This is not an essential assumption but it simplifies the presentation as we can then avoid to deal with the more complicated Jordan forms of the Hamiltonian matrices~\cite{AFIJ2003}. More important is the fact that the values of $\lambda$ will be restricted to a certain open interval  ${\cal D}_{H}=(\lambda_{\min},\lambda_{\max})$ for which $H^r_{\rm o}(\lambda)$ and $H^l_{\rm o}(\lambda)$ have no purely imaginary eigenvalues\footnote{One may have $\lambda_{\min}=-\infty$ but $\lambda_{\max}$ is  positive and finite.}. As will be shown in Sec.~\ref{SubsecIVB3}, this restriction is justified by the fact that the values of $\lambda$ outside ${\cal D}_H$ are irrelevant for the determination of the rate function $I(a)$.  
 
 The condition $\lambda\in {\cal D}_H$ has two significant consequences:  
 
 $\bullet$ The  Hamiltonian matrices are {\it dichotomically separable}~\cite{AFIJ2003}, with $n+2$  eigenvalues $s^+_1(\lambda), s^+_2(\lambda),...s^+_{n+2}(\lambda)$ having a positive   real part and $n+2$ eigenvalues $s_1^-(\lambda)=- s_1^+(\lambda), s_2^-(\lambda)=- s_2^+(\lambda),...s_{n+2}^-(\lambda)=-s^+_{n+2}(\lambda)$ having a negative real part (the eigenvalues are counted with their multiplicities and the ordering  is arbitrary unless otherwise specified).  
 
 $\bullet$ Exact mathematical statements ensure that the corresponding algebraic Riccati equations have real symmetric solutions (see the discussion in the next section).
   
 Let us consider the matrix $H_{\rm o}^r(\lambda)$.  As a result of the above assumptions, we  can introduce a basis-change  matrix  $W^r_{\rm o}(\lambda)$ that diagonalizes $H_{\rm o}^r(\lambda)$ such that 
\begin{align}
\label{EqHdiag}
H_{\rm o}^r(\lambda)=W^r_{\rm o}(\lambda)\
\begin{bmatrix}
 J(\lambda)&0\\
0&-J(\lambda)&
\end{bmatrix}
[W^r_{\rm o}(\lambda)]^{-1}\ ,
\end{align}
where  $J(\lambda)=\mbox{diag}\big(s^+_1(\lambda), s^+_2(\lambda),...s^+_{n+2}(\lambda)\big)$. Eq. (\ref{Eqsol1})  then   becomes
\begin{align}
\label{Eqsol2}
\begin{bmatrix}
 U^r_{\rm o}(\lambda,t)\\
V^r_{\rm o}(\lambda,t)
\end{bmatrix}
=W^r_{\rm o}(\lambda)\begin{bmatrix}
e^{ J(\lambda) t}&0\\
0&e^{-J(\lambda) t}&
\end{bmatrix}
[W^r_{\rm o}(\lambda)]^{-1}\begin{bmatrix}
I_{n+2}\\
0
\end{bmatrix}
\ .
\end{align}
We next  partition the matrix $W^r_{\rm o}(\lambda)$ into $4$ blocks of size $(n+2)\times(n+2)$,
\begin{align}
\label{EqWr}
W^r_{\rm o}(\lambda)=
\begin{bmatrix}
 W^{r,11}_{\rm o}(\lambda)& W^{r,12}_{\rm o}(\lambda)\\
 W^{r,21}_{\rm o}(\lambda)& W^{r,22}_{\rm o}(\lambda)&
\end{bmatrix}\ ,
\end{align}
where  the first $n+2$ columns are the eigenvectors  relative to the eigenvalues $s_i^+(\lambda)$ and the remaining columns are the eigenvectors relative to the eigenvalues $s_i^-(\lambda)$.

 Now, suppose that the submatrix $ W^{r,22}_{\rm o}(\lambda)$ is nonsingular, so that  the matrix 
 \begin{align}
\label{EqTr}
T^r_{\rm o}(\lambda)=&-[W^{r,22}_{\rm o}(\lambda)]^{-1}W^{r,21}_{\rm o}(\lambda)
\end{align}
exists. Simple manipulations, similar to those performed in Refs. \cite{K1973,FJ1996} and detailed in Appendix  \ref{AppendE}, then  lead to a  representation of the solution of the RDE  (\ref{EqRic:subeq1})  that only involves negative exponentials\footnote{We leave to the reader the proof that Eq. (\ref{Eqsol4})  does not depend on the choice of the basis-change matrix $W^r_{\rm o}(\lambda)$.}:
\begin{align}
\label{Eqsol4}
C^r_{\rm o}(\lambda,t)&=[W^{r,21}_{\rm o}(\lambda)+ W^{r,22}_{\rm o}(\lambda)P^r_{\rm o}(\lambda,t)][W^{r,11}_{\rm o}(\lambda)+W^{r,12}_{\rm o}(\lambda)P^r_{\rm o}(\lambda,t)]^{-1}\ ,
\end{align}
where 
\begin{align}
\label{EqPr}
P^r_{\rm o}(\lambda,t)=e^{-J(\lambda) t}T^r_{\rm o}(\lambda)e^{-J(\lambda) t}\ .
\end{align}
Two important features of the solution  are revealed by Eq. (\ref{Eqsol4}):

$\bullet$ $C^r_{\rm o}(\lambda,t)$ diverges at time $t$  if  the matrix $W^{r,11}_{\rm o}(\lambda)+W^{r,12}_{\rm o}(\lambda)P^r_{\rm o}(\lambda,t)$ is singular. Finite-time singularities in the solution are thus poles corresponding to  $\det[W^{r,11}_{\rm o}(\lambda)+W^{r,12}_{\rm o}(\lambda)P^r_{\rm o}(\lambda,t)]=0$. (We recall that the generating function $G_{\rm o,\lambda}(t)$ diverges {\it before} the solution or the RDE blows up.)

$\bullet$ $C^r_{\rm o}(\lambda,t)$ converges towards the matrix  $\hat C^{r,+}_{\rm o}(\lambda)\equiv W^{r,21}_{\rm o}(\lambda)[W^{r,11}_{\rm o}(\lambda)]^{-1}$ as $t\to \infty$. (This is true even if finite-time singularities are present\footnote{In fact, solving the linear system of ODEs (\ref{EqSys}) instead of the original  RDE is a practical method to bypass singularities, which is needed in certain applications, in particular for boundary-value problems~\cite{AMR1988}.}.)

As will be shown in the next subsection,  $\hat C^{r,+}_{\rm o}(\lambda)$ is the so-called ``maximal"  (real symmetric) solution of the corresponding algebraic equation (\ref{EqCARE:subeq1}) and its existence is certified~\cite{LR1995} (which means that the  matrix $W^{r,11}_{\rm o}(\lambda)$ is nonsingular). On the other hand, the assumption that $W^{r,22}_{\rm o}(\lambda)$  is nonsingular, which is required for  Eq. (\ref{Eqsol4}) to be meaningful, may not be satisfied  for some values of $\lambda$. In other words,  the condition $\det W^{r,22}_{\rm o}(\lambda)\ne 0$   ensures that the initial matrix $ C^{r}_{\rm o}(\lambda,0)=0$ belongs to the basin of attraction of  $\hat C^{r,+}_{\rm o}(\lambda)$.  Otherwise, Eq. (\ref{Eqsol4}) is no longer valid and the solution of the RDE  (\ref{EqRic:subeq1}) goes to another limit or may fail to converge (i.e., oscillates), as will be discussed in Sec. IV C.

Likewise, the solution of  Eq. (\ref{EqRic:subeq2}) can be represented as 
\begin{align}
\label{Eqsol4left}
C^l_{\rm o}(\lambda,t)&=[W^{l,21}_{\rm o}(\lambda)+ W^{l,22}_{\rm o}(\lambda)P^l_{\rm o}(\lambda,t)][W^{l,11}_{\rm o}(\lambda)+W^{l,12}_{\rm o}(\lambda)P^l_{\rm o}(\lambda,t)]^{-1}\ ,
\end{align}
where 
\begin{align}
\label{EqPl}
P^l_{\rm o}(\lambda,t)=e^{-J(\lambda) t}T^l_{\rm o}(\lambda)e^{-J(\lambda) t}\ ,
\end{align}
and 
\begin{align}
\label{EqTl}
T^l_{\rm o}(\lambda)=&-[W^{l,22}_{\rm o}(\lambda)-CW^{l,12}_{\rm o}(\lambda)]^{-1}[W^{l,21}_{\rm o}(\lambda)-CW^{l,11}_{\rm o}(\lambda)]\ .
\end{align}
This requires that the matrix $W^{l,22}_{\rm o}(\lambda)-CW^{l,12}_{\rm o}(\lambda)$ is nonsingular. If true, $C^l_{\rm o}(\lambda,t)$ converges  asymptotically toward the matrix $\hat C^{l,+}_{\rm o}(\lambda)\equiv W^{l,21}_{\rm o}(\lambda)[W^{l,11}_{\rm o}(\lambda)]^{-1}$ which is  the maximal (real symmetric) solution of the corresponding CARE and whose existence is also guaranteed (implying that $W^{l,11}_{\rm o}(\lambda)$ is nonsingular). 
Moreover, it is easily seen from the  structure of the Hamiltonian matrices $H_{\rm o}^r(\lambda)$ and $H_{\rm o}^l(\lambda)$ that 
\begin{align}
\label{EqWrWl}
\begin{bmatrix}
 W_{\rm o}^{l,11}&W_{\rm o}^{l,12}\\
W_{\rm o}^{l,21}&W_{\rm o}^{l,22}&
\end{bmatrix}
=\begin{bmatrix}
 -W_{\rm o}^{r,12}&-W_{\rm o}^{r,11}\\
W_{\rm o}^{r,22}&W_{\rm o}^{r,21}&
\end{bmatrix}\ .
\end{align}
As a result,  $\hat C^{l,+}_{\rm o}(\lambda)=-W^{r,22}_{\rm o}(\lambda)[W^{r,12}_{\rm o}(\lambda)]^{-1}$.

For future reference, it is instructive to rewrite the  conditions of convergence towards $\hat C^{r,+}_{\rm o}(\lambda)$ and $\hat C^{l,+}_{\rm o}(\lambda)$ as 
\begin{align}
\label{EqConv1}
\det W^{r,22}_{\rm o}(\lambda)&=\det W^{l,21}_{\rm o}(\lambda)=\det \hat C^{l,+}_{\rm o}(\lambda)\:\det W^{l,11}_{\rm o}(\lambda)\ne 0\ ,
\end{align}
and
\begin{align}
\label{EqConv2}
\det [W^{l,22}_{\rm o}(\lambda)-CW^{l,12}_{\rm o}(\lambda)]=\det [W^{r,21}_{\rm o}(\lambda)+CW^{r,11}_{\rm o}(\lambda)]=\det [\hat C^{r,+}_{\rm o}(\lambda)+C]\det W^{r,11}_{\rm o}(\lambda)\ne 0\ .
\end{align}
As $\det W^{l,11}_{\rm o}(\lambda)$ and $\det W^{r,11}_{\rm o}(\lambda)$ are $\neq 0$, this implies the following  equivalences: 
\begin{subequations}
\label{EqEquiv:subeqns}
\begin{align}
&\det \hat C^{l,+}_{\rm o}(\lambda)\ne0  \: \: \Leftrightarrow \: \:C^r_{\rm o}(\lambda,t)\to \hat C_{\rm o}^{r,+}(\lambda)\label{EqEquiv:subeq1}\\
&\det [\hat C^{r,+}_{\rm o}(\lambda)+C]\ne0  \: \: \Leftrightarrow \: \: C^l_{\rm o}(\lambda,t)\to \hat C_{\rm o}^{l,+}(\lambda)  \label{EqEquiv:subeq2}
\end{align}
\end{subequations}
These dual relations will be used again and again in the following. They are one of the main reasons for which it is fruitful  to study the generating functions $G^r_{\rm o,\lambda}({\bf u}_0,t)$  and $G^l_{\rm o,\lambda}({\bf u},t)$ together.

Finally, we stress that the representations  (\ref{Eqsol4}) and (\ref{Eqsol4left}) of the solutions of the RDEs  (\ref{EqRic:subeqns}) not only reveal the most significant features of the solutions but are also  useful  for  numerical calculations. Indeed, as shown in Appendix \ref{AppendD2}, we have explicit expressions of the basis-change matrices  $W^r_{w}(\lambda)$ and $W^l_{w}(\lambda)$ as a function of the eigenvalues of the Hamiltonian matrices.  We can then only consider the RDE corresponding the operator ${\cal R}_{w,\lambda}$ and  compute all matrices $C^r_{\rm o}(\lambda,t)$ and $C^l_{\rm o}(\lambda,t)$ by changing the initial conditions  and using relations  (\ref{EqRq1}) and  (\ref{EqRq2}).

\subsection{Fixed points of the Riccati flows} 
 
\label{SubsecRic3}

How does one know that the matrices $\hat C^{r,+}_{\rm o}(\lambda)$ and $\hat C^{l,+}_{\rm o}(\lambda)$ exist and what are their properties? To answer these  questions, we now consider the  stationary versions of the RDEs  (\ref{EqRic:subeqns}), 
\begin{subequations}
\label{EqCARE:subeqns}
\begin{align}
&{\cal R}_{\rm o,\lambda}[\hat C^r_{\rm o}(\lambda)]=0\label{EqCARE:subeq1}\\
&{\cal R}_{\rm o,\lambda}[-\hat C^l_{\rm o}(\lambda)]=0\label{EqCARE:subeq2}\ ,
\end{align}
\end{subequations}
which are referred to as continuous-time algebraic Riccati equations (CAREs) in the context of optimal control.  These equations may have no solutions at all or multiple solutions, including complex and non-symmetric ones, and we  first  recall how these solutions, in particular real symmetric ones, can be built. 

It is known that there is a one-to-one correspondence between the solutions of  a CARE and certain invariant subspaces of the associated Hamiltonian matrix  (see  Refs. \cite{BLW1991,LR1995,BIM2012} for reviews). Let us consider for instance  Eq. (\ref{EqCARE:subeq1}) and  denote a solution by $\hat C^{r,(\alpha)}_{\rm o}(\lambda)$ (then $-\hat C^{r,(\alpha)}_{\rm o}(\lambda)$ is   a solution of Eq. (\ref{EqCARE:subeq2})).
A direct calculation yields
\begin{align}
\label{Eqgraph}
H^r_{\rm o}(\lambda)\begin{bmatrix}
 I_{n+2}\\
 \hat C^{r,(\alpha)}_{\rm o}(\lambda)&
\end{bmatrix}=\begin{bmatrix}
 I_{n+2}\\
 \hat C^{r,(\alpha)}_{\rm o}(\lambda)&
\end{bmatrix}[-A_{\rm o}(\lambda)+D\hat C^{r,(\alpha)}_{\rm o}(\lambda)]\ ,
\end{align}
 which shows that  the columns of  the matrix $\left[\begin{smallmatrix}
I_{n+2}\\
\hat C^{r,(\alpha)}_{\rm o}(\lambda)&
\end{smallmatrix}\right]$  span a graph invariant subspace\footnote{The graph of  a matrix $X$ is defined  as the $(n+2)$-dimensional subspace 
\begin{align}
G(X)=\mbox{Im}\nonumber
\begin{bmatrix}
 I_{n+2}\\
 X&
\end{bmatrix} \in \mathbb{C}^{2(n+2)}
\end{align}
where   $\mbox{Im} $ denotes the image or column space of a matrix~\cite{LR1995}. A subspace of $\mathbb{C}^{2(n+2)}$ is  called a graph subspace if it has the form  $G(X)$ for some $X$.} of $H^r_{\rm o}(\lambda)$ and the eigenvalues of $-A_{\rm o}(\lambda)+D\hat C^{r,(\alpha)}_{\rm o}(\lambda)$ are a subset of the eigenvalues of $H^r_{\rm o}(\lambda)$.
This simple fact leads to the following characterization of the solutions~\cite{note20}: Each solution  $\hat C^{r,(\alpha)}_{\rm o}(\lambda)$  corresponds to a set ${\cal S}_{\lambda}^{(\alpha)}$ of $n+2$ eigenvalues  $s_{\alpha_1},s_{\alpha_2},...s_{\alpha_{n+2}}$ of $H^r_{\rm o}(\lambda)$ [specifically, the eigenvalues of $-A_{\rm o}(\lambda)+D\hat C^{r,(\alpha)}_{\rm o}(\lambda)$] and  $n+2$ associated eigenvectors  ${\bf e}^r_{\alpha_j}=\left[\begin{smallmatrix}
{\bf y}^r_{\alpha_j} \\
 {\bf z}^r_{\alpha_j}
\end{smallmatrix}\right]$  (with ${\bf y}^r_{\alpha_j}, {\bf z}^r_{\alpha_j}\in \mathbb{C}^{n+2}$),  such that 
\begin{align}
\label{EqXralpha} 
\hat C^{r,(\alpha)}_{\rm o}(\lambda)=Z^{r,(\alpha)}_{\rm o}(\lambda)[Y^{r,(\alpha)}_{\rm o}(\lambda)]^{-1}\ ,
\end{align}
where $Y^{r,(\alpha)}_{\rm o}(\lambda)=[{\bf y}^r_{\alpha_1},{\bf y}^r_{\alpha_2},...{\bf y}^r_{\alpha_{n+2}}]$ and $Z^{r,(\alpha)}_{\rm o}(\lambda)=[{\bf z}^r_{\alpha_1},{\bf z}^r_{\alpha_2},...{\bf z}^r_{\alpha_{n+2}}]$.  The invertibility of the $(n+2)\times (n+2)$ matrix $Y^{r,(\alpha)}_{\rm o}(\lambda)$ is the condition ensuring that the solution $\hat C^{r,(\alpha)}_{\rm o}(\lambda)$ exists, and conversely.
The solutions of  Eq. (\ref{EqCARE:subeq2}) can be characterized in the same manner, with the index $r$ replaced by $l$ and $A_{\rm o}(\lambda)$ replaced by $-A_{\rm o}(\lambda)$ in Eq. (\ref{Eqgraph}). (Recall  that $H_{\rm o}^r(\lambda)$ and $H_{\rm o}^r(\lambda)$ have the same eigenvalue spectrum so that $\hat C^{r,(\alpha)}_{\rm o}(\lambda)$ and $\hat C^{l,(\alpha)}_{\rm o}(\lambda)$ correspond  to the same subset ${\cal S}_{\lambda}^{(\alpha)}$ of eigenvalues of $H_{\rm o}^r(\lambda)$ and $H_{\rm o}^l(\lambda)$.)

Interestingly,  the solutions of Eqs. (\ref{EqCARE:subeqns}) for an observable $\rm o$' can be readily obtained from the solutions for the observable $\rm o$. This results from the  invariance property (\ref{Eqinvar1}) of the Riccati operator ${\cal R}_{\rm o,\lambda}$, which yields     
\begin{subequations}
\label{Eq:rel}
 \begin{align}
\hat C^{r,(\alpha)}_{\rm o'}(\lambda)=\hat C^{r,(\alpha)}_{\rm o}(\lambda)+\lambda(S_{\rm o}-S_{\rm o'})=\hat C^{r,(\alpha)}_{\rm o}(\lambda)+\lambda(B_{\rm o}-B_{\rm o'})\label{Eq:rel1}\\
\hat C^{l,(\alpha)}_{\rm o'}(\lambda)=\hat C^{l,(\alpha)}_{\rm o}(\lambda)-\lambda(S_{\rm o}-S_{\rm o'})=\hat C^{l,(\alpha)}_{\rm o}(\lambda)-\lambda(B_{\rm o}-B_{\rm o'})\label{Eq:rel2}\ .
\end{align}
  \end{subequations}
As a result, 
\begin{align}
\label{Eqrel2leftright}
\hat C^{r,(\alpha)}_{\rm o'}(\lambda)+\hat C^{l,(\alpha)}_{\rm o'}(\lambda)=\hat C^{r,(\alpha)}_{\rm o}(\lambda)+\hat C^{l,(\alpha)}_{\rm o}(\lambda)\ .
\end{align}
Note that $\hat C^{r,(\alpha)}_{\rm o'}(\lambda)$ and $\hat C^{l,(\alpha)}_{\rm o'}(\lambda)$  correspond  to the same set ${\cal S}_{\lambda}^{(\alpha)}$ of eigenvalues as $\hat C^{r,(\alpha)}_{\rm o}(\lambda)$ and $\hat C^{l,(\alpha)}_{\rm o}(\lambda)$, which is the set of eigenvalues of  the  matrices $-A_{\rm o}(\lambda)+D\hat C^{r,(\alpha)}_{\rm o}(\lambda)$ and $A_{\rm o}(\lambda)+D\hat C^{l,(\alpha)}_{\rm o}(\lambda)$.  Indeed,  from the definition of $A_{\rm o}(\lambda)$ [Eq. (\ref{EqtildeA})], one has 
\begin{subequations}
\label{Eqrel3}
\begin{align}
-A_{\rm o}(\lambda)+D\hat C^{r,(\alpha)}_{\rm o}(\lambda)&=-A+\lambda DB_{\rm o}+D[\hat C^{r,(\alpha)}_{\rm o'}(\lambda)-\lambda (B_{\rm o}-B_{\rm o'}]=-A_{\rm o'}(\lambda)+D\hat C^{r,(\alpha)}_{\rm o'}(\lambda)\label{Eqrel31}\\
A_{\rm o}(\lambda)+D\hat C^{l,(\alpha)}_{\rm o}(\lambda)&=A-\lambda DB_{\rm o}+D[\hat C^{l,(\alpha)}_{\rm o'}(\lambda)+\lambda (B_{\rm o}-B_{\rm o'})]=A_{\rm o'}(\lambda)+D\hat C^{l,(\alpha)}_{\rm o'}(\lambda)\label{Eqrel32}\ .
\end{align} 
\end{subequations}
Another interesting consequence of Eqs. (\ref{Eq:rel}) is that 
\begin{subequations}
\label{Eq:rel4}
 \begin{align}
\hat C^{r,(\alpha)}_{\rm o}(\lambda)&=\hat C^{r,(\alpha)}_{\rm o,antisym}(\lambda)-\lambda B_{\rm o,sym}\label{Eq:rel41}\\
\hat C^{l,(\alpha)}_{\rm o}(\lambda)&=\hat C^{l,(\alpha)}_{\rm o,antisym}(\lambda)+\lambda B_{\rm o,sym}\label{Eq:rel42}\ ,
\end{align}
  \end{subequations}
where $B_{\rm o,sym}\equiv(B_{\rm o}+B^T_{\rm o})/2$ is the symmetric part of the matrix $B_{\rm o}$ and $\hat C^{r,(\alpha)}_{\rm o,antisym}(\lambda)$ and $\hat C^{l,(\alpha)}_{\rm o,antisym}(\lambda)$ are the solutions of the CAREs  (\ref{EqCARE:subeqns}) associated with  $B_{\rm o,antisym}\equiv(B_{\rm o}-B^T_{\rm o})/2$.  Consequently,
\begin{align}
\label{Eq:rel5}
\hat C^{r,(\alpha)}_{\rm o}(\lambda)+\hat C^{l,(\alpha)}_{\rm o}(\lambda)=\hat C^{r,(\alpha)}_{\rm o,antisym}(\lambda)+\hat C^{l,(\alpha)}_{\rm o,antisym}(\lambda)\ .
\end{align}

In accordance with Eqs. (\ref{Eqmua1:subeqns}), we  associate to each solutions $\hat C^{r,(\alpha)}_{\rm o}(\lambda)$ and $\hat C^{l,(\alpha)}_{\rm o}(\lambda)$ of the CAREs (\ref{EqCARE:subeqns}) the scalar functions 
\begin{subequations}
\label{Eqmu:subeqns}
\begin{align}
f^{r,(\alpha)}_{\rm o}(\lambda)&=-\frac{1}{2}\mbox{Tr}[D(\hat C^{r,(\alpha)}_{\rm o}(\lambda)+\lambda B_{\rm o})]\label{Eqmu:subeq1}\\
f^{l,(\alpha)}_{\rm o}(\lambda)&=-\frac{1}{2}\mbox{Tr}[D(\hat C^{l,(\alpha)}_{\rm o}(\lambda)-\lambda B_{\rm o})]-\mbox{Tr}(A)\label{Eqmu:subeq2}\ .
\end{align}
\end{subequations}
These two functions are actually equal and independent of the observables. 
This follows from the fact that  ${\cal S}_{\lambda}^{(\alpha)}=\{s_{\alpha_i}\}_{i=1}^{n+2}$ is the spectrum of both $-A_{\rm o}(\lambda)+D\hat C^{r,(\alpha)}_{\rm o}(\lambda)$ and  $A_{\rm o}(\lambda)+D\hat C^{l,(\alpha)}_{\rm o}(\lambda)$, whatever $\rm o$, as  we just noticed. As a consequence, 
\begin{align}
\sum_{i=1}^{n+2} s_{\alpha_i}(\lambda)&=\mbox{Tr}[-A_{\rm o}(\lambda)+D\hat C^{r,(\alpha)}_{\rm o}(\lambda)]=\mbox{Tr}[A_{\rm o}(\lambda)+D\hat C^{l,(\alpha)}_{\rm o}(\lambda)]\ .
\end{align}
Using the definition of $A_{\rm o}(\lambda)$ [Eq. (\ref{EqtildeA})], this yields $f^{r,(\alpha)}_{\rm o}(\lambda)=f^{l,(\alpha)}_{\rm o}(\lambda)=f^{(\alpha)}(\lambda)$ with
\begin{align} 
\label{Eqmulmur}
f^{(\alpha)}(\lambda)=-\frac{1}{2}[\mbox{Tr}(A)+\sum_{i=1}^{n+2} s_{\alpha_i}(\lambda)]\ .
\end{align}

So far, we have considered all solutions of Eq. (\ref{EqCARE:subeqns}) (assuming that they exist). However,  in the present context we are only interested  in {\it real symmetric} solutions. It turns out that the existence of such solutions is ensured due to the following two properties of the matrices involved in the Riccati operator ${\cal R}_{\rm o,\lambda}$ defined by Eq. (\ref{EqRa1})~\cite{note6}:  

i)  the matrix $D$ is positive semidefinite, 

 ii) the pair of matrices $(A_{\rm o}(\lambda),D)$ is {\it controllable}\footnote{A standard result in control theory is that a pair $(A,B)$, where $A$ is a $n\times n$ matrix and $B$ is a $n\times m$ matrix,  is  {\it controllable} if the rank of the $n\times nm$ matrix $[B,AB,A^2B,...A^{n-1}B]$ is equal to $n$~\cite{B2021}. It is easy to show that  the pair $(A_{\rm o}(\lambda),D)$ satisfies this crucial property by using  the equivalent PBH (Popov-Belevitch-Hautus)  test which states that  there must be no nonzero left eigenvector ${\bf v}^T$ of $A_{\rm o}(\lambda)$ such that ${\bf v}^T\cdot D=0$. Since $D_{ij}=\delta_{i1}\delta_{j1}$,  the second condition imposes that ${\bf v}^T=(0,v_2,v_3,...v_{n+2})^T$. The first condition then yields $v_2=v_3...=v_{n+2}=0$, as can be readily checked by  inspection.}.  
 
 These properties  also ensure that~\cite{note7} :

$\bullet$  Each real symmetric solution $\hat C^{r,(\alpha)}_{\rm o}(\lambda)$ or $\hat C^{l,(\alpha)}_{\rm o}(\lambda)$ corresponds to a set ${\cal S}^{(\alpha)}_{\lambda}$ of eigenvalues of $H^r_{\rm o}(\lambda)$ and $H^l_{\rm o}(\lambda)$ such that $s \in {\cal S}^{(\alpha)}_{\lambda}$ implies $\bar s \in {\cal S}^{(\alpha)}_{\lambda}$ and $-\bar s \notin {\cal S}^{(\alpha)}_{\lambda}$.

 \vspace{0.2cm}

$\bullet$ The  matrices $\hat C^{r,+}_{\rm o}(\lambda)=W^{r,21}_{\rm o}(\lambda)[W^{r,11}_{\rm o}(\lambda)]^{-1}$ and $\hat C^{l,+}_{\rm o}(\lambda)=W^{l,21}_{\rm o}(\lambda)[W^{l,11}_{\rm o}(\lambda)]^{-1}$  introduced  previously  exist and are the {\it maximal}  real symmetric solutions of Eqs. (\ref{EqCARE:subeq1}) and (\ref{EqCARE:subeq2}) with respect to the  positive definiteness ordering. They are  obtained  by taking ${\cal S}^{(\alpha)}_{\lambda}$ to be the set ${\cal S}_{\lambda}^+=\{s^+_1,s^+_2,...s^+_{n+2}\}$ of eigenvalues with a positive real part (see the partitioning  of the basis change matrices  $W^r_{\rm o}(\lambda)$ and $W^l_{\rm o}(\lambda)$). Likewise,  the matrices $C^{r,-}_{\rm o}(\lambda)=W_{\rm o}^{r,22}(\lambda)[W_{\rm o}^{r,12}(\lambda)]^{-1}$ and $C^{l,-}_{\rm o}(\lambda)=W_{\rm o}^{l,22}(\lambda)[W_{\rm o}^{l,12}(\lambda)]^{-1}$ exist and are the {\it minimal} solutions obtained from  the set ${\cal S}_{\lambda}^-=\{s^-_1,s^-_2,...s^-_{n+2}\}$ of eigenvalues with a negative real part. This means that all other real symmetric solutions $\hat C^{r,(\alpha)}_{\rm o}(\lambda)$ (resp.  $\hat C^{l,(\alpha)}_{\rm o}(\lambda)$) are such that $\hat C^{r,-}_{\rm o}(\lambda)\le\hat C^{r,(\alpha)}_{\rm o}(\lambda)\le \hat C^{r,+}_{\rm o}(\lambda)$ (resp.  $\hat C^{l,-}_{\rm o}(\lambda)\le\hat C^{l,(\alpha)}_{\rm o}(\lambda)\le \hat C^{l,+}_{\rm o}(\lambda)$.

The maximal solutions  $\hat C^{r,+}_{\rm o}(\lambda)$ and $\hat C^{l,+}_{\rm o}(\lambda)$ can be shown to be analytic functions of $\lambda$~\cite{RR1988}, but  at variance with  the common situation in LQ optimal control~\cite{LR1995} these matrices are not necessarily positive semidefinite. On the other hand, since
\begin{align}
\label{Eqplusmoins}
\hat C^{l,+}_{\rm o}(\lambda)&=-\hat C^{r,-}_{\rm o}(\lambda)\nonumber\\
\hat C^{l,-}_{\rm o}(\lambda)&=-\hat C^{r,+}_{\rm o}(\lambda)
\end{align}
from the  relation between the basis change matrices $W_{\rm o}^r(\lambda)$ and $W_{\rm o}^l (\lambda)$ [Eq. (\ref{EqWrWl})], we have
\begin{align}
\label{Eqineq10}
\hat C^{r,+}_{\rm o}(\lambda)+\hat C^{l,+}_{\rm o}(\lambda)=\hat C^{r,+}_{\rm o}(\lambda)-\hat C^{r,-}_{\rm o}(\lambda)> 0\ . 
\end{align}
As shown in Appendix \ref{AppendF}, $\hat C^{r,+}_{\rm o}(\lambda)$ and $\hat C^{l,+}_{\rm o}(\lambda)$ have another important property: They are the {\it only} solutions of the CAREs (\ref{EqCARE:subeqns}) that satisfy  $\hat C^{r,+}_{\rm o}(0)=0$ and $\hat C^{l,+}_{\rm o}(0)=C$. Since $G^r_{\rm o,\lambda=0}({\bf u}_0,t)=1$ and $G^l_{\rm o,\lambda=0}({\bf u},t)=p({\bf u})$ for all ${\bf u}_0,{\bf u}$ and $t$ from the definition of the generating functions [Eqs. (\ref{EqGr}) and (\ref{EqGl})],  these conditions must be obeyed by the solutions of the RDEs (\ref{EqRic:subeq1}) and (\ref{EqRic:subeq2}), respectively\footnote{It is suggested in Ref. \cite{DBT2023} that the behavior at $\lambda=0$  can be used to select the correct asymptotic fixed point of the differential equation among all the solutions of the corresponding CARE. However, this is not a viable procedure in general as it assumes that these solutions are known analytically, which is only true for simple two-dimensional systems. By showing that  the maximal solution is the one that satisfies the exact condition at $\lambda=0$ we overcome this obstacle. Furthermore, we must keep in mind that the solution of the RDE may not be a continuous function of $\lambda$, as will be seen in Sec. \ref{SubsecIVC}.}. 

Finally, we introduce the function $f^+(\lambda)$ associated with the maximal solutions  of the CAREs which will play a prominent role in the following.  Since the maximal solutions  correspond to the set ${\cal S}_{\lambda}^+$, we  have 
\begin{align} 
\label{Eqmustar}
f^+(\lambda)=-\frac{1}{2}[\mbox{Tr}(A)+\sum_{i=1}^{n+2} s^+_i(\lambda)]
\end{align}
from Eq. (\ref{Eqmulmur}). $f^+(\lambda)$  stands out among  the functions $f^{(\alpha)}(\lambda)$ associated with the other solutions of  Eqs. (\ref{EqCARE:subeqns})  because of the following three properties:  

i) it is an analytic function of $\lambda$ in the interval ${\cal D}_H$ (since $\hat C^{r,+}_{\rm o}(\lambda)$ and $\hat C^{l,+}_{\rm o}(\lambda)$ are  analytic);

ii) it satisfies 
\begin{align} 
f^+(0)=0\ ,
\end{align} 
as can be readily seen by inserting $\hat C^{r,+}_{\rm o}(0)=0$ into  Eq. (\ref{Eqmu:subeq1})\footnote{Alternatively,  one can use the fact  that $p_H(0,s)=(-1)^np_A(s)p_A(-s)$ from Eq. (\ref{EqpH}). Therefore,  the  zeros of $p_H(0,s)$ with a positive real part are the zeros of $p_A(-s)$ and $\sum_{i=1}^{n+2} s^+_i(0)=-\mbox{Tr}(A)$.};

iii) it obeys the inequality
\begin{align} 
\label{EqInequal}
f^{(\alpha)}(\lambda)> f^+(\lambda),\; {\rm for}\; {\cal S}^{(\alpha)}_{\lambda}\neq {\cal S}^+_{\lambda}.
\end{align}
Indeed, a set ${\cal S}_{\lambda}^{(\alpha)}\ne {\cal S}_{\lambda}^+$ contains $m^{(\alpha)}\ne 0$ eigenvalues $s^-_{\alpha_i}$ with a negative real part and  $n+2-m^{(\alpha)}$ eigenvalues $s^+_{\alpha_i}$ with a positive real part. In consequence,
\begin{align} 
\label{Eqsum}
\sum_{i=1}^{n+2}s_{\alpha_i}(\lambda)&=\sum_{i=1}^{m^{(\alpha)}}s^-_{\alpha_i}(\lambda)+\sum_{i=m^{(\alpha)}+1}^{n+2-m^{(\alpha)}}s^+_{\alpha_i}(\lambda)=\sum_{i=1}^{n+2}s^+_i(\lambda)-2\sum_{i=1}^{m^{(\alpha)}}s^-_{\alpha_i}(\lambda)\ ,
\end{align}
where we have used the symmetry of the eigenvalues with respect to the imaginary axis. Inserting the above into Eq. (\ref{Eqmulmur}) and using Eq. (\ref{Eqmustar}), we then obtain
 \begin{align} 
\label{Eqmu2}
f^{(\alpha)}(\lambda)-f^+(\lambda)=-\sum_{i=1}^{m^{(\alpha)}}s^-_{\alpha_i}(\lambda)>0\ .
\end{align}
An alternative  expression of $f^+(\lambda)$ in terms of the spectral density of the process will be given  in Sec. \ref{SubsecIVB2}.

\section{Time evolution of the moment generating functions}
\label{SecIV}

Equipped with the explicit representations of the time-dependent solutions  $C^r_{\rm o}(\lambda, t)$ and  $C^l_{\rm o}(\lambda, t)$ of the RDEs (\ref{EqRic:subeqns}) [Eqs. (\ref{Eqsol4}) and (\ref{Eqsol4left})],  with the definition of the maximal solutions $ \hat C^{r,+}_{\rm o}(\lambda)$ and $ \hat C^{l,+}_{\rm o}(\lambda)$ of the CAREs (\ref{EqCARE:subeqns}), and with the expression of the function $f^+(\lambda)$, we are now in position to study  the time evolution  of the moment generating function $G_{\rm o,\lambda}(t)$ given by Eqs. (\ref{EqDynZ:subeqns}) and in particular their long-time behavior. We will then  explicitly compute the  SCGF $\mu_{\rm o}(\lambda)$ and the sub-exponential prefactors $g_{\rm o}(\lambda)$.  We recall that the values of $\lambda$ are restricted to the interval ${\cal D}_H=(\lambda_{\min},\lambda_{\max})$ for which  the Hamiltonian matrices have no purely imaginary eigenvalues and the existence of real symmetric solutions of the CAREs  is ensured\footnote{This does not mean that real symmetric solutions of the RDEs do not exist for $\lambda\not \in {\cal D}_H$, but they  blow up at some finite time. Note also that the extremal solutions  of the CAREs may also exist  when the Hamiltonian matrices have purely imaginary eigenvalues. However, this requires that the partial multiplicities of these eigenvalues are all even~\cite{note30}, and one can show that it is not the case for the model under study.}.

To make it more concrete, the forthcoming discussion is illustrated by  numerical results obtained for a gamma-distributed delay with  $n=5$ (which  generates a system of $n + 2 = 7$ dynamical variables).  Although this is still far from the case of a discrete delay, the gamma distribution (\ref{Eqgamma1}) is already markedly peaked around $t=\tau$ and  further investigations show that the qualitative behavior of the fluctuations does not change significantly for larger values of $n$. 
We choose (rather arbitrarily) $Q_0 = 2$, so that the harmonic oscillator is in a moderate underdamped regime, and a feedback gain $g=1.5$. We then  vary $\tau$ in order to illustrate different types of behavior. For these values of the parameters, the system  reaches a stable stationary state for all values of $\tau$\footnote{More generally, we recall that all our calculations assume that the system lies within the zones of stability.}.

\subsection{Finite-time divergences} 

\label{SubsecIVA}

As mentioned  in Sec. \ref{SubsecRic1},  the generating function $G_{\rm o,\lambda}(t)$ given by Eqs. (\ref{EqDynZ:subeqns}) may diverge in a finite time, depending on  the observable $\rm o$ and the value of $\lambda$.  
In some cases  one can prove from the outset that such divergence does not take place [see for instance inequalites (\ref{Eqineq3}) and (\ref{Eqineq4})].  In general, however,  one needs to numerically compute $C^r_{\rm o}(\lambda,t)$  from  Eq. (\ref{Eqsol4}) or  $C^l_{\rm o}(\lambda,t)$ from  Eq. (\ref{Eqsol4left}).  By increasing (resp. decreasing) $\lambda$  from $0$ to $\lambda_{\max}$ (resp. from $0$ to $\lambda_{\min}$) at a given $t$, one can  then determine the first value of $\lambda$ for which $\det(C^r_{\rm o}(\lambda,t)+C)=\det C^l_{\rm o}(\lambda,t)=0$ (we recall that the two determinants  vanish simultaneously since the  expressions  (\ref{EqDynZ:subeq1}) and (\ref{EqDynZ:subeq2}) of $G_{\rm o,\lambda}(t)$ are consistent as long as the matrices $C^r_{\rm o}(\lambda,t)$ and $C^l_{\rm o}(\lambda,t)$ exist).
This defines a time-dependent  interval ${\cal D}_{\rm o}(t)=\big(\lambda_{\rm o}^-(t),\lambda_{\rm o}^+(t)\big)$ such that $C+C^r_{\rm o}(\lambda,t)>0$ and $C^l_{\rm o}(\lambda,t)>0$ for all $\lambda\in {\cal D}_{\rm o}(t)$. Accordingly, $G_{\rm o,\lambda}(t)$ is finite for $\lambda\in {\cal D}_{\rm o}(t)$ and diverges at $\lambda=\lambda_{\rm o}^-(t)$ and $\lambda=\lambda_{\rm o}^+(t)$. Since the very definition of the SCGF requires that  $\lim_{t\to \infty}(1/t)\ln G_{\rm o,\lambda}(t)<\infty$,  the domain of existence of the SCGF is ${\cal D}_{\rm o}(\infty)$.

\begin{figure}[hbt]
\begin{center}
\includegraphics[trim={0cm 0cm 0cm 0cm},clip,width=10cm]{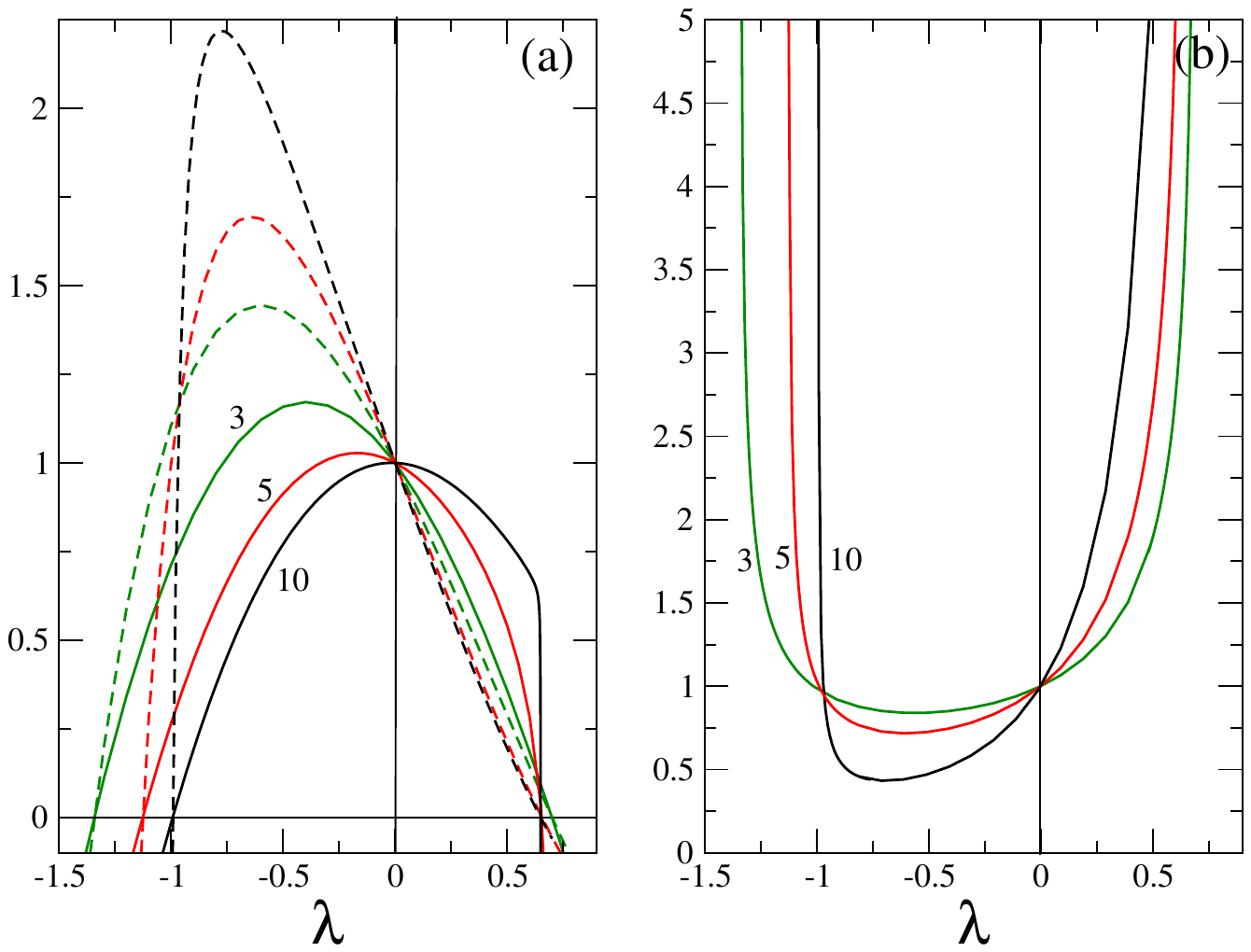}
\caption{ (Color on line) (a) Plots of  $\det C^l_w(\lambda,t)/ \det C$ (solid lines) and $\det(C^r_w(\lambda,t)+C)/\det C$ (dashed lines) versus $\lambda$ for  $t=3,5,10$ and $\tau=1$. (b) Corresponding moment generating function $G_{w,\lambda}(t)$.} 
\label{Fig2}
\end{center}
\end{figure}
 \begin{figure}[hbt]
\begin{center}
\includegraphics[trim={0cm 0cm 0cm 0cm},clip,width=10cm]{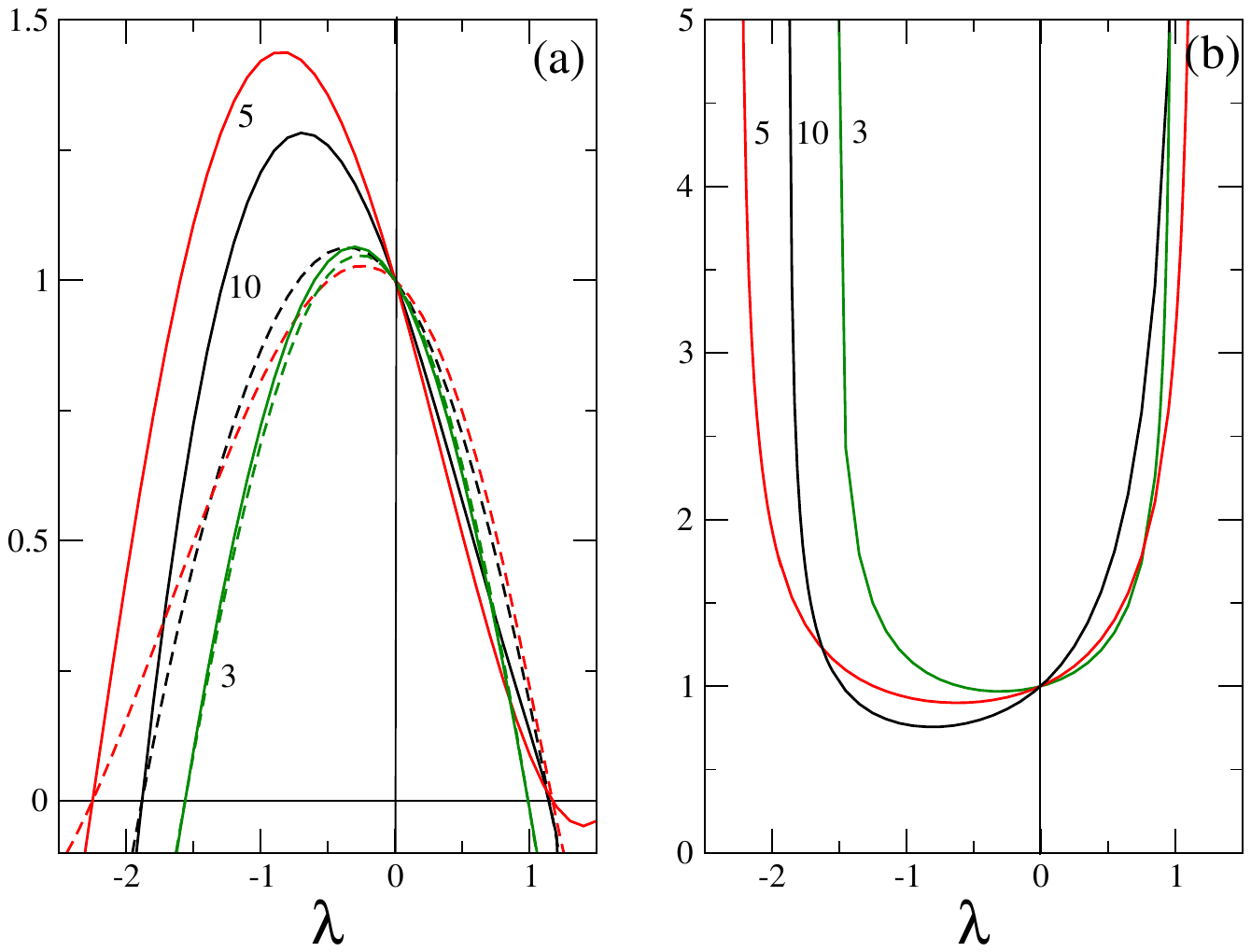}
\caption{ (Color on line) (a) Plots of  $\det C^l_w(\lambda,t)/ \det C$ (solid lines) and $\det(C^r_{\rm o}(\lambda,t)+C)/\det C$ (dashed lines) versus $\lambda$ for  $t=3,5,10$ and $\tau=3$. (b) Corresponding moment generating function $G_{w,\lambda}(t)$.} 
\label{Fig3}
\end{center}
\end{figure}

Two typical examples of the evolution of the determinants  with $\lambda$ at different times are shown in  Figs. \ref{Fig2} and \ref{Fig3} together with the corresponding moment generating functions. The calculations are performed for the observable ${\cal W}_t$  but the same type of results are obtained for the other observables. 
Inspection  of  the characteristic polynomial $p_H(s)$ reveals  that $\lambda_{\min}=-472.080$ and $\lambda_{\max}\simeq 1.019$ for $\tau=1$  and $\lambda_{\min}\simeq-5.118$ and $\lambda_{\max}\simeq 1.665$ for $\tau=3$. Moreover, it is found that $T_x/T\approx 1.663$, $T_v/T\approx 0.570$ for $\tau=1$ and $T_x/T\approx 1.152$, $T_v/T\approx 0.857$ for $\tau=3$. We can thus infer from Eq.  (\ref{Eqineq3})  that the matrices $C^r_w(\lambda,t)+C$ and $C^l_w(\lambda,t)$ are always positive definite and the generating function  $G_{w,\lambda}(t)$  is always finite for  values of $\lambda$ in the range $0\le \lambda \lesssim 0.601$ for $\tau=1$ and $0 \le \lambda \lesssim 0.868$ for $\tau=3$.
 
\begin{figure}[hbt]
\begin{center}
\includegraphics[trim={0cm 0cm 0cm 0cm},clip,width=10cm]{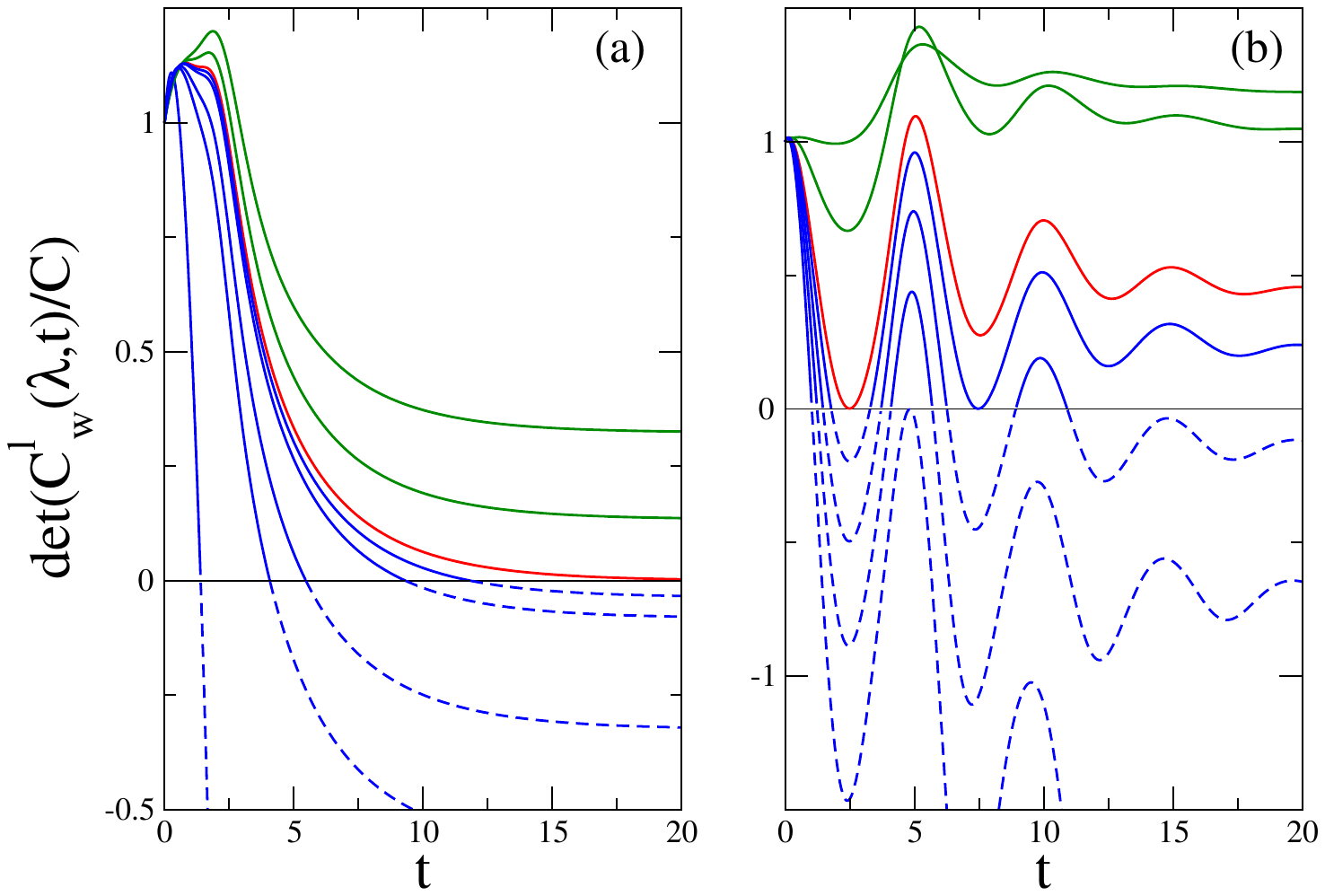}
\caption{ (Color on line) Time evolution of $\det C^l_w(\lambda,t)/ \det C$ for $\tau=1$ (a) and $\tau=3$ (b). The values of $\lambda$ decrease from top to bottom and $\det C^l_w(\lambda,t)$ is always positive for $\lambda\gtrsim-0.963$ (red curve in panel a) or $\lambda\gtrsim-1.51$ (red curve in panel b). The parts of the curves corresponding to $\det C^l_w(\lambda,t)<0$ are plotted as dashed lines.} 
\label{Fig4}
\end{center}
\end{figure}

By comparing Figs. \ref{Fig2} and \ref{Fig3}, we can  see at once that the width of the interval ${\cal D}_w(t)$ does not decrease monotonically with time for $\tau=3$, in contrast to the case $\tau=1$. (Note in passing that  $G_{w,\lambda}(t)$ does not satisfy the symmetry $\lambda \leftrightarrow 1-\lambda$; see the discussion in Ref. \cite{RTM2017}.) The difference between these two cases is even more manifest if we plot the  evolution of the determinants  with $t$ at fixed $\lambda$, as done  in Fig. \ref{Fig4} (for brevity,  we only show the behavior of $\det C^l_w(\lambda,t)$ and focus on negative values of $\lambda$). 
% $\tau=1$. The value of $\lambda$ ranges from $-0.8$ (top) to $-2$ (bottom)
% $tau=3$. From top to bottom: $-0.5,-1,-1.513,-1.633,-1.8,-2,-2.27$

In both cases, there is a critical value of $\lambda$ above which  $\det C^l_w(\lambda,t)>0$ at all times and the generating function is always finite. For smaller values of $\lambda$,  Figs. \ref{Fig4}(a) and \ref{Fig4}(b) depict two different scenarios. For $\tau=1$,  the determinant  vanishes at a {\it unique} time $t^-_w(\lambda)$ which  increases  monotonically to infinity as  $\lambda$ approaches the critical value. In this case $t^-_w(\lambda)$ is just the inverse function of $\lambda_w^-(t)$. For $\tau=3$, the curves  cross the $t$ axis a number of times (as the smallest eigenvalue of the matrix changes its sign), so that the determinant is positive in a certain time range, then negative, then positive again, etc. The function $t^-_w(\lambda)$ is thus {\it multivalued} and the fact that the positive parts of the curves disappear at different times explains the non-monotonic behavior observed in Fig. \ref{Fig3} as $t$ varies.

\begin{figure}[hbt]
\begin{center}
\includegraphics[trim={0cm 0cm 0cm 0cm},clip,width=10cm]{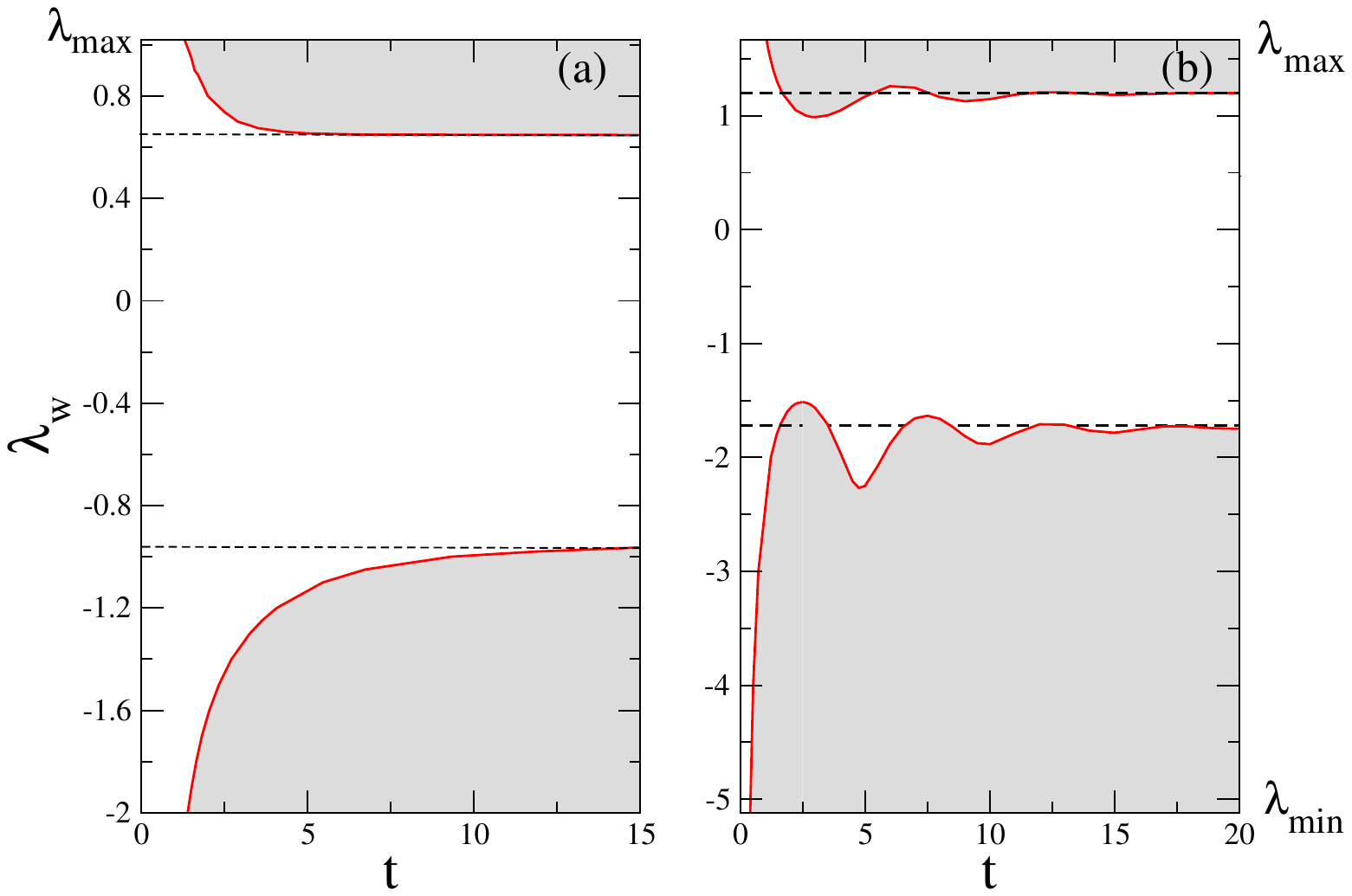}
\caption{ (Color on line) Evolution of the domain of existence of the generating function $G_{w,\lambda}(t)$, the interval ${\cal D}_w(t)=\big(\lambda^+_w(t),\lambda^-_w(t)\big)$, for $\tau=1$ (a) and $\tau=3$ (b). The matrices $C+C^r_w(\lambda,t)$ and $C_w^l(\lambda,t)$ are positive definite and $G_{w,\lambda}(t)$ is finite in the unshaded regions. At fixed $t$, $G_{w,\lambda}(t)$ diverges as $\lambda \to \lambda^+_w(t)$ and $\lambda \to \lambda^-_w(t)$ (solid red lines). The black dashed  lines indicate the limits $\lambda_{w1}$ and $\lambda_{w2}$ of the interval $\hat {\cal  D}_w$  (see  Fig. \ref{Fig7} below).}
\label{Fig5}
\end{center}
\end{figure}

A similar behavior is observed for $\lambda>0$ and this eventually leads to Figs. \ref{Fig5}(a) and  \ref{Fig5}(b) which describe how  the domain of existence of  $G_{w,\lambda}(t)$ evolves with time. Note that $\lambda_w^+(t)>0.601$ for $\tau=1$ and $\lambda_w^+(t)>0.868$ for $\tau=3$, in agreement with the predictions of Eq. (\ref{Eqineq3}).

We stress that the generating function $G_{w,\lambda}(t)$ is  finite at time $t$ if $\det C_w^l(\lambda,t)>0$ {\it whatever} the sign of the  determinant for $t'<t$. Otherwise, one would not obtain the same values of the thresholds $\lambda_w^-(t)$ and $\lambda_w^+(t)$  by varying $\lambda$ at fixed $t$ or  varying $t$ at fixed $\lambda$, and the results displayed in  Fig. \ref{Fig4}(b) would not be consistent with those in Fig.  \ref{Fig3}. We thus  disagree with  the statement made by C. Kwon {\it et al} in Ref. \cite{KNP2011} that  the  generating function\footnote{We recall that our expression (\ref{EqDynZ:subeq2}) of the generating function is similar to the one  derived in Ref. \cite{KNP2011}.} is well defined only when the determinant is positive for all $t'<t$, which implies that the threshold  is ``frozen"  beyond a critical time\footnote{We are indebted to Marco Zamparo for providing us with a simple example illustrating that  this assertion is erroneous.}. For $\tau=3$, this would amount to rejecting the non-monotonic dependence of  $\lambda^+_w(t)$ and $\lambda^-_w(t)$ on time displayed  in Fig. \ref{Fig5}(b). 
We therefore dispute the claim that  the generating function  jumps discontinuously  to infinity beyond this critical time (see Fig. 2(b) in Ref. \cite{KNP2011}) and we challenge the existence  of the so-called ``dynamic phase transitions" discussed in Refs. \cite{KNP2011,NKP2013,KNP2013}.
\begin{figure}[hbt]
\begin{center}
\includegraphics[trim={0cm 0cm 0cm 0cm},clip,width=10cm]{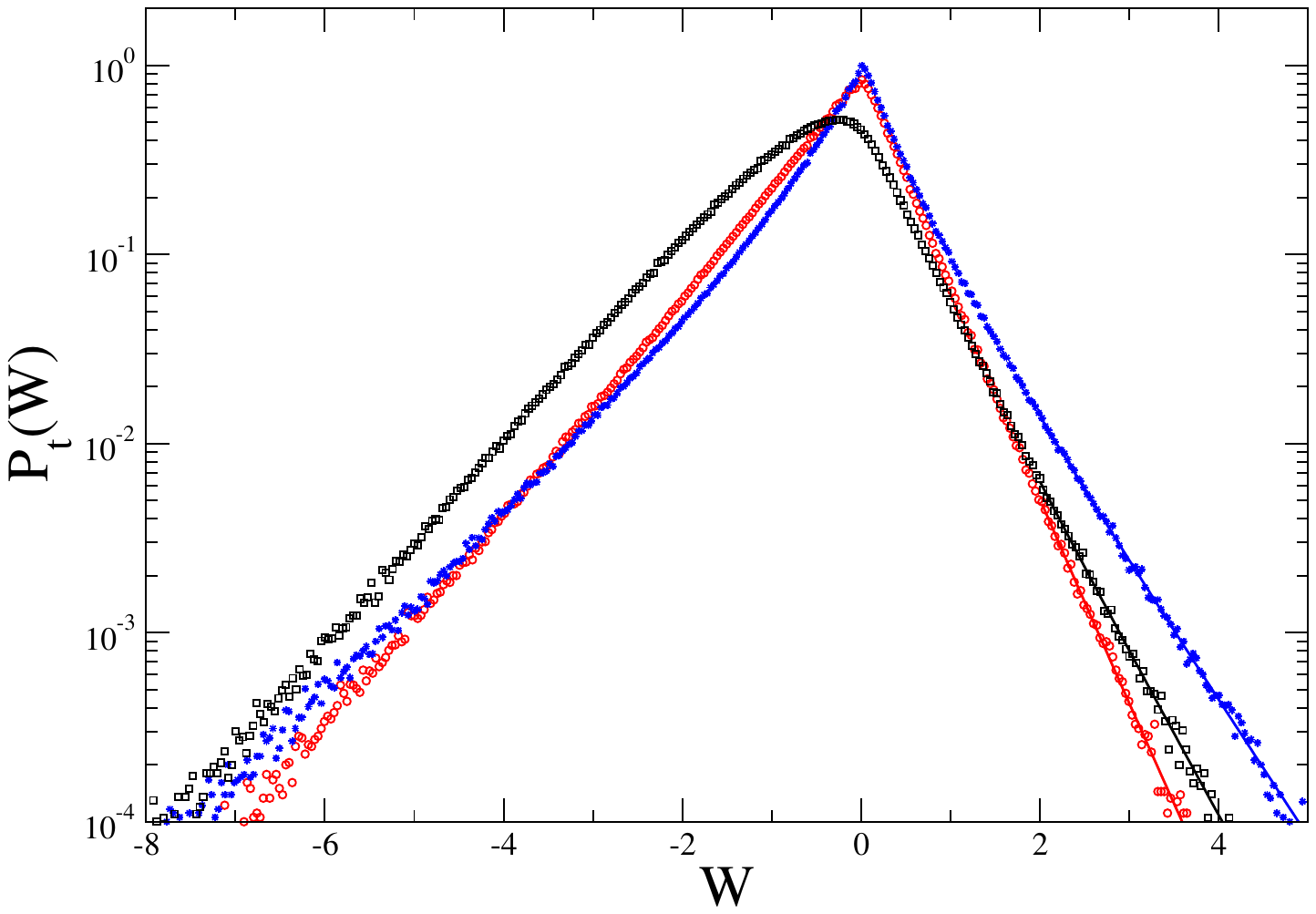}
\caption{ (Color on line) The pdf $P_t({\cal W})$ versus ${\cal W}$ for $\tau=3$ at different times: $t=3$ (blue stars), $5$ (red circles), and $10$ (black squares). $P_t({\cal W})$ is obtained from numerical simulations of $N=5.10^6$ samples. The straight lines on the right-hand side correspond to the asymptotic form (\ref{Eqtail1}) with the numerical values  of $\lambda_w^-(t)$ taken from Fig. \ref{Fig5}(b). }
\label{Fig5aux}
\end{center}
\end{figure}

In fact, the behavior of $\lambda_w^-(t)$ and $\lambda_w^+(t)$ with $t$ can be  directly checked by looking at the pdf $P_t(\beta {\cal W}_t={\cal W})$. Indeed, the divergences of $G_{w,\lambda}(t)$  at $\lambda^{\pm}_w(t)$ signal that $P_t({\cal W})$ has  exponential tails. 
More precisely,  the  numerical solution of the RDE (\ref{EqRic:subeq1}) reveals that  $\det C_w^l(\lambda,t)\sim \det (C_w^r(\lambda,t)+C)\sim (\lambda-\lambda^{\pm}_w(t))$ as $\lambda \to \lambda^{\pm}_w(t)$, so that, from Eqs. (\ref{EqDynZ:subeqns}),
\begin{align}
G_{w,\lambda}(t)\sim (\lambda-\lambda^{\pm}_w(t))^{-1/2}
\end{align}
and, thus~\cite{KNP2011},
 \begin{align}
\label{Eqtail1}
P_t({\cal W})\sim \frac{e^{\lambda^{\mp}_w(t) {\cal W}}}{\sqrt{\vert {\cal W} \vert}} \ , \: \: {\cal W}\to \pm \infty\ .
\end{align}

 In  Fig. \ref{Fig5aux}, the pdf obtained by numerically integrating  the set of dynamical equations (\ref{EqL}) for $n=5$ and $\tau=3$ is plotted in the semi-log scale. It is manifest that the slopes of the tails of the pdf  do not vary monotonically with time and they are very well described  by the  asymptotic  form (\ref{Eqtail1}) (see the solid lines on the right-hand side).
\begin{figure}[hbt]
\begin{center}
\includegraphics[trim={0cm 0cm 0cm 0cm},clip,width=10cm]{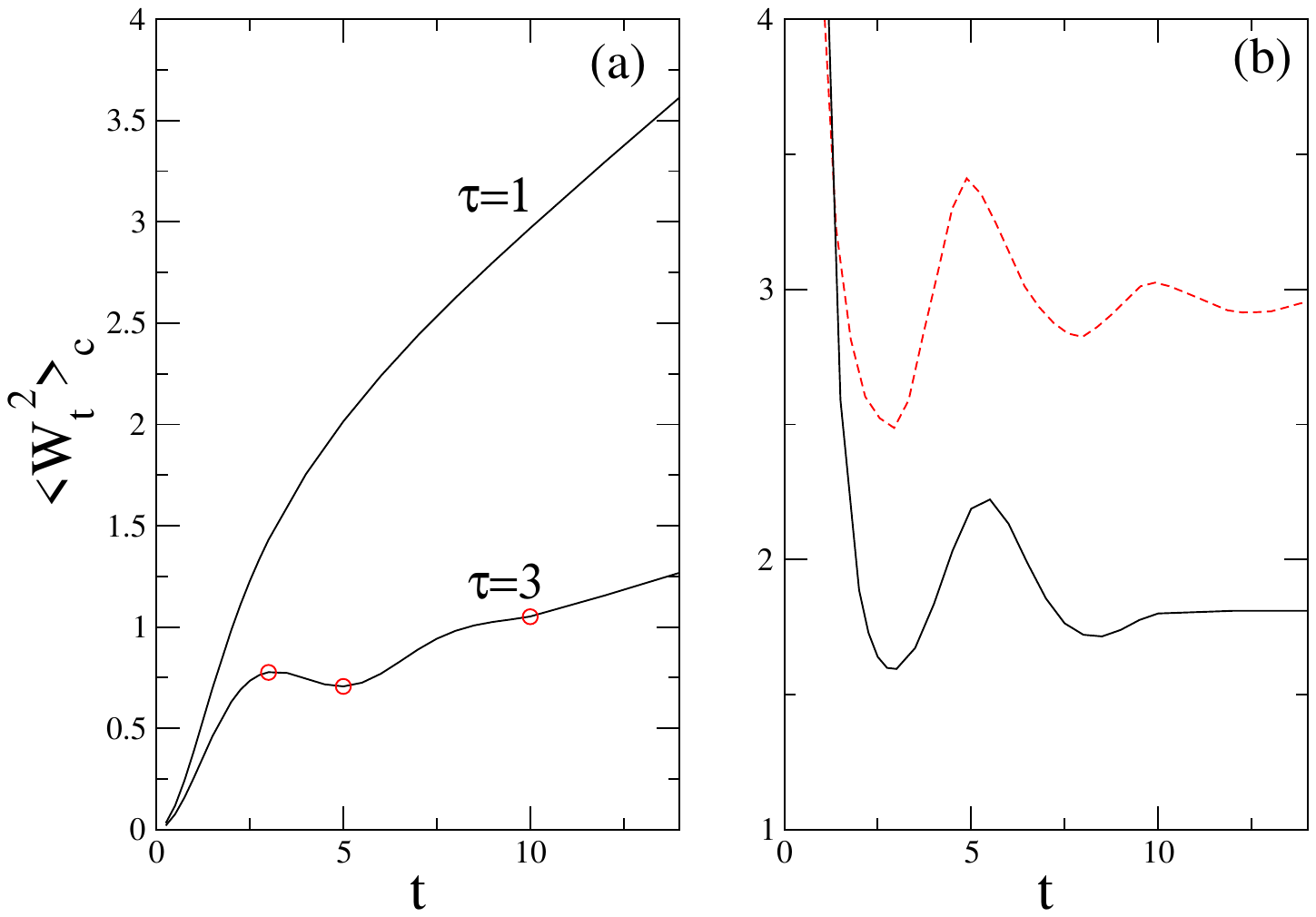}
\caption{(Color on line) (a) Time evolution of the variance $\langle W_t^2\rangle_c$ for $\tau=1$ and $\tau=3$. The circles represent the results of the numerical simulation.  (b) The width $l_w(t)=\lambda_w^+(t)-\lambda_w^-(t)$ of the interval ${\cal D}_w(t)$ for $\tau =3$ (red dashed line) is compared to $[\langle W_t^2\rangle_c-\mu_w''(0)t]^{-1}$ (solid black line).}
\label{Figvar}
\end{center}
\end{figure}

Remarkably,  the nontrivial behavior of  $\lambda_w^{\pm}(t)$ for $\tau=3$ also affects the variance $\langle W^2_t\rangle_c=\langle W_t^2\rangle -\langle W_t\rangle^2=\partial^2 \ln G_{w,\lambda}(t)/\partial^2 \lambda\vert_{\lambda=0}$.  As shown in  Fig. \ref{Figvar},  the  non-monotonic variation of $\langle W^2_t\rangle_c$  for $\tau=3$ is related  to  the evolution of the width $l_w(t)=\lambda_w^+(t)-\lambda_w^-(t)$ of the interval ${\cal D}_w(t)$. Since the variance is larger when $l_w(t)$ is smaller and $\langle W_t^2\rangle_c\sim \mu_w''(0)t$ as $t\to \infty$ (where $\mu_w(\lambda)=f^+(\lambda)$ is the SCGF - see Sec. \ref{SubsecIVB3} below), this relation is made more visible  by comparing $l_w(t)$ to the inverse of $\langle W_t^2\rangle_c- \mu_w''(0)t$, as done in Fig. \ref{Figvar}(b). So the pdf $P_t({\cal W})$ does not necessarily become more distributed as  $t$ increases, as could be naively expected.

\subsection{Long-time behavior} 
\label{SubsecIVB}

\subsubsection{Domain of existence of the scaled-cumulant generating function (SCGF)}
\label{SubsecIVB1}

We now focus on the behavior for $t\to \infty$. As can be seen  in Fig. {\ref{Fig5}, the thresholds  $\lambda_w^{\pm}(t)$  converge to finite  values $\lambda_w^{\pm}(\infty)$. More generally, for an observable $\rm o$, ${\cal D}_{\rm o}(t)\to {\cal D}_{\rm o}(\infty)=\big(\lambda_{\rm o}^-(\infty),\lambda_{\rm o}^+(\infty)\big)$ with ${\cal D}_{\rm o}(\infty)\subseteq {\cal D}_H$. On the other hand,  the analysis of Sec. \ref{SubsecRic2b} tells us that  $C^r_{\rm o}(\lambda,t)\to \hat C^{r,+}(\lambda)$  and $C^l_{\rm o}(\lambda,t)\to \hat C^{l,+}(\lambda)$ for generic values of $\lambda\in {\cal D}_H$ (i.e., for values of $\lambda$ such that $\det \hat C^{l,+}_{\rm o}(\lambda)\ne 0$ and $\det (\hat C^{r,+}_{\rm o}(\lambda)+C)\ne 0$). Therefore, ${\cal D}_{\rm o}(\infty)$, the domain of existence of the SCGF,  coincides with the interval $\hat {\cal D}_{\rm o}=(\lambda_{\rm o 1},\lambda_{\rm o 2})$ for which  the matrices  $\hat C^{r,+}_{\rm o}(\lambda)+C$  and $\hat C^{l,+}_{\rm o}(\lambda)$ are both positive definite. Generically,
\begin{equation}
{\cal D}_{\rm o}(\infty)=\hat {\cal D}_{\rm o}\subseteq {\cal D}_H,
\end{equation}
with $\hat {\cal D}_{\rm o}={\cal D}_H$ when these  matrices  $\hat C^{r,+}_{\rm o}(\lambda)+C$  and $\hat C^{l,+}_{\rm o}(\lambda)$ are positive definite for all $\lambda\in {\cal D}_H$.
\begin{figure}[hbt]
\begin{center}
\includegraphics[trim={0cm 0cm 0cm 0cm},clip,width=10cm]{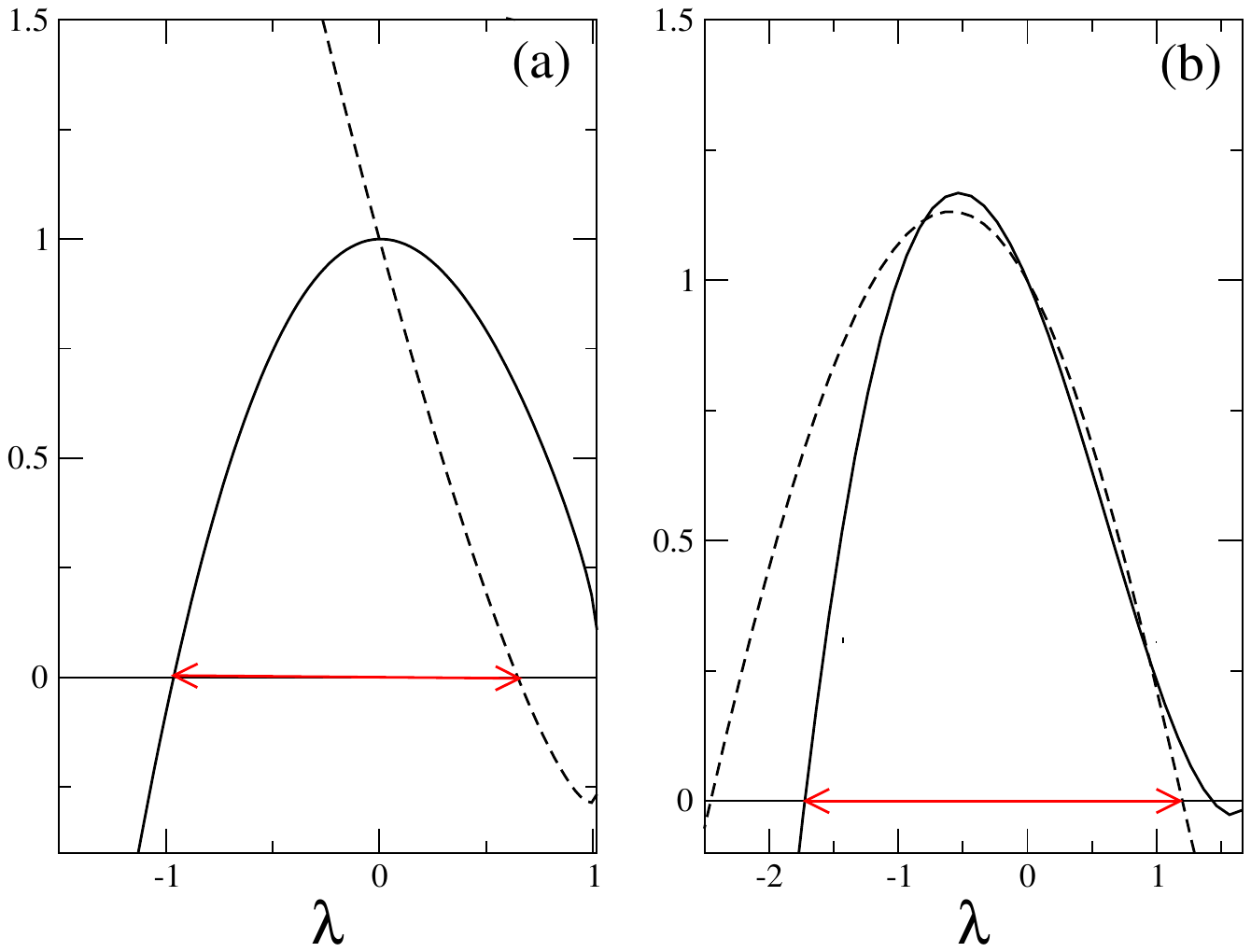}
\caption{ (Color on line) $\det \hat C^{l,+}_w(\lambda)/\det C$ (solid lines) and $\det (\hat C^{r,+}_w(\lambda)+C)/ \det C$ (dashed lines) as a function of $\lambda$ for $\tau=1$ (a) and $\tau=3$ (b) (in both panels, the largest value of $\lambda$ is $\lambda_{\max}$). The  arrows indicate the range $\hat {\cal D}_w$ of values of $\lambda$ for which the matrices  $\hat C^{r,+}_w(\lambda)+C$   and $\hat C^{l,+}_w(\lambda)$  are both positive definite.}
\label{Fig7}
\end{center}
\end{figure}

This is verified numerically in Fig. \ref{Fig7} which shows the variation of the determinants of $\hat C^{r,+}_w(\lambda)+C$  and $\hat C^{l,+}_w(\lambda)$ with $\lambda$ for the same values of the parameters as  in Figs. \ref{Fig2}-\ref{Figvar}. For $\tau=1$  and $\tau=3$,  it is found that the two  determinants are  positive  for $\lambda\in(-0.963,0.649)$  and $\lambda\in(-1.722,1.200)$, respectively, and these values are  in excellent agreement with the thresholds $\lambda_w^-(\infty)$ and $\lambda_w^+(\infty)$ that can be extrapolated from Fig. \ref{Fig5}.  Note that $\lambda_{w 1}$, the lower limit of the interval $\hat {\cal D}_w$, here corresponds to $\det \hat C^{l,+}_w(\lambda_{w1})=0$ (with $\det (\hat C^{r,+}_w(\lambda_{w 1})+C)\neq 0$) whereas $\lambda_{w 2}$, the upper limit, corresponds to $\det (\hat C^{r,+}_w(\lambda_{w 2})+C)=0$ (with $\det \hat C^{l,+}_w(\lambda_{w2})\neq 0$)\footnote{This is not a general feature and the opposite behavior is  observed for other values of the parameters (for instance $g=1$ and $\tau=3$) or for  the other observables.}.

At first sight, the latter observation may seem to contradict the fact that the determinants of the matrices $C^l_w(\lambda,t)$ and $C^r_w(\lambda,t)+C$  vanish for the {\it same} values of $\lambda$, i.e., $\lambda_w^-(t)$ or $\lambda_w^+(t)$ (see Figs. \ref{Fig2} and \ref{Fig3}). The solution to this puzzle requires one to look more closely into the long-time behavior of the determinants at the boundaries of the interval $\hat {\cal D}_w=(\lambda_{w1},\lambda_{w2})$. This is done in Appendix \ref{AppendG}. 

\subsubsection{Expressions of the SCGF and of the pre-exponential prefactors} 
\label{SubsecIVB2}

Now that the domain of existence of the SCGF has been identified as $\hat {\cal D}_{\rm o}=(\lambda_{\rm o1},\lambda_{\rm o,2})$, we derive explicit expressions for the SCGF $\mu_{\rm o}(\lambda)$ and the pre-exponential factors $g_{\rm o}(\lambda)$ when $\lambda_{\rm o1}<\lambda<\lambda_{\rm o2}$. Since $C^r_{\rm o}(\lambda,t)+C \to \hat C^{r,+}_{\rm o}(\lambda)+C>0$ and $C^l_{\rm o}(\lambda,t) \to \hat C^{l,+}_{\rm o}(\lambda)>0$, we have
\begin{align} 
\label{Eqflim}
&f^r_{\rm o}(\lambda,t)\to f^+(\lambda)\nonumber\\
&f^l_{\rm o}(\lambda,t) \to f^+(\lambda)\ ,
\end{align}
and   from Eqs. (\ref{EqSol1:subeqns})  we obtain the two asymptotic expressions
\begin{subequations}
\label{EqAsym:subeqns}
\begin{align}
&G^r_{\rm o,\lambda}({\bf u}_0,t)\sim g^r_{\rm o}(\lambda)\exp \big(-\frac{1}{2}{\bf u}_0^T\hat C^{r,+}_{\rm o}(\lambda){\bf u}_0\big)e^{f^+(\lambda)t}\label{EqAsym:subeq1}\\
&G^l_{\rm o,\lambda}({\bf u},t)\sim g^l_{\rm o}(\lambda)\exp \big(-\frac{1}{2}{\bf u}^T\hat C^{l,+}_{\rm o}(\lambda){\bf u}\big)e^{f^+(\lambda)t} \label{EqAsym:subeq2}\ ,
\end{align}
\end{subequations}
with
 \begin{subequations}
\label{EqAsym1:subeqns}
\begin{align}
g^r_{\rm o}(\lambda)&= \exp \left(\int_0^{\infty} dt\:[f^r_{\rm o}(\lambda,t)-f^+(\lambda)]\right) \label{EqAsym1:subeq1}\\
g^l_{\rm o}(\lambda)&=\left(\frac{\det C}{(2\pi)^{n+2}}\right)^{1/2} \exp \left(\int_0^{\infty} dt\:[f^l_{\rm o}(\lambda,t)-f^+(\lambda)]\right)\label{EqAsym1:subeq2}\ .
\end{align}
\end{subequations} 
Integrating  $G^r_{\rm o,\lambda}({\bf u}_0,t)p({\bf u}_0)$ over the initial state ${\bf u}_0$ and  $G^l_{\rm o,\lambda}({\bf u},t)$ over the final state ${\bf u}$, we finally get 
\begin{align} 
\label{EqZAasymp}
G_{\rm o,\lambda}(t)\sim g_{\rm o}(\lambda)e^{f^+(\lambda)t} \ ,
\end{align} 
with
\begin{subequations}
\label{EqgA1:subeqns}
\begin{align}
g_{\rm o}(\lambda)&=\Big[\frac{\det [C+\hat C^{r,+}_{\rm o}(\lambda)]}{\det C}\Big]^{-1/2}\exp \left(\int_0^{\cal \infty} dt\:[f^r_{\rm o}(\lambda,t)-f^+(\lambda)]\right)\label{EqgA1:subeq1}\\
&=\Big[ \frac{\det \hat C^{l,+}_{\rm o}(\lambda)}{\det C}\Big]^{-1/2}\exp\left(\int_0^{\cal \infty} dt\:[f^l_{\rm o}(\lambda,t)-f^+(\lambda)]\right)\label{EqgA1:subeq2}\ .
\end{align}
\end{subequations}
Eq. (\ref{EqZAasymp}) is a central result of our work as it shows that for $\lambda\in \hat {\cal D}_{\rm o}$ the SCGF $\mu_{\rm o}(\lambda)$ defined in Eq. (\ref{EqSCGF}) is equal to the function  $f^+(\lambda)$ which is associated with the maximal solutions  of the CAREs (\ref{EqCARE:subeqns}). The SCGF is therefore the same for the three observables ${\cal W}_t,{\cal Q}_t$ and $\Sigma_t$ since the corresponding  matrices $B_{\rm o}$ have the same anti-symmetric part, which implies that the Hamiltonian matrices have the same  spectrum. On the other hand,  the interval $\hat {\cal D}_{\rm o}$ and the pre-exponential factor $g_{\rm o}(\lambda)$  depend on the observable. In particular, the expressions (\ref{EqgA1:subeqns}) of $g_{\rm o}(\lambda)$ diverge at the boundaries of $\hat {\cal D}_{\rm o}$\footnote{However, this does not mean that the {\it actual} prefactor diverges at $\lambda_{\rm o1}$ or $\lambda_{\rm o 2}$, as discussed in Sec. \ref{SubsecIVC}.}.

The SCGF can  be computed from Eq. (\ref{Eqmustar}), i.e., from the eigenvalues of the Hamiltonian matrices. This is a standard numerical task, even for large $n$, but one can derive an equivalent and even more convenient expression as an integral over frequency. As shown in Appendix \ref{AppendH}, it reads
\begin{align} 
\label{Eqmunlambda}
f^+(\lambda)&= -\frac{1}{2} \int \frac{d\omega}{2\pi}\: \ln \Big[1-\frac{4\lambda g}{Q_0^2}\omega \sin_n(\omega \tau) \vert \chi_n(\omega)\vert ^2\Big]\ ,
\end{align}
where  $\chi_n(\omega)$ is the response function of the system defined  in Fourier space by $x(\omega)= \chi_n(\omega) \xi(\omega)$, and  
\begin{align}
\label{Eqsinn}
\sin_n(x)=\frac{1}{2i}[(1-\frac{ix}{n})^{-n}-(1+\frac{ix}{n})^{-n})]
\end{align}
is a $n$-dependent sine-like function which  satisfies $\lim_{n\to \infty} \sin_n(x)=\sin(x)$. It is  readily found from Eq. (\ref{EqlinearTrick1}) that  
\begin{align}
\label{Eqchi}
\chi_n(\omega)=\Big(-\omega^2-\frac{i\omega}{Q_0}+1-\frac{g}{Q_0}(1-\frac{i\omega \tau}{n})^{-n}\Big)^{-1}\ ,
\end{align}
which can be also written as 
\begin{align}
\label{Eqchi1}
\chi_n(\omega)&=(\frac{n}{\tau})^n\frac{(1-\frac{i\omega \tau}{n})^{n}}{p_A(s=-i\omega)}
\end{align}
by using Eq. (\ref{Eqchar}). Therefore, $f^+(\lambda)$ can be simply expressed in terms  of the spectral density $\vert \chi_n(\omega)\vert ^2$. This  shows the connection between the Riccati formalism and the  results in the mathematical literature~\cite{BD1997,BGR1997,GRZ1999,ZS2023} for the SCGF of stationary Gaussian processes\footnote{These results  concern quadratic observables instead of linear currents, but one can easily adapt the present derivation to such observables  and obtain an integral representation of  the SCGF similar to one given in the first line of Eq. (\ref{Eqmu2}).}.

A plot of $f^+(\lambda)$ vs $\lambda$ for $\tau=1$ and $\tau=3$ is shown in Fig. \ref{Figfplus}(a). This illustrates two important properties of the function:

 i) $f^+(\lambda)$ has  infinite slopes at the boundaries of ${\cal D}_H$, 
 
  ii) $f^+(\lambda)$ is convex in ${\cal D}_H$ (we already know that this function is differentiable). Accordingly, the  Legendre-Fenchel transform (\ref{EqLF}) that relates the SCGF to the rate function  reduces to the usual Legendre transform. Specifically, defining $\lambda^*(a)$  as the (unique) solution of  $df^+(\lambda)/d\lambda\vert_{\lambda=\lambda^*}=-a$, one has~\cite{T2009}
 \begin{align}
\label{EqLegendre}
I(a)=-a\lambda^*-f^+(\lambda^*)\: \: \: \mbox{for}\: \: \lambda^*(a)\in \hat {\cal D}_{\rm o}\ .
\end{align} 
If $\hat {\cal D}_{\rm o}= {\cal D}_H$ the function $I(a)$, which does not depend on the observable, is asymptotically linear as $a\to -\infty$ (resp. $a\to +\infty$) with slope $-\lambda_{\max}$ (resp. $-\lambda_{\min}$)\footnote{Due to the minus sign in our definition of $G_{\rm o,\lambda}(t)$ (which is the same as in Ref. \cite{J2020}),  the behavior of the SCGF for $\lambda>0$ (resp. $\lambda<0$) is relevant for the rate function for $a<\langle a\rangle$ (resp. $a>\langle a\rangle$).}. This means that the values of  $\lambda$ outside ${\cal D}_H$ are irrelevant to the determination of the rate function,  which justifies our initial choice to restrict the study of the solutions of the Riccati equations to this interval. 
\begin{figure}[hbt]
\begin{center}
\includegraphics[trim={0cm 0cm 0cm 0cm},clip,width=10cm]{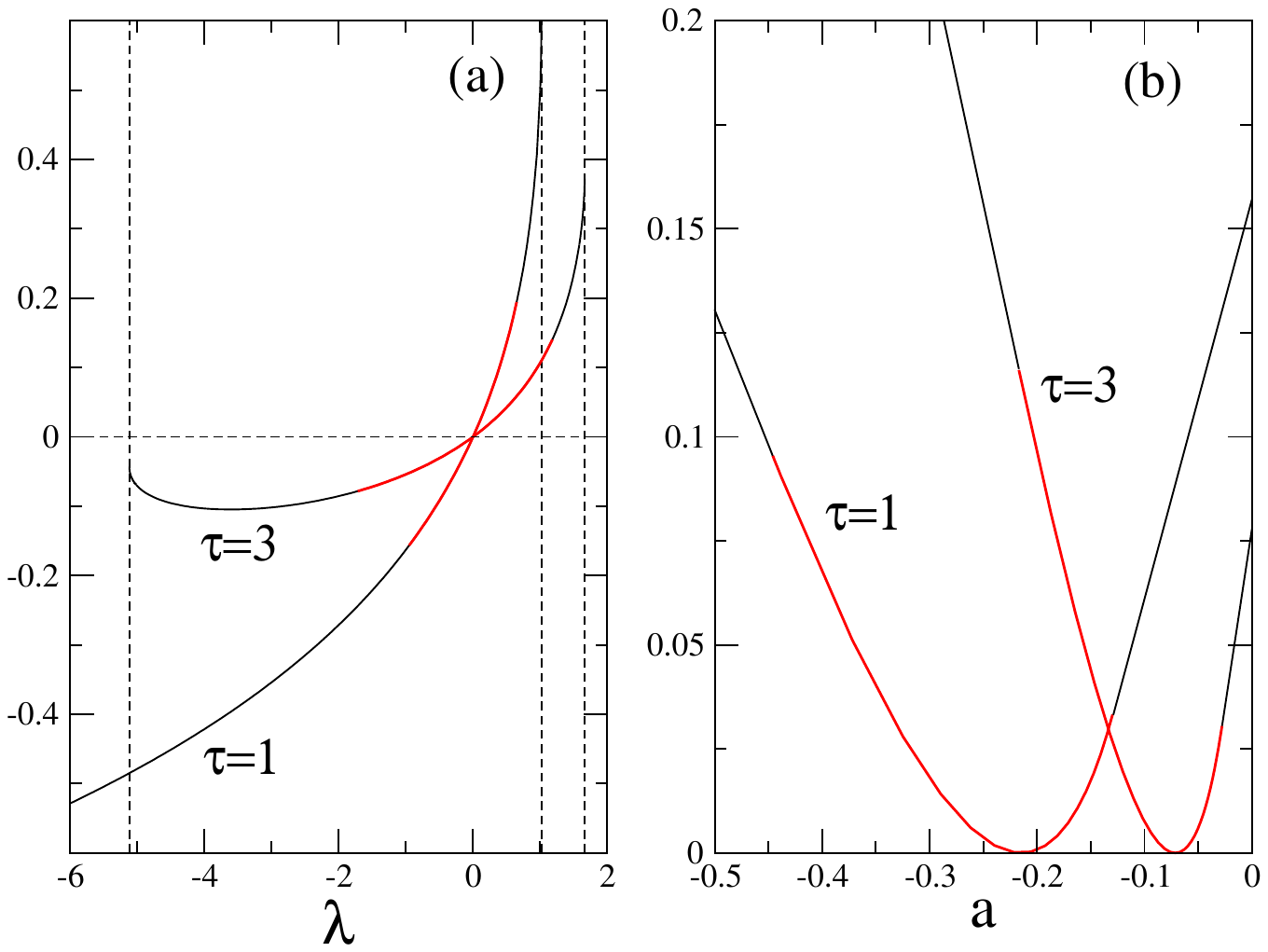}
\caption{ (Color on line) (a) Function $f^+(\lambda)$  computed from Eq. (\ref{Eqmunlambda}) for $n=5$, $\tau=1$ and $\tau=3$. The  parts of the curves  in red correspond to the interval $\hat {\cal D}_w$ for which  the  SCGF $\mu_w(\lambda)$  is finite and equal to $f^+(\lambda)$  (see Fig. \ref{Fig7}) and the vertical dashed lines indicate the limits $\lambda_{\min}$ and $\lambda_{\max}$ of ${\cal D}_H$, the interval of definition of  $f^+(\lambda)$ ($\lambda_{\min}\approx -472.8$ for $\tau=1$, which is outside the scale of the figure).  (b) Corresponding rate function $I_w(a)$. The parts of the curves  in red are obtained from the Legendre transform (\ref{EqLegendre}) and the linear parts in black are obtained from Eqs. (\ref{EqRate:subeqns}). }
\label{Figfplus}
\end{center}
\end{figure}

On the other hand, if $\hat {\cal D}_{\rm o}\subset {\cal D}_H$ (where here and below $\subset $ means a strict inclusion), the condition $\lambda^*(a)\in \hat {\cal D}_{\rm o}$  defines two  special values $a_{\rm o}^-=-df^+(\lambda)/d\lambda\vert_{\lambda_{\rm o 2}}$ and $a_{\rm o}^+=-df^+(\lambda)/d\lambda\vert_{\lambda_{\rm o1}}$ beyond which the rate function  is no longer given by  Eq. (\ref{EqLegendre}). Since $f^+(\lambda)$ has finite slopes at the two cutoff  $\lambda_{\rm o 1}$ and $\lambda_{\rm o 2}$,  the properties of Legendre transforms~\cite{T2009} imply that the rate function (which now depends on the observable) exhibits linear branches beyond $a_{\rm o}^{\pm}$,
\begin{subequations}
\label{EqRate:subeqns}
 \begin{align}
I_{\rm o}(a)&=-a\lambda_{\rm o 2}-f^+(\lambda_{\rm o 2})\: \: \: \: \: a\le a_{\rm o}^-\label{EqRate:subeq1}\\
I_{\rm o}(a)&=-a\lambda_{\rm o 1}-f^+(\lambda_{\rm o 1}) \: \: \: \: \: a\ge a_{\rm o}^+\label{EqRate:subeq2}\ .
\end{align} 
\end{subequations}
This yields  the curves $I_w(a)$ plotted in Fig. \ref{Figfplus}(b). 

Finally, we give the expression of the SCGF in the limit $n\to \infty$. It can be obtained either from the expression of  $f^+(\lambda)$ in Laplace space\footnote{This amounts to  replacing $(1\pm s\tau/n)^n$ by $ e^{\pm s\tau}$ in Eq. (\ref{Eqmu2}) and  using Eq. (\ref{Eqchar1}). This yields
\begin{equation*}
\label{EqpH2}
\lim_{n\to \infty} \frac{(-1)^np_H(\lambda,s)}{p_A(s)p_A(-s)}= 1-\frac{2\lambda g}{Q_0^2}\frac{s(e^{-s\tau}-e^{s\tau})}{(s^2+\frac{s}{Q_0}+1 -\frac{g}{Q_0}e^{-s\tau})(s^2-\frac{s}{Q_0}+1 -\frac{g}{Q_0}e^{s\tau})}\ .
\end{equation*} }
or directly  from Eq. (\ref{Eqmunlambda}), which yields
\begin{align} 
\label{Eqmuinf}
f^+_{\infty}(\lambda)\equiv \lim_{n \to \infty}f^+(\lambda)&= - \frac{1}{2}\int \frac{d\omega}{2\pi}\: \ln \Big[1-\frac{4\lambda g}{Q_0^2}\omega \sin(\omega \tau) \vert \chi_{\infty}(\omega)\vert ^2\Big]\ ,
\end{align}
with 
\begin{align}
\label{Eqchiinf}
\chi_{\infty}(\omega)=(-\omega^2 -\frac{i\omega}{Q_0}+1-\frac{g}{Q_0}e^{i\omega \tau})^{-1}\ .
\end{align}
As could be expected, this result coincides with the expression of the SCGF for the discrete delay obtained in our previous work~\cite{RTM2017} by imposing  periodic boundary conditions on the solution of the Langevin equation (i.e., $x(0)=x(t))$ and expanding  $x(t)$  in a Fourier series (this amounts to assuming that   boundary conditions can be neglected to leading order in $t$ in the calculation of  $\langle e^{-\lambda {\cal A}_t}\rangle$; see, e.g., Ref. \cite{ZBCK2005} for a similar calculation). It is clear that  Eq. (\ref{Eqmunlambda}) can be also derived by using the same method for $n$ finite. 

\subsubsection{Relation with the spectral problem for the tilted generators}
\label{SubsecIVB3}

As is well known, the SCGF of additive functionals such as the ones considered in this work is  given (in its domain of definition) by  the  dominant  eigenvalue $\mu_{\rm o}(\lambda)$ of the tilted generators ${\cal L}_{\rm o,\lambda}$ and ${\cal L}^\dag_{\rm o,\lambda}$~\cite{T2009,J2020,CT2015,T2018}. This results from  the expansion of the restricted generating function $G_{\rm o,\lambda}({\bf u},t\vert {\bf u}_0)$ in a complete basis of bi-orthogonal eigenfunctions, which yields asymptotically~\cite{CT2015}
\begin{align}
\label{EqZAasym}
G_{\rm o,\lambda}({\bf u},t\vert {\bf u}_0)\sim r_{\rm o,\lambda}({\bf u}_0)l_{\rm o,\lambda}({\bf u})e^{\mu_{\rm o}(\lambda)t}\ ,
\end{align}
where $r_{\rm o,\lambda}({\bf u}_0)$ and $l_{\rm o,\lambda}({\bf u})$ are the right and left eigenfunctions associated with $\mu_{\rm o}(\lambda)$. 
Finding the SCGF thus requires solving the spectral problem
\begin{subequations}
\label{EqSP1:subeqns}
\begin{align} 
{\cal L}_{\rm o,\lambda}\,  r_{\rm o,\lambda}({\bf u}_0)&=\mu_{\rm o}(\lambda)\, r_{\rm o,\lambda}({\bf u}_0)\label{EqSP1:subeq1}\\
{\cal L}_{\rm o,\lambda}^{\dag}\, l_{\rm o,\lambda}({\bf u})&=\mu_{\rm o}(\lambda)\, l_{\rm o,\lambda}({\bf u})\label{EqSP1:subeq2}\ , 
\end{align}
\end{subequations}
with  the eigenfunctions  commonly normalized according to~\cite{CT2015} 
\begin{subequations}
\label{Eqnorm:subeqns}
\begin{align}
&\int d{\bf u} \:  l_{\rm o,\lambda}({\bf u})r_{\rm o,\lambda}({\bf u})=1\label{Eqnorm:subeq1}\\
&\int d{\bf u} \:  l_{\rm o,\lambda}({\bf u})=1\label{Eqnorm:subeq2}\ .
\end{align}
\end{subequations}
(In particular, Eq. (\ref{Eqnorm:subeq1}) is inherited from the duality  between the operators ${\cal L}_{\rm o,\lambda}$ and ${\cal L}^\dag_{\rm o,\lambda}$ which imposes the boundary condition $l_{\rm o,\lambda}({\bf u})r_{\rm o,\lambda}({\bf u})=0$ at infinity~\cite{T2018}.)
Since we have shown above that the SCGF $\mu_{\rm o}(\lambda)$ is equal to $f^+(\lambda)$ inside its domain of definition $\hat {\cal D}_{\rm o}$, we already know the solution of the spectral problem for the dominant eigenvalue,\footnote{It may be confusing that $f^+(\lambda)$ is the largest eigenvalue of the operators ${\cal L}_{\rm o,\lambda}$ and ${\cal L}^\dag_{\rm o,\lambda}$  but is smaller than of all other functions $f^{(\alpha)}(\lambda)$ according to inequality  (\ref{EqInequal}). There is no contradiction though  because the corresponding matrices $\hat C^{r,(\alpha)}_{\rm o}(\lambda)$  and $\hat C^{l,(\alpha)}_{\rm o}(\lambda)$  are not valid solutions of the spectral problem (\ref{EqSP1:subeqns}).  Indeed, the condition $\hat C^{r,(\alpha)}_{\rm o}(\lambda)+\hat C^{l,(\alpha)}_{\rm o}(\lambda)>0$ is only satisfied by the maximal  solutions of the CAREs [Eq. \ref{Eqineq10})]. Take for instance the matrices $\hat C^{r,-}_{\rm o}(\lambda)$ and $\hat C^{l,-}_{\rm o}(\lambda)$ that are the minimal solutions of the CAREs. From Eq. (\ref{Eqmulmur}), $f^-(\lambda)$ is the {\it largest} of all functions $f^{(\alpha)}(\lambda)$ since all eigenvalues of the Hamiltonian matrices in the set  ${\cal S}_{\lambda}^-$ have a negative real part. On the other hand, relations (\ref{Eqplusmoins}) imply that $\hat C^{r,-}_{\rm o}(\lambda)+ \hat C^{l,-}_{\rm o}(\lambda)=-[\hat C^{r,+}_{\rm o}(\lambda)+\hat C^{l,+}_{\rm o}(\lambda)]$ and therefore all the eigenvalues of this matrix are negative. As a result, the  condition (\ref{Eqnorm:subeq1}) is only satisfied by the eigenfunctions given by  Eqs. (\ref{EqEigen1:subeqns}).}
and the  eigenfunctions corresponding to $ f^+(\lambda)$ are readily obtained from the asymptotic expressions  (\ref{EqAsym:subeqns}) as
\begin{subequations}
\label{EqEigen1:subeqns}
\begin{align} 
  r_{\rm o,\lambda}({\bf u}_0)&=g_{\rm o}^r(\lambda)\exp\left(-\frac{1}{2}{\bf u}_0^T\hat C^{r,+}_{\rm o}(\lambda){\bf u}_0\right)\label{EqEigen1:subeq1}\\
l_{\rm o,\lambda}({\bf u})&=\frac{g_{\rm o}^l(\lambda)}{\int d{\bf u}_0\: p({\bf u}_0)r_{\rm o,\lambda}({\bf u}_0)}\exp\left(-\frac{1}{2}{\bf u}^T\hat C^{l,+}_{\rm o}(\lambda){\bf u}\right)\label{EqEigen1:subeq2} \ ,
\end{align}
 \end{subequations}
with $g_{\rm o}^r(\lambda)$ and $g_{\rm o}^l(\lambda)$ given by Eqs. (\ref{EqAsym1:subeqns}).
These expressions  can  be used to derive a simple expression of the prefactor $g_{\rm o}(\lambda)$. Inserting  Eqs. (\ref{EqEigen1:subeqns}) into Eq. (\ref{Eqnorm:subeq1}) and performing a few  manipulations,  we  obtain the two relations
\begin{subequations}
\label{Eqfarl:subeqns}
\begin{align} 
\exp\left(\int_0^{\cal \infty} dt\:[f^r_{\rm o}(\lambda,t)-f^+(\lambda)]\right)=\Big[\frac{\det[\hat C^{r,+}_{\rm o}(\lambda)+\hat C^{l,+}_{\rm o}(\lambda)]}{\det \hat C^{l,+}_{\rm o}(\lambda)}\Big]^{1/2}\label{Eqfarl:subeq1}\\
\exp \left(\int_0^{\cal \infty} dt\:[f^l_{\rm o}(\lambda,t)-f^+(\lambda)]\right)=\Big[\frac{\det[\hat C^{r,+}_{\rm o}(\lambda)+\hat C^{l,+}_{\rm o}(\lambda)]}{\det [C+\hat C^{r,+}_{\rm o}(\lambda)]}\Big]^{1/2}\label{Eqfarl:subeq2}\ ,
\end{align}
\end{subequations}
 and using Eqs. (\ref{EqgA1:subeqns}) we find 
 \begin{align}
\label{EqgA}
g_{\rm o}(\lambda)=\left (\frac{\det C\:\det [\hat C^{l,+}_{\rm o}(\lambda)+\hat C^{r,+}_{\rm o}(\lambda)]}{\det \hat C^{l,+}_{\rm o}(\lambda)\: \det [\hat C^{r,+}_{\rm o}(\lambda)+C]}\right )^{1/2}\ .
\end{align}
This latter expression is another significant result of this work. It has the great advantage of no longer involving an integration over time;  it suffices to compute the maximal solutions of the CAREs (\ref{EqCARE:subeqns}) from Eq. (\ref{EqXralpha}), which is a  simple task\footnote{As those in Eqs. (\ref{EqgA1:subeqns}), this expression diverges when $C^{l,+}_{\rm o}(\lambda)$ and/or $\hat C^{r,+}_{\rm o}(\lambda)+C$ are no longer positive definite, i.e., at the limits of the interval $\hat {\cal D}_{\rm o}$. However, we stress again that this does not imply that the {\it actual} prefactor is infinite at $\lambda_{\rm o 1}$ or $\lambda_{\rm o 2}$, as will be shown in Sec. \ref{SubsecIVC}.}.

Finally, after using Eqs. (\ref{Eqfarl:subeqns}) and the normalization integral (\ref{Eqnorm:subeq2}), the expressions of $ r_{\rm o,\lambda}({\bf u}_0)$ and $ l_{\rm o,\lambda}({\bf u})$  can  be rewritten as
\begin{subequations}
\label{EqEigen2:subeqns}
\begin{align} 
 r_{\rm o,\lambda}({\bf u}_0)&=\Big[\frac{\det [\hat C^{r,+}_{\rm o}(\lambda)+\hat C^{l,+}_{\rm o}(\lambda)]}{\det \hat C^{l,+}_{\rm o}(\lambda)}\Big]^{1/2}\exp\left(-\frac{1}{2}{\bf u}_0^T\hat C^{r,+}_{\rm o}(\lambda){\bf u}_0\right)\label{EqEigen2:subeq1}\\
l_{\rm o,\lambda}({\bf u})&=\Big[\frac{\det \hat C^{l,+}_{\rm o}(\lambda)}{(2\pi)^{n+2}}\Big]^{1/2}\exp\left(-\frac{1}{2}{\bf u}^T\hat C^{l,+}_{\rm o}(\lambda){\bf u}\right)\label{EqEigen2:subeq2} \ .
\end{align}
 \end{subequations}

One can notice the similarity of these equations with Eqs. (55) and (56) in Ref. \cite{KSD2011}. In this paper,  the large-time expression of the generating function for the heat flow in harmonic chains was computed by using finite-time Fourier transforms.  As already mentioned, the expression  (\ref{Eqmunlambda}) of the SCGF is easily obtained by this method. With more efforts, one can also compute the eigenfunctions, i.e., the matrices $\hat C^{r,+}_{\rm o}(\lambda)$ and $\hat C^{l,+}_{\rm o}(\lambda)$, and in turn the pre-exponential factors $g_{\rm o}(\lambda)$ (Eq. (\ref{EqgA}) is  similar to Eq. (59) in Ref. \cite{KSD2011}). However, this method has a major drawback which makes it impracticable for multidimensional systems: Each component of the  matrices is given by an integral over frequency, so that  the numerical computation  becomes more and more burdensome as the dimensionality increases\footnote{As a matter of fact, the method has only been applied to one-dimensional systems~\cite{Sab2011}.}. Furthermore, one does not have access to the finite-time behavior  of the generating functions and one cannot  study the special cases where the solutions of the RDEs do not converge to $\hat C^{r,+}_{\rm o}(\lambda)$ and $\hat C^{l,+}_{\rm o}(\lambda)$ (see Sec. \ref{SecV}).  
For all these reasons, the  Riccati approach  is much  more illuminating and numerically effective.

\subsubsection{Effective process}
\label{SubsecIVB4}

From the knowledge of the eigenfunction $r_{\rm o,\lambda}({\bf u})$, one can  build  the so-called effective or driven process  that describes how fluctuations of the observables are created dynamically in the long-time limit~\cite{CT2015}. By construction, a given fluctuation $a$ of the time-intensive observable ${\cal A}_t/t$ in the original process  is realized as a typical value in the effective process. 

Since $r_{\rm o,\lambda}({\bf u})$ is a multivariate Gaussian, this process is again a linear diffusion, 
\begin{align}
\label{EqLeff}
\dot{{\bf u}}_t=\hat A(\lambda) {\bf u}_t+{\boldsymbol \xi}_t\,,
\end{align}
with the same diffusion matrix $D$ as the original process but with a modified drift  $\hat {\bf F}_{\lambda}=\hat A(\lambda){\bf u}$ given by~\cite{CT2015} 
\begin{align}
\label{Eqeffdrift}
\hat {\bf F}_{\lambda}&={\bf F}+D(\nabla \ln r_{\rm o,\lambda}-\lambda {\bf g}_{\rm o})\nonumber\\
&=[A-D(\hat C^{r,+}_{\rm o}(\lambda)+\lambda B_{\rm o})]{\bf u}\ .
\end{align}
Hence, from the definition of the matrix $A_{\rm o}(\lambda)$ [Eq. (\ref{EqtildeA})],
\begin{align}
\label{EqAeff}
\hat A(\lambda)=A_{\rm o}(\lambda)-D\hat C^{r,+}_{\rm o}(\lambda).
\end{align}
By definition of the maximal solution $\hat C^{r,+}_{\rm o}(\lambda)$, the eigenvalues of $-\hat A(\lambda)$ belong to the subset ${\cal S}_{\lambda}^+$ of  eigenvalues of the Hamiltonian matrices [see Eq. (\ref{Eqgraph})] and therefore $-\hat A(\lambda)>0$. In addition, owing to Eqs. (\ref{Eqrel31}) and   (\ref{Eq:rel41}), $\hat A(\lambda)$ only depends on the antisymmetric part of the  matrix $B_{\rm o}$. As  a result, the driven process is  the same for all observables having the same antisymmetric part (but it  is only defined for  $\lambda\in\hat {\cal D}_{\rm o}$, the domain of definition of the SCGF, which  depends on the observable). 

The corresponding invariant density is then given by~\cite{CT2015}
\begin{align}
\label{Eqinvdens}
\hat p_{\lambda}({\bf u})&=r_{\rm o,\lambda}({\bf u})l_{\rm o,\lambda}({\bf u})=\big[ (2\pi)^{n+2}\det \hat \Sigma(\lambda)\big]^{-1/2}\exp\left(-\frac{1}{2}{\bf u}^T \hat \Sigma^{-1}(\lambda){\bf u}\right)\ ,
\end{align}
with 
\begin{align}
\label{EqSigmaeff}
\hat \Sigma(\lambda)= [\hat C^{r,+}_{\rm o}(\lambda)+\hat C^{l,+}_{\rm o}(\lambda)]^{-1}\ .
\end{align}
Using the fact that  $\hat C^{r,+}_{\rm o}(\lambda)$ and $\hat C^{l,+}_{\rm o}(\lambda)$ are solutions of the CAREs (\ref{EqCARE:subeq1}) and ({\ref{EqCARE:subeq2}), respectively, one can easily verify that  $\hat A(\lambda)$ and  $\hat \Sigma(\lambda)$ are related via the Lyapunov equation 
\begin{align}
\label{EqLyap}
\hat A(\lambda) \hat \Sigma(\lambda)+\hat \Sigma(\lambda)\hat A^T(\lambda)=-D\ ,
\end{align}
as it must be. We recall that the sum $\hat C^{r,+}_{\rm o}(\lambda)+\hat C^{l,+}_{\rm o}(\lambda)$ does not depend on the observable due to Eq.  (\ref{Eqrel2leftright}) and is positive definite, which guarantees that the invariant density exists and the driven process is ergodic. 

The above equations are quite general and apply to any  multidimensional linear  diffusions\footnote{Eqs. (\ref{EqAeff}) and  (\ref{Eqinvdens})  correspond to  Eqs. (65) and  (54) in Ref. \cite{DBT2023} via the changes  $\hat A(\lambda)\to- M_k$ and  $\hat \Sigma(\lambda)\to C_k$.} but they take a simpler form for  the   model  under study owing to the fact that $D_{ij}=\delta_{i1}\delta_{j1}$. The elements of the matrix $\hat A(\lambda)$ are then given by
\begin{align}
\hat A_{ij}(\lambda)=[A_{\rm o}(\lambda)]_{ij}-[\hat C^{r,+}_{\rm o}(\lambda)]_{1j}\delta_{i1}. 
\end{align}
Moreover, since  $\hat A(\lambda)$ does not depend on the observable, we can choose  $\rm o=w$, which yields 
\begin{align}
\label{EqAhat}
\hat A_{ij}(\lambda)=A_{ij}-[\hat C^{r,+}_w(\lambda)]_{1j}\delta_{i1} \ .
\end{align}
Therefore, only the first equation of the set of equations (\ref{EqL}) is modified and replaced by
\begin{align}
\label{Eqeff}
\dot  v(t)&=-\big(\frac{1}{Q_0} +[\hat C_w^{r,+}(\lambda)]_{11}\big)v(t)-x(t)+ \frac{g}{Q_0}x_n(t)-\sum_{j=0}^{n}\:[\hat C_w^{r,+}(\lambda)]_{1(j+2)}\: x_j(t)+\xi(t)\nonumber\\
&=-\big[\frac{1}{Q_0} -2f^+(\lambda)\big]v(t)-\big[1 +2f^+(\lambda)(f^+(\lambda)-\frac{1}{Q_0})\big]x(t)+ \frac{g}{Q_0}x_n(t)-\sum_{j=1}^{n}\:[\hat C_w^{r,+}(\lambda)]_{1(j+2)}\: x_j(t)+\xi(t) \ ,
\end{align}
where we have used that $[\hat C_w^{r,+}(\lambda)]_{11}=-2f^+(\lambda)$ [Eq. (\ref{Eqmu:subeq1})] and $[\hat C_w^{r,+}(\lambda)]_{12}= 2f^+(\lambda)(f^+(\lambda)-1/Q_0)$ to derive the second line\footnote{More generally, the element $(12)$ of a  general solution  of the CARE ${\cal R}_{w,\lambda}[\hat C_w^{r,(\alpha)}(\lambda)]=0$  is equal to $2f^{r,(\alpha)}(\lambda)[f^{r,(\alpha)}(\lambda)-1/Q_0]$. One also has $[\hat C_w^{r,(\alpha)}(\lambda)]_{23}=(2\tau/n) f^{r,(\alpha)}(\lambda)\big[(f^{r,(\alpha)}(\lambda))^2- f^{r,(\alpha)}(\lambda)/Q_0+1\big]\big[ f^{r,(\alpha)}(\lambda)-1/Q_0\big]$. The other elements of the matrix $\hat C_w^{r,(\alpha)}(\lambda)$ are not simply expressed in terms of the function $f^{r,(\alpha)}(\lambda)$.}.  
After inserting  the definition of the auxiliary variables  $x_j(t)$ [Eq. (\ref{EqlinearTrick2})],   the equation is finally recast as
\begin{align}
\label{Eqeffdrift2}
&\dot  v(t)=-\big[\frac{1}{Q_0} -2f^+(\lambda)\big]v(t)-\big[1 +2f^+(\lambda)(f^+(\lambda)-\frac{1}{Q_0})\big]x(t)+\int_{-\infty}^t ds\: [\frac{g}{Q_0} \: g_n(t-s,n/\tau)-h_n(\lambda,t-s)]x(s)+\xi(t)\ ,
\end{align}
where 
\begin{align}
h_n(\lambda, t)=\sum_{j=1}^{n}\:[\hat C_w^{r,+}]_{1(j+2)}\:g_j(t,n/\tau)\ .
\end{align}
 Eq. (\ref{Eqeffdrift2}) is another major result of this work. To our knowledge, this is the first  explicit example of  an effective process for a non-Markovian Langevin dynamics.  Comparing with the original process governed by Eq. (\ref{EqlinearTrick1}), we  see that atypical fluctuations of the observable are created in the long-time limit by modifying not only the friction and the spring constants but also the memory kernel.  
  
The  stationary density  associated with Eq. (\ref{Eqeffdrift2}) is  obtained by tracing out the auxiliary variables $u_j$ ($j\ge 2$) in Eq. (\ref{Eqinvdens}), which leads to the bivariate Gaussian distribution
 \begin{align}
 \label{Eqplambda}
\hat p_{\lambda}(x,v)=\frac{1}{\pi Q_0\sqrt{\frac{\hat T_x(\lambda)}{T}\frac{\hat T_v(\lambda)}{T}}}e^{-\frac{1}{Q_0}(\frac{T}{\hat T_x(\lambda)}x^2+\frac{T}{\hat T_v(\lambda)}v^2)} 
\end{align}
 characterized by the two $\lambda$-dependent  temperatures  $\hat T_x(\lambda)=(2T/Q_0)\langle x^2\rangle_{\lambda}=(2T/Q_0)\hat \Sigma_{22}(\lambda)$ and $\hat T_v(\lambda)=(2T/Q_0)\langle v^2\rangle_{\lambda}=(2T/Q_0)\hat \Sigma_{11}(\lambda)$  (here  $\langle ...\rangle_{\lambda}$ denotes an average over stochastic trajectories generated by the effective process in the stationary limit).
Interestingly,  the variances  $\langle x^2\rangle_{\lambda}$ and $\langle v^2\rangle_{\lambda}$, and more generally all elements of the covariance matrix $\hat \Sigma (\lambda)$, can be expressed  in terms of the spectral density $\vert \chi_n(\omega)\vert ^2$, as the SCGF $f^+(\lambda)$.  

To show this, we write the solution of the Lyapunov equation (\ref{EqLyap})  as an integral  over frequency of the spectrum matrix. Since $D_{ij}=\delta_{i1}\delta_{j1}$, we have~\cite{G2004}
 \begin{align} 
\label{EqSigma}
\hat \Sigma_{ij}(\lambda)=\int \frac{d\omega}{2\pi} \langle {\bf u}(\omega){\bf u}^T(-\omega)\rangle_{\lambda}=\int \frac{d\omega}{2\pi}\hat R_{i1}(\lambda,\omega)\hat R_{j1} (\lambda,-\omega) \ ,
\end{align}
where   $\hat R(\lambda,t)=e^{\hat A(\lambda)t}$ ($t>0$) is the response or Green's function which  reads in the Fourier space
\begin{align} 
\label{Eqhatchi}
\hat R(\lambda,\omega)=\int_0^{\infty} e^{\hat A(\lambda)t}e^{i\omega t} dt=-[\hat A(\lambda)+i\omega I_{n+2}]^{-1}\ .
\end{align}
Thanks to Eq. (\ref{EqAhat}), the first column of  $\hat R (\lambda,\omega)$ does not depend on $[\hat C^{r,+}_{\rm o}(\lambda)]_{1j}$ ($j=1,...,n+2)$, and from the expression of the drift matrix $A$ [Eq. (\ref{EqDrift})] we find
\begin{align}
\label{Eqhatchi1}
\hat R_{11}(\lambda,\omega)=\frac{i\omega(i\omega-\frac{n}{\tau})^n}{q^+(\lambda,s=i\omega)}
\end{align}
and
\begin{align}
\label{Eqhatchi2}
\hat R_{i1}(\lambda,\omega)=\frac{(-1)^{i-1}(\frac{n}{\tau})^{i-2}(i\omega-\frac{n}{\tau})^{n+2-i}}{q^+(\lambda,s=i\omega)}\ , \: \: \: i=2,3,...,n+2\ ,
\end{align}
where $q^+(\lambda,s)=\prod_{i=1}^{n+2}[s-s_i^+(\lambda)]$ [Eq. (\ref{Eqqplus})].
Inserting the above expressions into Eq. (\ref{EqSigma}) and using Eqs. (\ref{EqpH}), (\ref{Eqchi}), (\ref{Eqchi1}) and (\ref{Eqsinn}), we then obtain
\begin{subequations}
\label{EqTemp:subeqns}
 \begin{align}
\frac{\hat T_x(\lambda)}{T}=\frac{2}{Q_0}\hat \Sigma_{22}(\lambda)&=\frac{2}{Q_0}\int \frac{d\omega}{2\pi}\frac{(\omega^2+\frac{n^2}{\tau^2})^n}{(-1)^np_{H,\lambda}(s=i\omega)}\nonumber\\
&=\frac{2}{Q_0}\int\frac{d\omega}{2\pi}\frac{1}{\vert \chi_n(\omega)\vert^{-2}-\frac{4\lambda g}{Q_0^2}\omega\sin_n(\omega \tau)}\label{EqTemp:subeq1}\\
\frac{\hat T_v(\lambda)}{T}=\frac{2}{Q_0}\hat \Sigma_{11}(\lambda)&=\frac{2}{Q_0}\int \frac{d\omega}{2\pi}\frac{\omega^2(\omega^2+\frac{n^2}{\tau^2})^n}{(-1)^np_{H,\lambda}(s=i\omega)}\nonumber\\
&=\frac{2}{Q_0}\int \frac{d\omega}{2\pi}\frac{\omega^2}{\vert \chi_n(\omega)\vert^{-2}-\frac{4\lambda g}{Q_0^2}\omega\sin_n(\omega \tau)}\label{EqTemp:subeq2}\ .
\end{align}
\end{subequations}
Likewise, 
\begin{align} 
\label{EqSig21}
\hat \Sigma_{n+2,1}(\lambda)=\langle x_n(t)v(t)\rangle_{\lambda} &=\int \frac{d\omega}{2\pi}\frac{ (\frac{n}{\tau})^ni\omega (i\omega +\frac{n}{\tau})^n}{(-1)^np_H(\lambda,s=i\omega)}\nonumber\\
&=-\int \frac{d\omega}{2\pi}\frac{\omega \sin_n(\omega \tau)}{\vert \chi_n(\omega)\vert^{-2}-\frac{4\lambda g}{Q_0^2}\omega\sin_n(\omega \tau)}\ , 
\end{align}
and by comparing with the expression of $f^+(\lambda)$ [Eq. (\ref{Eqmunlambda})], we find that  $(2g/Q_0^2)\langle x_n(t)v(t)\rangle_{\lambda}=- df^+(\lambda)/d\lambda$. This equality was expected. It  expresses  that if one is interested by a particular fluctuation in which the time-intensive observable $(2g/Q_0^2)(1/t)\int_0^t dt' x_n(t')\circ dx(t')$ takes the value $a$,  this fluctuation is realized  in the long-time limit as a typical value in the effective process, with  $\lambda$ given by $a=-df^+(\lambda)/d\lambda$. (However, while the three observables ${\cal W}_t/t$, ${\cal Q}_t/t$ and $\Sigma_t/t$ are identical in the long-time limit, one must not forget that the  Legendre duality only holds for $\lambda$ in the interval ${\hat D}_{\rm o}$ which depends on the observable.)
\begin{figure}[hbt]
\begin{center}
\includegraphics[trim={0cm 0cm 0cm 0cm},clip,width=10cm]{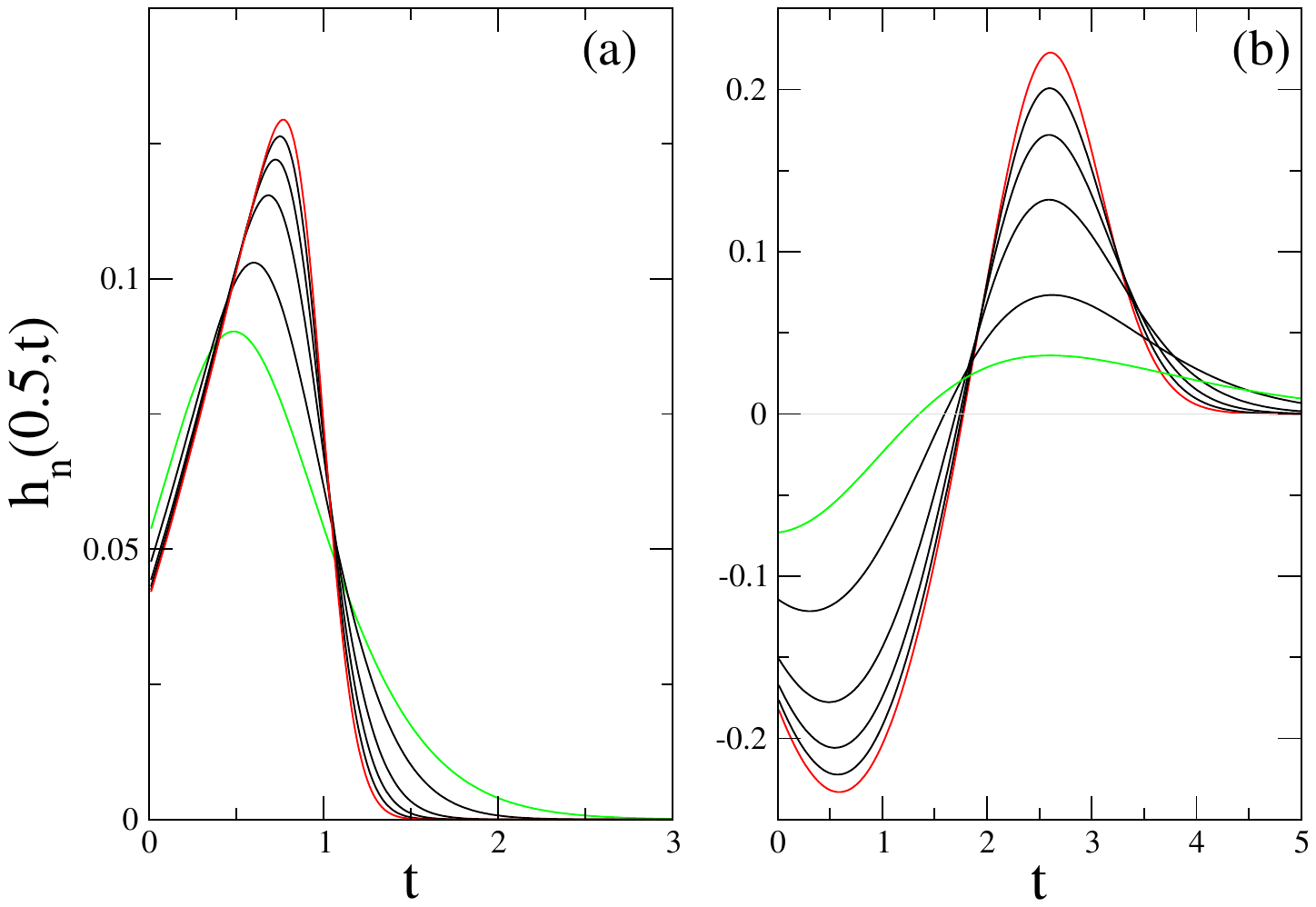}
\caption{ (Color on line) The function $h_n(\lambda, t)$ for  $\lambda=0.5$, $\tau=1$ (a), $\tau=3$ (b), and increasing values of $n$:  $n=5$ (green) $10,20,30,40,50$ (red).}
\label{Fig15}
\end{center}
\end{figure}

Finally, we consider the limit $n\to \infty$. As illustrated in Fig. \ref{Fig15}, the function $h_n(\lambda, t)$  appears to converge to a well-defined limit $h_{\infty}(\lambda, t)$. (We remind the reader that the SCGF $f^+(\lambda)$ and the matrix $\hat C_{\rm o}^{r,+}(\lambda)$ depend on $n$.)  We thus expect  the driven process for the discrete delay to be governed by the equation
\begin{align}
\label{Eqeffdrift3}
\dot  v(t)&=-\big[\frac{1}{Q_0} -2f^+_{\infty}(\lambda)\big]v(t)-\big[1 +2f^+_{\infty}(\lambda)(f^+_{\infty}(\lambda)-\frac{1}{Q_0})\big]x(t)+\frac{g}{Q_0}x(t-\tau)-\int_{-\infty}^t ds\: h_{\infty}(\lambda,t-s)x(s)+\xi(t)\ ,
\end{align}
where $f^+_{\infty}(\lambda)$ is given by Eq. (\ref{Eqmuinf}) (see also Eq. (53) in Ref. \cite{RTM2017}).  The memory kernel does not reduce to a delta function in this limit. This interesting issue deserves a more complete study that we leave to future work.

Furthermore, we obtain from Eqs. (\ref{EqTemp:subeqns})
\begin{subequations}
\label{EqTemp1:subeqns}
 \begin{align}
\frac{\hat T_x(\lambda)}{T}\to\frac{2}{Q_0}\int \frac{d\omega}{2\pi}\frac{1}{\vert \chi_{\infty}(\omega)\vert^{-2}-\frac{4\lambda g}{Q_0^2}\omega\sin(\omega \tau)}\label{EqTemp1:subeq1}\\
\frac{\hat T_v(\lambda)}{T}\to\frac{2}{Q_0}\int\frac{d\omega}{2\pi}\frac{\omega^2}{\vert \chi_{\infty}(\omega)\vert^{-2}-\frac{4\lambda g}{Q_0^2}\omega\sin(\omega \tau)}\label{EqTemp1:subeq2}\ ,
\end{align}
\end{subequations}
 with $\chi_{\infty}(\omega)$ given by Eq. (\ref{Eqchiinf}).

\subsubsection{Role of (temporal) boundary terms}
\label{SubsecIVB5}

As we have seen in the previous sections, the average over the initial and final points of the stochastic trajectories may induce  finite-time divergences in the moment generating functions  and a reduction of the  domain of existence of the SCGF from ${\cal D}_H$ to $\hat {\cal D}_{\rm o}$, which in turn induces linear branches in the rate function $I_{\rm o}(a)$. 

It is natural to put this in relation to the role of  the so-called ``boundary"  terms in the dynamical observables, i.e., terms that are not extensive in time such as $\Delta E$, the change in the internal energy of the system, which differentiates  the  heat from the work.  This issue  is  well documented in the literature, both theoretically~\cite{VC2003,Fa2002,V2006,BJMS2006,PRV2006,HRS2006,TC2007,Sab2011,N2012,NP2012} and experimentally (see  Ref. \cite{C2017} and references therein).  In particular, the breakdown of the Gallavotti-Cohen fluctuation relation~\cite{GC1995,K1998,LS1999}  can be attributed to such boundary terms which become relevant in the case of an unbounded potential. Things are more complicated in the presence of a continuous (non-Markovian) feedback, but  fluctuations of work, heat, and entropy production are indeed different, as discussed in our previous work~\cite{RTM2017}.

However, at odds with the assumption made in Ref. \cite{RTM2017}, the present numerical calculations  show that $\hat {\cal D}_w\subset {\cal D}_H$ and the rate function $I_w(a)$ has linear branches (see Fig. \ref{Figfplus}) even though the stochastic work  defined by Eq. (\ref{EqW}) does not contain an explicit  boundary term. This may come as a surprise, and one may   argue that the definition (\ref{EqW}) is misleading and that a boundary term does exist by decomposing $\beta {\cal W}_t=\int_0^t (B_w {\bf u}_{t'})\circ d{\bf u}_{t'}$ as
\begin{align}
\label{EqWantisym}
\beta {\cal W}_t= \frac{g}{Q_0^2} [x_n(t)x(t)-x_n(0)x(0)]+\frac{g}{Q_0^2}\int_0^t [x_n(t')\circ dx(t')-x(t')\circ dx_n(t')]\ .
\end{align}  
This amounts to splitting the matrix $B_w$ defined by  Eq.  (\ref{EqBw}) into its symmetric and antisymmetric components, i.e., 
\begin{align}
B_w=B_{w,sym}+B_{w,antisym}= \frac{g}{Q_0^2} \left(
 \begin{bmatrix}
 0&0&0 &...&0\\
 0&0&0&...& 1 \\
  0&0&0 &...& 0\\
  \vdots & \vdots  & \vdots & \vdots& \vdots \\
0&1&0 &...& 0&
\end{bmatrix}
+\begin{bmatrix}
 0&0&0 &...&0\\
 0&0&0&...& 1 \\
  0&0&0 &...& 0\\
  \vdots & \vdots  & \vdots & \vdots& \vdots \\
0&-1&0 &...& 0&
\end{bmatrix}\right)\ .
\end{align}
The symmetric component $B_{w,sym}$ yields the  boundary term in Eq. (\ref{EqWantisym})  by direct integration over time. 
\begin{figure}[hbt]
\begin{center}
\includegraphics[trim={0cm 0cm 0cm 0cm},clip,width=10cm]{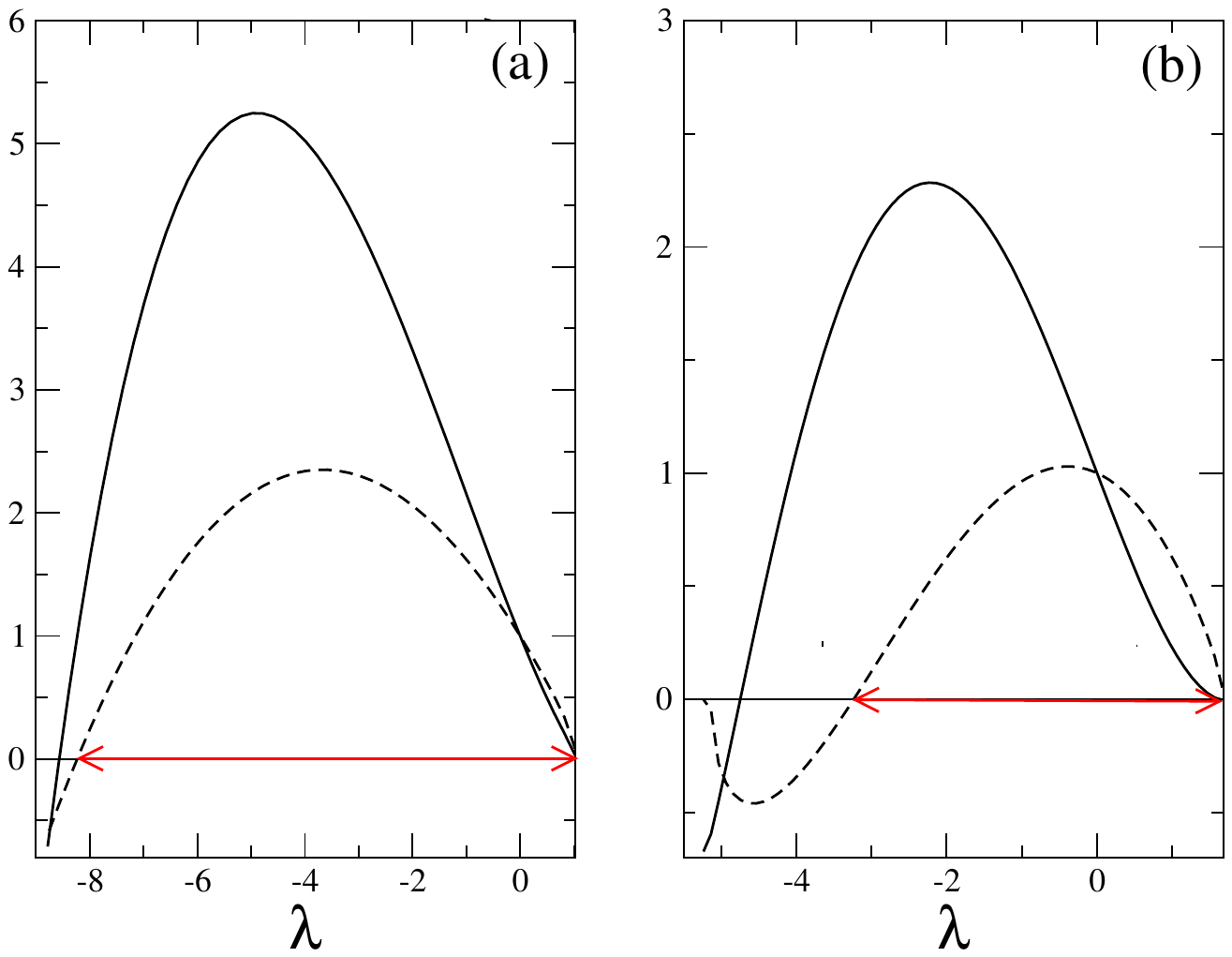}
\caption{ (Color on line) $\det \hat C^{l,+}_{w,antisym}(\lambda)/\det C$ (solid lines) and $\det (\hat C^{r,+}_{w,antisym}(\lambda)+C)/ \det C$ (dashed lines) as a function of $\lambda$. The  arrows indicate the range $\hat {\cal D}_w$ of values of $\lambda$ for which the matrices  $\hat C^{l,+}_{w,antisym}(\lambda)$ and $\hat C^{r,+}_{w,antisym}(\lambda)+C$   are both positive definite (compare with  Fig. \ref{Fig7}). (a) $\tau =1$: ${\cal D}_H=(-472.08,1.019)$, $\hat {\cal D}_w=(-8.232,1.019)$.  (b) $\tau=3$: ${\cal D}_H=(-5.118,1.665)$, $\hat {\cal D}_w (-3.235,1.612)$.}
\label{Fig10}
\end{center}
\end{figure}

According to the recent work of du Buisson and Touchette (see Section III C of Ref. \cite{DBT2023} or Sec.  4.1.3 of Ref. \cite{DB2023} devoted to  linear current-type observables), this symmetric component should be responsible for the reduction of the  domain of existence of the SCGF from ${\cal D}_H$ to $\hat {\cal D}_w$\footnote{It may be that we misinterpret the analysis done in Ref. \cite{DBT2023} and that it only states  that the symmetric part of the matrix $B_{\rm o}$ leads to an {\it additional} reduction of the domain of existence of the SCGF. However, this is not what comes from the reading of the paper.}. The results presented in Fig. \ref{Fig10}, in which  only the second term of Eq. (\ref{EqWantisym}) is taken into account, show that this is not true. The interval  for which the matrices $\hat C^{l,+}_{w,antisym}(\lambda)$ and  $\hat C^{r,+}_{w,antisym}(\lambda)+C$  are both positive definite is enlarged compared with the one associated with  the full work $\beta {\cal W}_t$ (see Fig. \ref{Fig7}) but it is still {\it smaller} than ${\cal D}_H$. 

We believe that the problem with the analysis done in Ref. \cite{DBT2023}, which treats separately the  cases where the matrix  $B_{\rm o}$ is purely antisymmetric and that where it also has a nonzero symmetric part, is that it only  focuses on the time evolution of the generating function $G^r_{\rm o,\lambda}({\bf u}_0,t)$.  As a result, the role of the matrix $\hat C^{l,+}_{\rm o}(\lambda)$ is not clearly recognized, as we now briefly explain. 

In the long-time limit, $G^r_{\rm o,\lambda}({\bf u}_0,t)$ is given by Eq. (\ref{EqAsym:subeq1}) which only involves the matrix $\hat C^{r,+}_{\rm o}(\lambda)$. However,   $G^r_{\rm o,\lambda}({\bf u}_0,t)$ results from the average of $G_{\rm o,\lambda}({\bf u},t\vert {\bf u}_0)$ over the terminal state ${\bf u}$. Therefore, from Eq. (\ref{EqZAasym}), $G^r_{\rm o,\lambda}({\bf u}_0,t)$ is finite if the  left eigenfunction $l_{\rm o,\lambda}({\bf u})$ is integrable, which  requires the matrix $\hat C^{l,+}_{\rm o}(\lambda)$ to be positive definite (this is why the determinant of  $\hat C^{l,+}_{\rm o}(\lambda)$ appears in the expression (\ref{EqEigen2:subeq1}) of the right eigenfunction). 

Surprisingly, the integration of  $G_{\rm o,\lambda}({\bf u},t\vert {\bf u}_0)$ over ${\bf u}$  is not taken into account  in Sec. III C.1 of Ref. \cite{DBT2023} which considers the case of a purely antisymmetric matrix $B_{\rm o}$. It is only stated that the SCGF exists if the drift matrix  of the effective process (cf. Eq. (65) in Ref. \cite{DBT2023}) is positive definite. As  we noted  after Eq. (\ref{EqAeff}), this condition is automatically satisfied when $\lambda\in {\cal D}_H$.

On the other hand, for a general matrix $B_{\rm o}$ (Sec. III C.2 of Ref. \cite{DBT2023}),  an additional condition is derived from the integral of $G_{\rm o,\lambda}({\bf u},t\vert {\bf u}_0)$ over  ${\bf u}$. A certain matrix, defined by Eq. (76), must be positive definite. Translated into our notations\footnote{See footnotes 5 and 27 for the correspondence between the notations of Ref. \cite{DBT2023} and those used in this work. In addition,   the matrix $B^*_k$  in Eq. (76) corresponds to $-(1/2) \hat C^{r,+}_{\rm o,antisym}(\lambda)$.}, this condition reads 
 \begin{align}
\label{EqcondCl}
[\hat C^{r,+}_{\rm o,antisym}(\lambda)+\hat C^{l,+}_{\rm o,antisym}(\lambda)]-\hat C^{r,+}_{\rm o,antisym}(\lambda)+\lambda B_{\rm o, sym}>0\ .
\end{align}
Using  Eq. (\ref{Eq:rel42}), we see that it is just  the condition $\hat C_{\rm o}^{l,+}(\lambda)>0$! 
So the positive definiteness of the matrix  $\hat C_{\rm o}^{l,+}(\lambda)$ is always required, regardless of the symmetry of the matrix $B_{\rm o}$. Moreover, one may have $\hat {\cal D}_w\subset {\cal D}_H$ even  if $B_{\rm o}$ is purely antisymmetric and no explicit boundary term is present. Note that this behavior is not specific to the present model: see, e.g., the two-dimensional model  studied in Ref. \cite{NKP2013}  where the matrix $B_w$ is purely antisymmetric or the model of a confined active particle studied in Ref. \cite{SGSZ2023} and revisited in Appendix \ref{AppendJ}\footnote{In the latter model, the matrix $B_w$ contains a symmetric part, but we have performed numerical calculations that show that $\hat {\cal D}_w\subset {\cal D}_H$ even when taking into account the antisymmetric part only.}.

\subsection{Behavior of the SCGF at the boundaries of its interval of definition} 
\label{SubsecIVC}

As signaled by Eqs. (\ref{EqEquiv:subeqns}) and illustrated numerically  by Figs. \ref{Fig8} and \ref{Fig9}, something special happens for the solutions of the RDEs at the boundaries of the interval  $\hat {\cal D}_{\rm o}$ (throughout this section we assume that $\hat {\cal D}_{\rm o}\subset {\cal D}_H$ where we recall that $\subset$ denotes a strict inclusion). This may look as a minor issue, but we will see in Sec. V that it cannot   be ignored in order to understand the behavior of the fluctuations of heat and entropy production for $\lambda=1$.

For concreteness we suppose hereafter that $\det \hat C^{l,+}_{\rm o}(\lambda_{\rm o1})=0$ while  $\det (\hat C^{r,+}_{\rm o}(\lambda_{\rm o2})+C)=0$ like in  the example of Fig. \ref{Fig7} (a similar discussion would take place in the opposite case).  Eqs. (\ref{EqEquiv:subeqns}) then tell us that  $C^r_{\rm o}(\lambda_{\rm o1},t)\not \to \hat C^{r,+}_{\rm o}(\lambda_{\rm o1})$ and $C^l_{\rm o}(\lambda_{\rm o1},t) \to \hat C^{l,+}_{\rm o}(\lambda_{\rm o1})$ as $t\to \infty$ whereas  $C^l_{\rm o}(\lambda_{\rm o2},t)\not \to \hat C^{l,+}_{\rm o}(\lambda_{\rm o2})$ and $C^r_{\rm o}(\lambda_{\rm o2},t) \to \hat C^{r,+}_{\rm o}(\lambda_{\rm o2})$. Therefore, there is a discontinuity in the solution of the RDE (\ref{EqRic:subeq1}) or (\ref{EqRic:subeq2}) and this naturally raises the question: What is the value of  the SCGF for $\lambda=\lambda_{\rm o1}$ and $\lambda=\lambda_{\rm o2}$? We note that it is taken for granted in Ref. \cite{DBT2023} that the SCGF diverges (see Eq. (78); see also  Eq. (4.104) in Ref. \cite{DB2023}).

As we now discuss, two different scenarios may occur.

 \subsubsection{Attraction to a non-maximal solution of the CARE}
\label{SubsecIVC1}

The first possible scenario is that the solution of the RDE is attracted to a fixed point that is not the maximal solution of the  CARE. This is the behavior  observed  in  Fig. \ref{Fig9} where  $\lim_{t \to \infty}\det(C^r_w(\lambda_{w1},t)+C)/\det C\approx 1.17$ whereas $\det (\hat C^{r,+}_w(\lambda_{w1})+C)/\det C\approx 2.98$. Hence 
 $C^r_w(\lambda_{w1},t)$  is attracted to  $\hat C_w^{r,(\alpha)}(\lambda_{w1})\ne \hat C_w^{r,+}(\lambda_{w1})$.
 
 Accordingly, the asymptotic expression (\ref{EqAsym:subeq1}) of the  generating function $G^r_{\rm o,\lambda_{\rm o1}}({\bf u}_0,t)$  is   replaced by
\begin{align}
G^r_{\rm o,\lambda_{\rm o1}}({\bf u}_0,t)\sim g^{r,(\alpha)}_{\rm o}(\lambda_{\rm o1})\exp\left(-\frac{1}{2}{\bf u}_0^T\hat C_{\rm o}^{r,(\alpha)}(\lambda_{\rm o1}){\bf u}_0\right)e^{f^{(\alpha)}(\lambda_{\rm o1})t}\ ,
\end{align}
where $f^{(\alpha)}(\lambda_{\rm o1})$ is given by Eq. (\ref{Eqmulmur}) and 
\begin{align} 
\label{Eqprefgr}
g^{r,(\alpha)}_{\rm o}(\lambda_{\rm o1})=\exp\left(\int_0^{\cal \infty} dt\:[f^r_{\rm o}(\lambda_{\rm o1},t)-f^{(\alpha)}(\lambda_{\rm o1})]\right)\ .
\end{align}
It remains to integrate $G^r_{\rm o,\lambda_{\rm o1}}({\bf u}_0,t)p({\bf u}_0)$ over ${\bf u}_0$.  Whereas $\hat C^{r,+}_{\rm o}(\lambda_{\rm o1})+C>0$ by assumption, it is not guaranteed that   $\hat C^{r,(\alpha)}_{\rm o}(\lambda_{\rm o1})+C>0$  because $\hat C^{r,(\alpha)}_{\rm o}(\lambda_{\rm o,1})<\hat C^{r,+}_{\rm o}(\lambda_{\rm o,1})$.  If this positivity condition is not satisfied, the integral diverges and the SCGF is {\it  infinite} for $\lambda= \lambda_{\rm o1}$.
On the other hand,  if $\hat C^{r,(\alpha)}_{\rm o}(\lambda_{\rm o1})+C>0$, as is the case in the example of Fig. \ref{Fig9}, one finds
 \begin{align} 
G_{\rm o,\lambda_{\rm o1}}(t)\sim g_{\rm o}(\lambda_{\rm o1})e^{f^{(\alpha)}(\lambda_{\rm o1})t} \ ,
\end{align} 
with
\begin{align}
\label{Eqga2}
g_{\rm o}(\lambda_{\rm o1})= \Big[\frac{\det \big(C+\hat C^{r,(\alpha)}_{\rm o}(\lambda_{\rm o1})\big)}{\det C}\Big]^{-1/2}\exp\left(\int_0^{\cal \infty} dt\:[f^r_{\rm o}(\lambda_{\rm o1},t)-f^{(\alpha)}(\lambda_{\rm o1})]\right)\ .
\end{align}
In consequence,  the SCGF $\mu_{\rm o}(\lambda_{\rm o1})$ is  {\it finite} and equal to $ f^{(\alpha)}(\lambda_{\rm o1})$. Furthermore, since $f^{(\alpha)}(\lambda_{\rm o1})>f^+(\lambda_{\rm o1})$ [Eq. (\ref{EqInequal})],  the SCGF exhibits a {\it positive jump discontinuity} at $\lambda=\lambda_{\rm o1}$. Note also that the actual pre-exponential factor $g_{\rm o}(\lambda_{\rm o1})$ is finite.
\begin{figure}[hbt]
\begin{center}
\includegraphics[trim={0cm 0cm 0cm 0cm},clip,width=10cm]{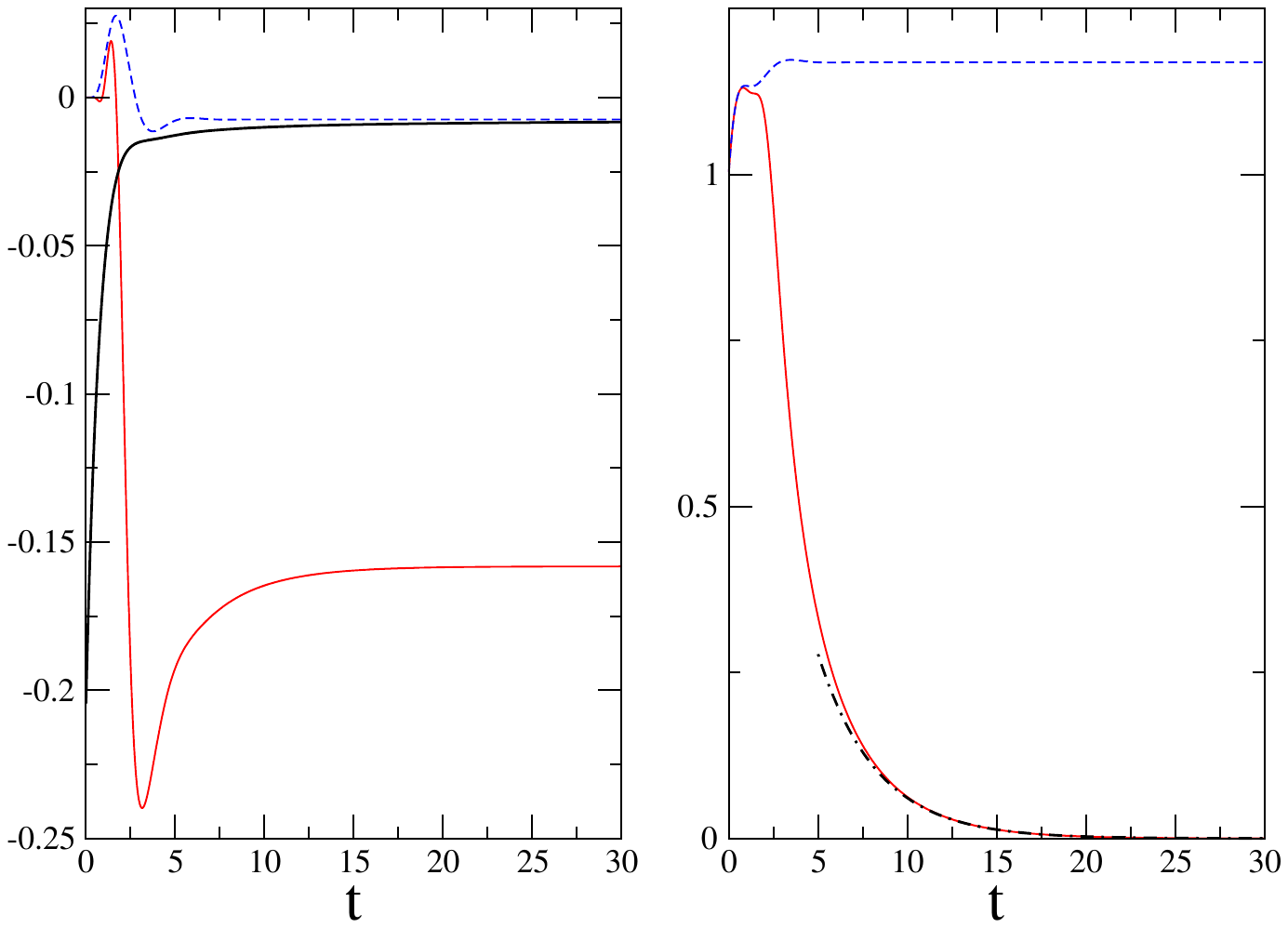}
\caption{ (Color on line) Time evolution of the solutions of the RDEs (\ref{EqRic:subeqns}) for the observable ${\cal W}_t$ for $\tau=1$ and $\lambda=\lambda_{w1}\approx -0.9635155087944421810$.  (a) $f_w^l(\lambda_{w1},t)$ (red solid line), $f_w^r(\lambda_{w1},t)$ (blue dashed line), $(1/t)\ln G_w(\lambda_{w1},t)$ (black solid  line).    (b) $\det C_w^l(\lambda_{w1},t)/\det C$ (red solid line) and $\det (C_w^r(\lambda_{w,1},t)+C)/\det C$ (blue dashed line).  The dashed-dotted black line represents the fit $1.26 \exp(-0.302 t)$ to $\det C_w^l(\lambda_{w1},t)/\det C$ for $t>10$.}
\label{Fig11}
\end{center}
\end{figure}

How can this result be compatible with that obtained by integrating  $G^l_{\rm o,\lambda_{\rm o1}}({\bf u},t)$ over the final state ${\bf u}$? The long-time behavior of $G^l_{\rm o,\lambda_{\rm o1}}({\bf u},t)$  is still described  by Eq. (\ref{EqAsym:subeq2}) but $\det C^l_{\rm o}(\lambda_{\rm o1},t)\to \det \hat C^{l,+}_{\rm o}(\lambda_{\rm o1})=0$, which makes the expression  (\ref{EqgA1:subeq2})  of the pre-exponential factor  divergent. Therefore, the direct integration of  Eq. (\ref{EqAsym:subeq2}) over ${\bf u}$ does not give the proper result. To resolve the puzzle one must go back to the expression (\ref{EqDynZ:subeq2}) of the generating function and carefully inspect the  long-time behavior  of the solution of the RDE (\ref{EqRic:subeq2}).  It is then found that $\det C^l_{\rm o}(\lambda_{\rm o1},t)\sim e^{-2\delta t}$  with  $\delta =f^{(\alpha)}(\lambda_{\rm o1})-f^+(\lambda_{\rm o1})>0$.  
As $t \to \infty$,  the  factor $e^{f^+(\lambda_{\rm o1})t}$ in the denominator of  Eq. (\ref{EqDynZ:subeq2}) then cancels the same factor in the numerator that comes from the limit of $e^{\int_0^t f^l_{\rm o}(\lambda_{\rm o1},t)dt}$, and one eventually  recovers that $\mu_{\rm o}(\lambda_{\rm o1})=f^{(\alpha)}(\lambda_{\rm o1})$, as it must be.

This nontrivial behavior  is illustrated numerically in Fig. \ref{Fig11} for the observable ${\cal W}_t$ and the same  parameters  as in Figs. \ref{Fig7}(a),  \ref{Fig8}, and \ref{Fig9}. Note that there is no explicit  expression of $\lambda_{w1}$ so  its numerical value is  only known approximately   by solving  the equation $\det C_w^{l+}(\lambda_{w1})=0$. As a result, the solution of the RDE (\ref{EqRic:subeq1}) is ultimately attracted to $\hat C_w^{r,+}(\lambda_{w1})$ and a very high  precision in the determination of $\lambda_{w1}$ is required  to observe the convergence to $\hat C_w^{r,(\alpha)}(\lambda_{w1})$ within a  sufficiently large time window. Then, we see in Fig. \ref{Fig11}(a) that the function $f_w^l(\lambda_{w1},t)$   converges to $f^+(\lambda_{w1})\simeq -0.158$ whereas both $f_w^r(\lambda_{w1},t)$ and  $(1/t)\ln G_{w,\lambda_{w1}}(t)$ converge to  $f^{(\alpha)}(\lambda_{w1})\simeq -0.007$ which is thus the actual value of the SCGF $\mu_w(\lambda_{w1})$. We also observe that $2[f^+(\lambda_{w1})-f^{(\alpha)}(\lambda_{w1}]\approx -0.302$, which is in perfect agreement with the exponential fit to $\det C_w^l(\lambda_{w1},t)/\det C$ shown in Fig. \ref{Fig11}(b) and confirms the theoretical analysis. The discontinuity in the SCGF for $\lambda=\lambda_{w1}$ (which does not contradict the required convexity of the SCGF~\cite{T2009} since $\mu_w(\lambda)=+\infty$ for $\lambda<\lambda_{w1}$)) signals an abrupt change in the mechanism responsible for the fluctuations of ${\cal W}_t$ which can interpreted as a {\it dynamical phase transition}. 

More generally, one may wonder whether the exceptional value $f^{(\alpha)}(\lambda_{\rm o1})\ne f^+(\lambda_{\rm o1})$ of the SCGF could also be obtained by solving the spectral problem
\begin{subequations}
\label{EqSP2:subeqns}
\begin{align} 
{\cal L}_{\rm o,\lambda_{\rm o1}}r_{\rm o,\lambda_{\rm o1}}({\bf u}_0)&=f^{(\alpha)}(\lambda_{\rm o1}) r_{\rm o,\lambda_{\rm o1}}({\bf u}_0)\label{EqSP2:subeq1}\\
{\cal L}_{\rm o,\lambda_{\rm o1}}^{\dag}\, l_{\rm o,\lambda_{\rm o1}}({\bf u})&=f^{(\alpha)}(\lambda_{\rm o1}) l_{\rm o,\lambda_{\rm o1}}({\bf u})\label{EqSP2:subeq2}\ ,
\end{align}
\end{subequations}
with $r_{\rm o,\lambda}({\bf u}_0)=g_{\rm o}^r(\lambda_{\rm o1})\exp\left(-\frac{1}{2}{\bf u}_0^T\hat C^{r,(\alpha)}_{\rm o}(\lambda_{\rm o1}){\bf u}_0\right)$  and   $g_{\rm o}^r(\lambda_{\rm o1})$ given by Eq.  (\ref{Eqprefgr}). This requires  the existence of a left eigenfunction $l_{\rm o,\lambda_{\rm o1}}({\bf u})$  such that the normalization conditions (\ref{Eqnorm:subeqns}) are satisfied. If so, the driven process is again a linear diffusion with a drift matrix given by Eq. (\ref{EqAeff}).  The invariant density is thus a multidimensional Gaussian  and  the left eigenfunction is also Gaussian from Eq. (\ref{Eqinvdens}). In other words, one must have $l_{a,\lambda}({\bf u}_0)=g_{\rm o}^l(\lambda_{\rm o1})\exp\left(-\frac{1}{2}{\bf u}^T\hat C^{l,(\alpha)}_{\rm o}(\lambda_{\rm o1}){\bf u}\right)$ and the problem thus amounts to computing the matrix $\hat C^{l,(\alpha)}_{\rm o}(\lambda_{\rm o1})$ and checking whether the normalization conditions are satisfied. 

One first needs to identify the corresponding set ${\cal S}_{\lambda_{\rm o1}}^{(\alpha)}$ of eigenvalues of the Hamiltonian matrices. To be concrete, let us take the same example as in Fig. \ref{Fig11}. For these values of the parameters, we find that all eigenvalues of the Hamiltonian matrices are simple with $3$ pairs of complex conjugate eigenvalues $(s_1^+,\bar s_1^+)$, $(s_2^+,\bar s_2^+)$, $(s_3^+,\bar s_3^+$)  and one real eigenvalue $s_4^+$ on the r.h.s. of the complex plane ($s_4^+$ is the eigenvalue with the smallest real part). The  CARE (\ref{EqCARE:subeq1}) has thus $16$ real, symmetric solutions~\cite{note7}, and by using  Eq. (\ref{Eqmulmur}) we find that  the value $f^{(\alpha)}(\lambda_{w1})\approx -0.007$ is obtained from the set ${\cal S}_{\lambda_{w,1}}^{(\alpha)}=\{s_1^+,\bar s_1^+,s_2^+,\bar s_2^+,s_3^+,\bar s_3^+,s_4^-\}$ with  $s_4^-=-s_4^+$.  This allows us to compute the corresponding matrix  $\hat C^{l,(\alpha)}_w(\lambda_{w1})$. We then observe that neither $\hat C^{l,(\alpha)}_w(\lambda_{\rm o,1})$ nor  $\hat C^{r,(\alpha)}_w(\lambda_{w1})+\hat C^{l,(\alpha)}_w(\lambda_{w1})$ are positive definite. Hence $\hat C^{l,(\alpha)}_w(\lambda_{w1})$ is not an acceptable eigenfunction and we  conclude that the SCGF cannot be obtained from the spectral problem. We believe that this is a general feature although we have not been able to prove this conjecture so far\footnote{Note in passing that the matrix $\hat C_w^{r,(\alpha)}(\lambda_{w1})$ is  the only solution of the CARE (\ref{EqCARE:subeq1}) besides $C_w^{r+}(\lambda_{w1})$ that satisfies the condition $\hat C_w^{r,(\alpha)}(\lambda_{w1})+C>0$. However,  this is not a general feature. For instance, if we now focus on the upper limit of the interval $\hat {\cal D}_w$  where  $\det (\hat C_w^{r,+}(\lambda_{w2})+C)=0$ [see Fig. \ref{Fig7}(a)],  we find that two solutions of the  CARE (\ref{EqCARE:subeq2}) besides $C_w^{l+}(\lambda_{w2})$  are positive definite and that the solution of the RDE (\ref{EqRic:subeq2}) converges to the solution $\hat C^{l,(\alpha)}_w(\lambda_{w2})$ which is  the ``closest"   to  the matrix $\hat C^{l,+}_w(\lambda_{w2})$ with respect to  standard matrix norms (it also gives a smaller value of $f^{(\alpha)}(\lambda_{w2})$ than the other  fixed point).}.

\subsubsection{Oscillatory behavior}
 \label{SubsecIVC2}
 
Another possible scenario is that the solution of the RDE does not converge at all for $\lambda=\lambda_{\rm o1}$ or $\lambda=\lambda_{\rm o2}$. Suppose for instance that $\det (\hat C^{l,+}_{\rm o}(\lambda_{\rm o2})\ne 0$ and  $\det (\hat C^{r,+}_{\rm o}(\lambda_{\rm o2})+C)=0$. As illustrated in Fig. \ref{Fig13} for $\rm o=w$ and $\tau=0.2$, $C^r_w(\lambda_{w2},t) \to \hat C^{r,+}(\lambda_{w2})$ with $f^r(\lambda_{w2},t)\to f^+(\lambda_{w2})\approx 0.047$ whereas $C^l_w(\lambda_{w2},t)$, after some short transient, oscillates  between $\hat C^{l,+}(\lambda_{w2})$ and another fixed point $\hat C^{l,(\alpha)}(\lambda_{w2})$  corresponding to $f^{(\alpha)}(\lambda_{w2})\approx 0.607$ [see Fig.  \ref{Fig13}(a)]. More precisely, it is observed that each element of the matrix $C^l_w(\lambda_{w2},t)$ behaves  as  
\begin{align}
\label{EqOsc1}
[C^l_w(\lambda_{w2},t)]_{ij}\sim [\hat C^{l,+}_w(\lambda_{w2})]_{ij}\frac{1+h(t+\delta_{ij})}{2}+[\hat C^{l,(\alpha)}_w(\lambda_{w2})]_{ij}\frac{1-h(t+\delta_{ij})}{2} \ , 
\end{align}
where $h(t)$ is a periodic function varying between $-1$ and $+1$ and $\delta_{ij}$ is a phase shift. This yields from Eq. (\ref{EqSol1:subeq2})
\begin{align}
\label{EqOsc2}
G^l_{w,\lambda_{\rm o2}}({\bf u},t)\sim g^l_w(\lambda_{w2})\exp\left(-\frac{1}{2}{\bf u}^TC^l_w(\lambda_{w2},t){\bf u}\right)e^{\bar f(\lambda_{w2})t}\ ,
\end{align}
with   
\begin{align}
g^l_w(\lambda_{w2})=\exp\left(\int_0^{\cal \infty} dt\:[f^l_w(\lambda_{w2},t)-\bar f(\lambda_{w2})]\right)\ ,
\end{align}
and 
\begin{align}
\bar f(\lambda_{w2})=f^+(\lambda_{w2})\frac{1+\bar h}{2}+f^{(\alpha)}(\lambda_{w2})\frac{1-\bar h}{2}\ ,
\end{align}
where $\bar h$ is the average of $h(t)$ over a single period.  Since $\det  C^l_w(\lambda_{w2},t)> 0$ [see Fig. \ref{Fig13}(b)], the  integral of  $G^l_{w,\lambda_{w2}}({\bf u},t)$ over ${\bf u}$  then yields 
\begin{align} 
G_{w,\lambda_{w2}}(t)\sim g_{w}(\lambda_{w2},t)e^{\bar f(\lambda_{w2})t} \ ,
\end{align} 
with
\begin{align}
g_{w}(\lambda_{w2},t)=\big[\frac{\det C^l_w(\lambda_{w2},t)}{\det C}\Big]^{-1/2}\exp\left(\int_0^{\infty} dt\:[f^l_w(\lambda_{w2},t)-\bar f(\lambda_{w2})]\right)\ .
\end{align}
\begin{figure}[hbt]
\begin{center}
\includegraphics[trim={0cm 0cm 0cm 0cm},clip,width=10cm]{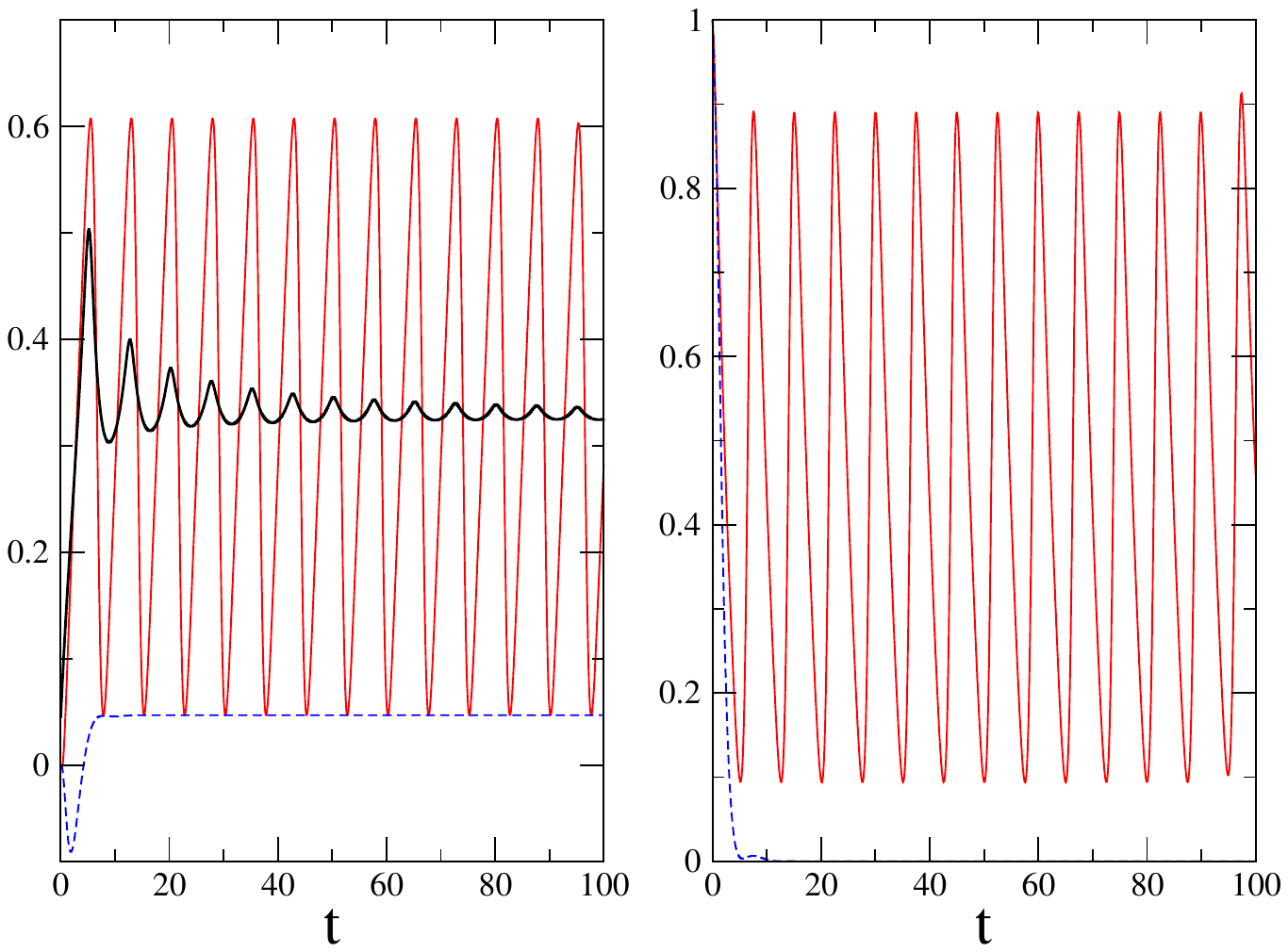}
\caption{ (Color on line) Time evolution of the solutions of the RDEs (\ref{EqRic:subeqns}) for the observable ${\cal W}_t$ for $\tau=0.2$ and $\lambda=\lambda_{w2}\approx 0.3998261896653990830031041$. (a) $f_w^l(\lambda_{w2},t)$ (red solid line), $f_w^r(\lambda_{w2},t)$ (blue dashed line), $(1/t)\ln G_w(\lambda_{w2},t)$ (black solid  line).    (b) $\det C_w^l(\lambda_{w2},t)/\det C$ (red solid line) and $\det (C_w^r(\lambda_{w2},t)+C)/\det C$ (blue dashed line). }
\label{Fig13}
\end{center}
\end{figure}

As can be seen in Fig. \ref{Fig13}(a),  the generating function  exhibits decreasing oscillations and $(1/t)\ln G_{w,\lambda_{w2}}(t)\to \bar f(\lambda_{w2})\approx 0.33$, which is  the genuine value of the SCGF. Moreover, the sub-exponential factor is time-dependent, which is an unusual feature. As it must be, the same result is obtained from the RDE (\ref{EqDynZ:subeq1}), with $C^r_w(\lambda_{w2},t)\to \hat C^{r,+}(\lambda_{w2})$. This is due again to the fact that  the determinant of $C^r_w(\lambda_{w2},t)+C$  decreases exponentially to $0$ as $t \to \infty$ in a very specific way.

\section{The special case $\lambda=1$}
\label{SecV}

In this final section, we  study the  special case $\lambda=1$ in relation with the fluctuation relations for the heat ${\cal Q}_t$ and the (apparent) entropy production $\Sigma_t$.  This is an important application of the general formalism described in the preceding sections and it also motivates our focus on the  behavior of  the SCGF at the limits of its domain of definition. For concreteness, we  restrict our attention to the model with distributed  time delay  described by Eq. (\ref{EqlinearTrick1}).

\subsection{Preliminaries}
\label{SubsecVA}

A first observation is that  the characteristic polynomial  of the Hamiltonian matrices for $\lambda=1$ admits  the  factorization 
\begin{align} 
\label{EqRelation4}
p_H(1,s)=(-1)^n q^*(s)q^*(-s),
\end{align} 
with
\begin{align} 
\label{EqpAtilde}
q^*(s)=(-\frac{n}{\tau})^n[(s^2+\frac{s}{Q_0}+1)(1-\frac{s\tau}{n})^n-\frac{g}{Q_0}] \ . 
\end{align}
This results from simple manipulations of Eq. (\ref{EqpH}).
From the general property of algebraic Riccati equations [Eq. (\ref{Eqfactorisation})], this factorization indicates  that there exist  a symmetric matrix $\hat C^{r,*}_{\rm o}(1)$, solution of the CARE (\ref{EqCARE:subeq1}), and a symmetric matrix $\hat C^{l,*}_{\rm o}(1)$, solution of the CARE (\ref{EqCARE:subeq2}), such that $q^*(s)$ is the characteristic  polynomial of  $-A_{\rm o}(1)+D\hat C^{r,*}_{\rm o}(1)$ and $A_{\rm o}(1)+D\hat C^{l,*}_{\rm o}(1)$. The matrices $\hat C^{r,*}_{\rm o}(1)$ and $\hat C^{l,*}_{\rm o}(1)$ are associated with  the subset ${\cal S}^*=\{s^*_1, s^*_2,...s^*_{n+2}\}$ of  eigenvalues of the Hamiltonian matrices  composed of the $n+2$ roots  of $q^*(s)$. 

From Eq. (\ref{EqpAtilde}) we have
\begin{align} 
q^*(s)=s^{n+2} +s^{n+1}(\frac{1}{Q_0}-\frac{n^2}{\tau})+...\ ,
\end{align}
so that 
\begin{align} 
\label{Eqsumsi}
\sum_{i}^{n+2}s^*_i=\frac{n^2}{\tau}-\frac{1}{Q_0}
\end{align}
and from  Eq. (\ref{Eqmulmur}),
\begin{align} 
\label{Eqftilde1}
f^*(1)=-\frac{1}{2}[\mbox{Tr}(A)+\sum_{i}^{n+2}s_i^*]=\frac{1}{Q_0}\,.
\end{align}
It then follows from Eqs. (\ref{Eqmu:subeqns}) that
\begin{subequations}
\label{CrCl1:subeqns}
\begin{align} 
[\hat C^{r,*}_{\rm o}(1)]_{11}=-\frac{2}{Q_0}-[S_{\rm o}]_{11}\label{CrCl1:subeq1}\\
[\hat C^{l,*}_{\rm o}(1)]_{11}=\frac{2n^2}{\tau}+[S_{\rm o}]_{11}\label{CrCl1:subeq2}\ ,
\end{align}
\end{subequations}
and  by inspecting the CARE  (\ref{EqCARE:subeq1}) we discover that the only solution  satisfying Eq. (\ref{CrCl1:subeq1}) is  the matrix
\begin{align} 
\label{EqXrplus1}
\hat C^{r,*}_{\rm o}(1)=S_q -S_{\rm o}\ .
\end{align} 
Hence  $q^*(s)$ is just the characteristic polynomial of $-A_{\rm o}(1)+D(S_q-S_{\rm o})=-A+DS_q$, as can be verified explicitly. On the other hand, no  simple expression of the  matrix $\hat C^{l,*}_{\rm o}(1)$ is available.  

In consequence two cases may occur: 

$\bullet$ Case (a):   All roots of $q^*(s)$ have a positive real part, i.e,  ${\cal S}^*_1={\cal S}_1^+$ and $q^+(1,s)=q^*(s)$ (note that Eq. (\ref{Eqsumsi}) implies that this case does not exist if $\tau/Q_0>n^2$). Then,
\begin{align} 
\label{EqCrplus1}
\hat C_{\rm o}^{r,+}(1)=\hat C^{r,*}_{\rm o}(1)
\end{align} 
and 
\begin{align} 
f^+(1)=f^*(1)=\frac{1}{Q_0}\ .
\end{align}

Furthermore, if $\hat C^{r,+}_{\rm o}(1)+C>0$ and $\hat C^{l,+}_{\rm o}(1)>0$ (i.e., $\lambda=1\in\hat {\cal D}_{\rm o}$) so that the effective process exists for the observable $\rm o$, this process  takes a remarkable form. Since $\hat C_w^{r,+}(1)=S_q$, Eq. (\ref{Eqeffdrift2}) indeed becomes
\begin{align}
\label{EqDriven1}
&\dot  v(t)=\frac{1}{Q_0}v(t)-x(t)+\frac{g}{Q_0} \int_{-\infty}^t ds\:g_n(t-s,n/\tau)x_s+\xi(t)\ .
\end{align}
Therefore, atypical fluctuations  of the observable are created in the long-time limit by simply changing the sign of the friction coefficient in the original Langevin dynamics (see also Ref. \cite{RTM2017}). 
 Moreover, for $\rm o=w$, an analysis similar to the one performed in Sec. IIIB  and Appendix A of Ref. \cite{RTM2017} shows that the pre-exponential factor $g_w(1)$ is given by\footnote{Formula (\ref{Eqgw}) is obtained by comparing the Onsager-Machlup action functionals associated with the path probabilities generated by the dynamics in Eq. (\ref{EqlinearTrick1}) on the one hand and the dynamics in Eq. (\ref{EqDriven1}) on the other hand. The analysis is similar to that leading to Eq. (33) and Eqs. (A4)-(A8) in Ref. \cite{RTM2017}.}
 \begin{align}
\label{Eqgw}
g_w(1)=\frac{1}{\sqrt{(1-\frac{T_x}{T})(1-\frac{T_v}{T})(1+\frac{\hat T_x(1)}{T})(1+\frac{\hat T_v(1)}{T})}}\ .
\end{align}
 According to the identity (\ref{EqDet1}) the condition $\hat C^{r,+}_w(1)+C=S_q+C>0$  requires that the  temperatures $T_x$ and $T_v$  are both strictly smaller  or both strictly larger than the bath temperature $T$. This ensures that Eq. (\ref{Eqgw}) gives a real result\footnote{This condition was overlooked in Ref. \cite{RTM2017}.}. We stress again that this simple and remarkable expression  only holds  in case (a).

 $\bullet$ Case (b):  Some roots of $q^*(s)$ have a negative real part, i.e., ${\cal S}^{(\tilde \alpha)}_1\ne{\cal S}_1^+$ and $q^+(1,s)\ne q^*(s)$ . Then, 
 \begin{align} 
\label{EqCrplus2}
\hat C_{\rm o}^{r,+}(1)>\hat C^{r,*}_{\rm o}(1)
\end{align}
since $\hat C_{\rm o}^{r,+}(1)$ is the maximal solution of the CARE (\ref{EqCARE:subeq1}), and 
\begin{align} 
\label{Eqfstar1}
f^+(1)=\frac{1}{Q_0}+ \sum_i s_i^{*-}<\frac{1}{Q_0}\ ,
\end{align}
where the sum runs over all roots of $q^*(s)$ with a negative real part [cf. Eq. (\ref{Eqmu2})]. 

For given values of $Q_0$ and $g$, case (a)  typically takes place in a limited range of values of $\tau$. This is illustrated  in Fig. \ref{Fig14}(a) where $f^+(1)$ is plotted as a function of $\tau$ (we recall that for this choice of the parameters the system reaches a stable stationary state for all values of $\tau$). Here, $f^+(1)=1/Q_0$ for $\tau_1\le \tau \le \tau_2$ with $\tau_1\approx 0.71$ and $\tau_2\approx 1.77$. We also display in Fig. \ref{Fig14}(b) the corresponding  kinetic temperature $\hat T_v(1)=2/Q_0\langle v^2\rangle_{\lambda=1}$ computed from  Eq. (\ref{EqTemp:subeq2}) which characterizes the stationary distribution of the effective process together with $\hat T_x(1)$. Note that these temperatures diverge for $\tau=\tau_1$ and $\tau=\tau_2$  since a pair of roots of $p_{H,1}(s)$ becomes purely imaginary and thus  $\lambda_{\max}=1$ (this is also true for all components of the covariance matrix $\hat \Sigma(1)$ except $[\hat \Sigma(1)]_{12}=\langle v x\rangle_{\lambda=1}=0$).   
\begin{figure}[hbt]
\begin{center}
\includegraphics[trim={0cm 0cm 0cm 0cm},clip,width=10cm]{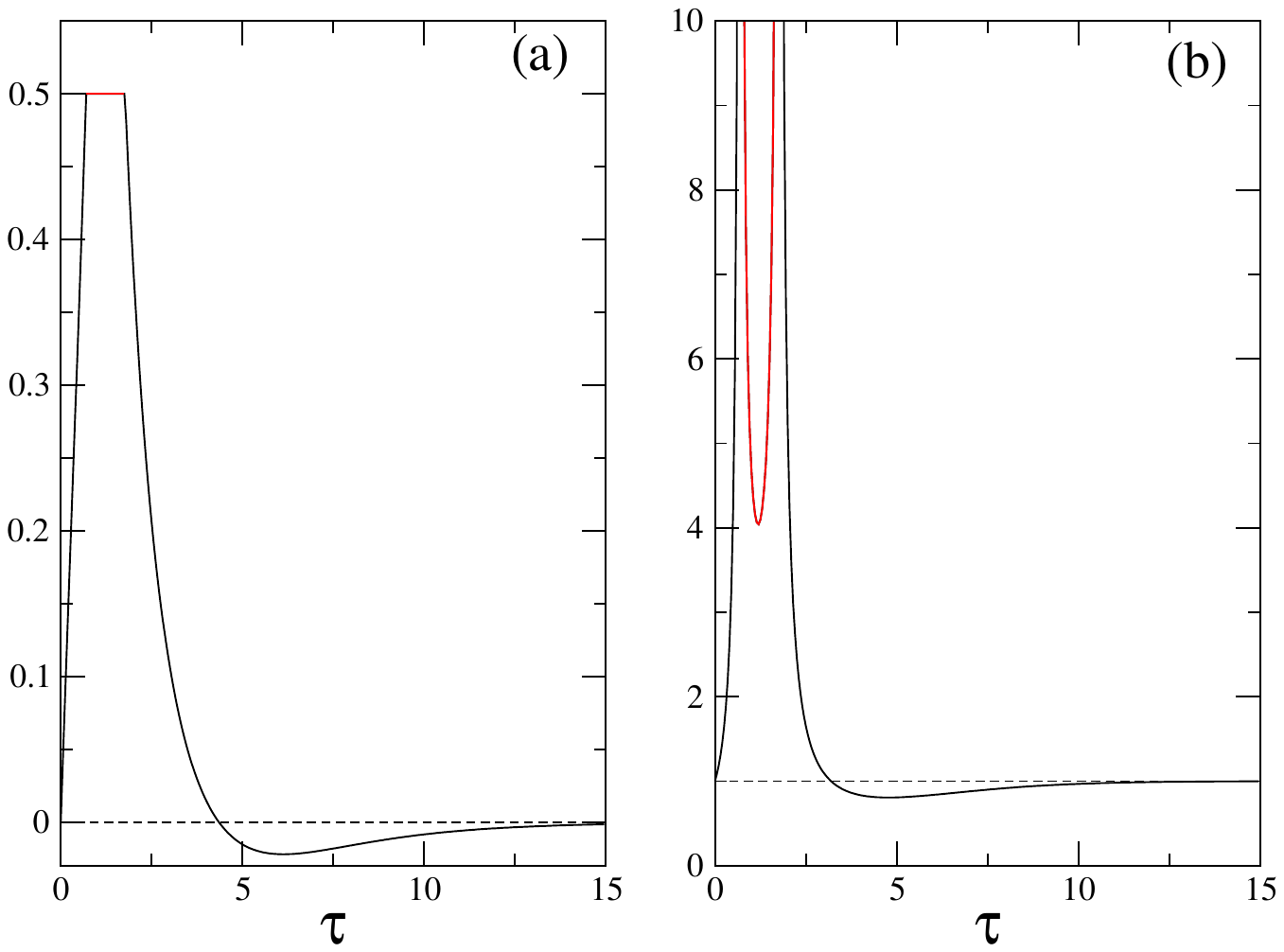}
\caption{ (Color on line) (a) $f^+(1)$ as a function of $\tau$ for $Q_0=2,g=1.5$ and $n=5$. One has $f^+(1)=1/Q_0$ for $0.70\lesssim\tau\lesssim1.77$ (in red). (b) Corresponding stationary kinetic temperature $\hat T_v(1)$ of the effective process.}
\label{Fig14}
\end{center}
\end{figure}

\subsection{Integral fluctuation theorem (IFT) for the heat}
\label{SubsecVB}

 As pointed out at the end of Sec. \ref{SubsecRic1}, the solution of the RDE (\ref{EqRic:subeq1})  with initial condition $C^r_q(1,0)=0$ is  $C^r_q(1,t)=0$ owing to the fact that $K_q(1)=0$.  As a result, $f_q^r(1,t)=1/Q_0$ from Eq. (\ref{Eqmua2:subeq1}) and one  recovers the universal IFT $\langle e^{-\beta {\cal Q}_t}\rangle=e^{t/Q_0}$ for the fluctuating heat in underdamped Langevin processes~\cite{RTM2016}. In particular, the SCGF is equal to $1/Q_0$ and is associated with the matrix $\hat C_q^{r,*}(1)=0$ in agreement with Eq. (\ref{EqXrplus1}) above. In addition, $g_q(1)=1$.

It is instructive to examine how this exact result is recovered from the solution of the RDE (\ref{EqRic:subeq2}) for the matrix $C^l_q(1,t)$. 
According to Eqs. (\ref{EqCrplus1}) and (\ref{EqCrplus2}),  the matrix $\hat C^{r,+}_q(1)$ is positive semidefinite. Hence,  $C+\hat C^{r,+}_q(1)> 0$, which implies that $C^l_q(1,t)\to \hat C_q^{l,+}(1)$ in both cases (a) and (b). However, the corresponding scenarios are different.

$\bullet$ Case (a): One has $\hat C_q^{r,+}(1)=\hat C_q^{r,*}(1)=0$ from Eq. (\ref{EqCrplus1}) and, thus, $\hat C_q^{r,+}(1)$ is  the trivial (i.e., zero) asymptotic fixed point of the RDE (\ref{EqRic:subeq1}). Then  $\hat C^{l,+}_q(1)$ is also associated with the SCGF $ f^+(1)=1/Q_0$ and  $\det \hat C^{l,+}_q(1)\ne 0$.  Furthermore,  from Eq. (\ref{EqgA}), $g_q(1)=1$. In short, $\lambda=1$ is in interior of the interval $\hat{\cal D}_q$ and the SCGF $\mu_q(\lambda)$ is regular at $\lambda=1$.

$\bullet$ Case (b): $\hat C_q^{r,+}(1)>\hat C_q^{r,*}(1)=0$ and $f^+(1)<1/Q_0$ according to Eqs. (\ref{EqCrplus2}) and (\ref{Eqfstar1}). Since $\hat C_q^l(1,t)\to  \hat C_q^{l,+}(1)$ but $\hat C_q^r(1,t)=0\not \to  \hat C_q^{r,+}(1)$, one must have $\det \hat C^{l,+}_q(1)=0$, and  the scenario is the one described in Sec. \ref{SubsecIVC1} with  $\lambda_{q2}=1$.  In particular,  $\det C^l_q(1,t)\sim e^{-2t[1/Q_0-f^+(1)]}$ as $t\to \infty$  so that the actual  value $1/Q_0$ of the SCGF  is recovered. The SCGF $\mu_q(\lambda)$ then displays a positive jump discontinuity at $\lambda=1$, which is the upper limit of its domain of definition, and the rate function $I_q(a)$ has a linear tail for $a\le a^-=-df^+(\lambda)/d\lambda\vert_{\lambda=1}$.

\subsection{Asymptotic fluctuation relation for the entropy production}
\label{SubsecVB2}

\subsubsection{Acausal  dynamics and Jacobian}

As pointed out in Ref. \cite{RMT2015}, the microscopic reversibility condition (or local detailed balance) that links dissipation to time-reversal symmetry breaking~\cite{S2012,K1998, C1998,LS1999,MN2003,G2004a} is not satisfied by a Langevin dynamics with a time-delayed continuous feedback. On the other hand, local detailed balance is recovered if the time-reversal operation is combined with the change $\tau \to -\tau$. Although this procedure is purely mathematical,  it was used in our previous studies to derive a second-law-like inequality for the extracted work rate in the stationary state (see Eq. (84) in Ref. \cite{RMT2015}) and to predict that
\begin{align} 
\label{EqmusigmaJacobian}
\mu_{\sigma}(1)\equiv\lim_{t\to \infty}(1/t)\ln \langle e^{-\Sigma_t}\rangle=\dot {\cal S}_{\cal J}\equiv  \lim_{t\to \infty} \frac{1}{t}\langle \ln \frac{{\cal J}_t}{\widetilde {\cal J}_t}\rangle\ ,
\end{align}
where ${\cal J}_{t}$ is the Jacobian of the transformation $\xi(t) \to x(t)$ associated with Eq. (\ref{EqLlinred}) and  $\widetilde {\cal J}_{t}$ is the Jacobian\footnote{As discussed in Ref. \cite{RMT2015},  the ``acausal" Jacobian  is in general a nontrivial functional of the path $\{x(t')\}_0^t$. However, it  is just a function of $t$ when the dynamics is linear.} associated with the corresponding {\it acausal}  Langevin equation in which $\tau$ is changed into $-\tau$~\cite{MR2014,RMT2015,RTM2017}. If true, Eq. (\ref{EqmusigmaJacobian})  is remarkable  since the left-hand side is a quantity that can be extracted from numerical simulations of Eq. (\ref{EqLlinred})~\cite{RTM2017,MR2014} or  from experiments~\cite{DRLK2022} whereas the quantity on the right-hand side is associated with a  dynamics that is not physically realizable.  
Our objective in the following is to compute the explicit expression of $\mu_{\sigma}(1)$ for an arbitrary value of $n$ via the Riccati-based approach and  to show that it is indeed identical to the expression of $\dot {\cal S}_{\cal J}$ when this latter quantity exists.   

 For the sake of conciseness, we will not repeat  here the lengthy analysis performed in Ref.~\cite{RMT2015} that yielded the expression of $\dot {\cal S}_{\cal J}$ for the discrete delay (i.e., for $n=\infty$). The main steps of the calculation for $n$ finite are similar and we refer the interested reader to this previous article. The outcome is the integral formula
\begin{align} 
\label{EqSdotJ}
\dot {\cal S}_{\cal J}=\frac{1}{2\pi i}\int_{c-i\infty}^{c+i\infty} ds\: \ln \frac{\widetilde \chi_n(s)}{\chi_{g=0}(s)}\ ,
\end{align}
where 
\begin{align}
\chi_{g=0}(s)=[s^2+ \frac{s}{Q_0}+1]^{-1}
\end{align}
is the response function  of the system in the absence of feedback and
\begin{align}
\label{Eqchitilde1}
\widetilde \chi_n(s)=[s^2+ \frac{s}{Q_0}+1-\frac{g}{Q_0}(1-\frac{s\tau}{n})^{-n}]^{-1}\ .
\end{align}
 The integration in Eq. (\ref{EqSdotJ})  is performed along the vertical line $\mbox{Re}(s) = c$ and a careful study of the branch cuts of the multivalued function $\ln[\widetilde \chi_n(s)/\chi_{g=0}(s)]$ in the complex $s$-plane shows that $c$ must be chosen such that {\it  two and only two poles} of $\widetilde \chi_n(s)$, hereafter denoted by $s^*_1$ and $s^*_2$, are located on the left side of the line $\mbox{Re}(s) = c$.  An  integration contour similar to the one displayed in Fig. 4 of Ref. \cite{RMT2015} then leads to the  simple expression
\begin{align} 
\label{EqSdotJ1}
\dot {\cal S}_{\cal J}=\frac{1}{Q_0}+(s^*_1+s^*_2) \ .
\end{align}
As discussed in Refs. \cite{RMT2015,RTM2017}, $\widetilde \chi_n(s)$  may be viewed as the bilateral Laplace transform of the response function  associated with the acausal dynamics\footnote{Following the notations in Refs. \cite{RTM2017,RMT2015}  the   symbol ``tilde"   refers to quantities associated with the acausal dynamics.}, and in analogy with  Eq. (\ref{Eqchi1}) it can be also expressed as
\begin{align}
\label{Eqchitilde2}
\widetilde \chi_n(s)=\frac{(s-\frac{n}{\tau})^n}{\widetilde p_A(s)}\ ,
\end{align}
where $\widetilde p_A(s)\equiv p_A^{\tau\to -\tau}(s)$ [see Eq. (\ref{Eqchar})]. In fact, $\widetilde p_A(s)$ is nothing but $q^*(s)$, the polynomial introduced in Sec. \ref{SubsecVA} [Eq. (\ref{EqpAtilde})]. In consequence,  $s^*_1$ and $s^*_2$ are roots of $q^*(s)$, and by comparing with Eq. (\ref{Eqfstar1})   we  deduce that 
\begin{subequations}
\label{EqSJ:subeqns}
\begin{align}
&\bullet \: \mbox{(a)}\: \:\:\dot {\cal S}_{\cal J}  >f^+(1) =1/Q_0  \: \: \mbox{if all roots of $q^*(s)$  have a positive real part ($c>0$)}\label{EqSJ:subeq1}\\
&\bullet  \: \mbox{(b1)}\: \: \: \dot {\cal S}_{\cal J}  =f^+(1)< 1/Q_0\: \: \mbox{if exactly two roots of $q^*(s)$  have a negative real part ($c=0$)} \label{EqSJ:subeq2}\\
&\bullet  \: \mbox{(b2)}\: \: \: f^+(1)<\dot {\cal S}_{\cal J}<1/Q_0\: \: \mbox{if more than two roots of $q^*(s)$  have a negative real part ($c<0$)}\label{EqSJ:subeq3}\ .
\end{align}
\end{subequations}
As  will be seen in the following, there is also a fourth case that was missed in Ref. \cite{RMT2015}: If the root of $q^*(s)$ with the smallest real part is real and the next roots are complex conjugate,  the Bromwich contour $\mbox{Re}(s) = c$ cannot be defined and  Eq. (\ref{EqSdotJ}) breaks down (which simply means that $\mu_{\sigma}(1)$ does not exist). 

Before embarking into the (rather lengthy) proof that $\mu_{\sigma}(1)$ (when it exists) is also given by Eq. (\ref{EqSdotJ1}), let us make a short remark about the acausal dynamics. Despite the  unphysical character of this dynamics, a steady state can still be defined  if the response function $\widetilde \chi_n(t)$ in the time domain decreases sufficiently fast to zero for both $t\to +\infty$ and $t\to -\infty$ (see the discussion   in Appendix A of Ref. \cite{RTM2017}). This requires that   $q^*(s)$ has only two roots  with a negative real part  (case (b1) above). The steady state is then characterized by a multivariate Gaussian distribution with a covariance matrix $\widetilde \Sigma$ whose elements are given by  [see Eqs. (\ref{EqSigma}) and (\ref{Eqhatchi})]
\begin{align} 
\widetilde \Sigma_{ij}=\int \frac{d\omega}{2\pi}\widetilde R_{i,1}(\omega)\widetilde R_{j,1} (-\omega)\ ,
\end{align}
where
\begin{align} 
\widetilde R(\omega)=-[\widetilde A+i\omega I_{n+2}]^{-1}
\end{align}
and  $\widetilde A\equiv A^{\tau \to -\tau}$. In particular, one has 
\begin{subequations}
\label{EqTemptilde:subeqns}
 \begin{align}
\frac{\widetilde T_x}{T}&=\frac{2}{Q_0} \widetilde  \Sigma_{22}= \frac{2}{Q_0}\int\frac{d\omega}{2\pi}\vert \widetilde\chi_n(\omega)\vert^2\label{EqTemptilde:subeq1}\\
\frac{\widetilde T_v}{T}&=\frac{2}{Q_0} \widetilde  \Sigma_{11}=\frac{2}{Q_0}\int \frac{d\omega}{2\pi}\omega^2\vert \widetilde\chi_n(\omega)\vert^2\label{EqTemptilde:subeq2}\ .
\end{align}
\end{subequations}
Using Eq. (\ref{Eqchitilde2}) and the fact that $\widetilde p_A(s)\widetilde p_A(-s)=q^*(s)q^*(-s)=(-1)^np_H(1,s)$, it is readily seen that these temperatures coincide with the temperatures $\hat T_x(1)$ and $\hat T_v(1)$  characterizing the stationary density associated with the effective process [see Eqs. (\ref{EqTemp:subeqns})]. On the other hand, it is found that  $\widetilde \Sigma_{n+2,1}=-\hat  \Sigma_{n+2,1}(1)$ [with $\hat  \Sigma_{n+2,1}(1)$ given by Eq. (\ref{EqSig21})]. The negative sign  indicates that  the acausal dynamics  does not generate the trajectories leading to a given fluctuation of the time-intensive observable $(2g/Q_0^2)(1/t)\int_0^t dt' x_n(t')\circ dx(t')$. Therefore, this dynamics must not be misinterpreted as being the effective process, as was done in  Sec. IVB4 of Ref. \cite{RTM2017}. 

\subsubsection{Calculation of the SCGF}

Let us  assume that  the matrix  $C^l_{\sigma}(1,t)$ is  nonsingular at all times so that its inverse $\Sigma^l_{\sigma}(1,t)$ exists\footnote{The analysis  performed in Appendix \ref{AppendC} shows that the  matrices $C+C^r_{\sigma,\lambda}(t)$ and $C^l_{\sigma,\lambda}(t)$ are positive definite for $0\le \lambda<1$ but only positive {\it semidefinite} for $\lambda=1$.}.
To prove that $\mu_{\sigma}(1)=\dot {\cal S}_{\cal J}$, it  will be convenient to work with the expression of $G_{\sigma,1}(t)$ given by Eq.  (\ref{EqZAnew}), that is 
\begin{align} 
\label{EqZsigma}
G_{\sigma,1}(t)&=\exp \left(-\frac{1}{2}\int_0^t \mbox{Tr}\big [K_{\sigma}(1)\Sigma^l_{\sigma}(1,t')+DB_{\sigma}\big ]dt'\right)\ ,
\end{align}
where $\Sigma^l_{\sigma}(1,t)$ satisfies  the complementary RDE 
\begin{align} 
\label{EqRDESigma}
\dot \Sigma^l_{\sigma}(1,t)= A_{\sigma}(1) \Sigma^l_{\sigma}(1,t)+ \Sigma^l_{\sigma}(1,t)A^T_{\sigma}(1)- \Sigma^l_{\sigma}(1,t) K_{\sigma}(1)\Sigma^l_{\sigma}(1,t)+ D
\end{align}
with initial condition $\Sigma^l_{\sigma}(1,0)=\Sigma$. It turns out that Eq. (\ref{EqZsigma}) can be greatly simplified by introducing the inverse of the matrix $X_{\sigma}^l(1,t)\equiv C_{\sigma}^l(1,t)-S_{\sigma}$. Due to the invariance property (\ref{Eqinvar1}), $X_{\sigma}^l(1,t)$ is solution of the RDE 
\begin{align} 
\label{EqXsigma}
\dot X_{\sigma}^l(1,t)= -AX_{\sigma}^l(1,t)-X_{\sigma}^l(1,t)A^T- X_{\sigma}^l(1,t) DX_{\sigma}^l(1,t)+K_w(1)
\end{align}
with initial condition $X_{\sigma}^l(1,0)=C-S_{\sigma}$ while its inverse, which is simply  denoted $Y(t)$ hereafter, is solution of  the   complementary RDE 
\begin{align} 
\label{EqCRDE}
\dot Y(t)= AY(t)+Y(t)A^T- Y(t) K_w(1)Y(t)+D
\end{align}
with  initial condition $Y(0)=(C- S_{\sigma})^{-1}$. Note that the fact  that $X_{\sigma}^l(1,t)$ is invertible results from the assumption that $\Sigma_{\sigma}^l(1,t)$ exists. Indeed, inspection of  Eq. (\ref{EqRDESigma}) shows that  $[\Sigma^l_{\sigma}(1,t)]_{11}=\Sigma_{11}=(Q_0/2)T_v, [\Sigma^l_{\sigma}(1,t)]_{22}=\Sigma_{22}=(Q_0/2)T_x$ and $[\Sigma^l_{\sigma}(1,t)]_{12}=\Sigma_{12}=0$, which, together with the definition of the matrix $S_{\sigma}$, yields 
\begin{align}
\label{EqrelSsigma}
\det\big( I_{n+2}-S_{\sigma}\Sigma^l_{\sigma}(1,t)\big)=\frac{T_xT_v}{T^2}\,.
\end{align}
As a result,
\begin{align}
 \det (C_{\sigma}^l(1,t)-S_{\sigma})=\frac{T_xT_v}{T^2}\det C^l_{\sigma}(1,t)\ne 0\ .
\end{align}
 The  matrix $Y(t)$, solution of Eq. (\ref{EqCRDE}), has  two remarkable properties:

(i) the elements $Y_{11}(t),Y_{22}(t)$ and $Y_{12}(t)$  are constant, with $Y_{11}(t)=Y_{22}(t)=Q_0/2$  and $Y_{12}(t)=0$,

(ii)  the $2n$  elements $Y_{13}(t),Y_{14}(t)...Y_{1n+2}(t)$ and $Y_{23}(t),Y_{24}(t)...Y_{2n+2}(t)$  evolve  independently of all other elements.

These  properties considerably simplify the calculation of  $\mu_{\sigma}(1)$. Indeed,   after inserting  the expressions of the matrices  $B_{\sigma}$ and $K_{\sigma,1}$ [Eqs. (\ref{EqBa}) and (\ref{EqKsigma})] into Eq.  (\ref{EqZsigma}), we  obtain
\begin{align}
\label{EqZsigma1} 
G_{\sigma,1}(t)=\exp[-\frac{2g}{Q_0^2} \int_0^t dt' \: Y_{1n+2}(t')]\ ,
\end{align}
which means that we only need to focus on  the $n\times 2$ matrix  
\begin{align} 
L(t) \equiv \begin{bmatrix}
Y_{13}(t)&Y_{23}(t)\\
Y_{14}(t)&Y_{24}(t)\\
.&.\\
.&.\\
.&.\\
Y_{1n+2}(t)&Y_{2n+2}(t)
\end{bmatrix}
\end{align}
instead of the full $(n+2)\times(n+2)$ matrix $Y(t)$. Specifically,
\begin{align} 
\label{Eqmu1b}
\mu_{\sigma}(1) =-\frac{2g}{Q_0^2} \lim_{t\to \infty}Y_{2,n+2}(t)=-\frac{2g}{Q_0^2} \lim_{t\to \infty} L_{n1}(t)\ .
\end{align}
From Eq. (\ref{EqCRDE}),  it is found that  the submatrix $L(t)$  satisfies the  non-symmetric RDE 
\begin{align} 
\label{EqRDE1}
\dot L(t)=- L(t)F_{11}+F_{22}L(t)- L(t)F_{12} L(t)+F_{21}\ ,
\end{align}
with initial conditions $L_{i1}(0)=Y_{1,i+2}(0)=(T/T_v)\Sigma_{1,i+2}$ and $L_{i2}(0)=Y_{2,i+2}(0)=(T/T_x)\Sigma_{2,i+2}$ ($i=1,2,..,n$). The expressions of the matrices $F_{11}$, $F_{12}$, $F_{21}$ and $F_{22}$  of dimensions $2\times 2$, $2\times n$, $n\times 2$, and $n\times n$, respectively,  are given in Appendix \ref{AppendI}. This defines the $(n+2)\times (n+2)$ square matrix
\begin{align}
F=
\begin{bmatrix}
F_{11}&F_{12}\\
  F_{21}&F_{22}&
\end{bmatrix}
\end{align}
which plays a similar role to that of the Hamiltonian matrices $H_{\rm o}^r(\lambda)$ and $H_{\rm o}^l(\lambda)$. In particular, each real solution  $\hat L^{(\alpha)}$ of the  CARE
\begin{align} 
\label{EqARE1}
 LF_{11}-F_{22}L+ LF_{12} L-F_{21}=0
\end{align}
is associated with some admissible set ${\cal S}^{(\alpha)}$ of eigenvalues of $F$ and can be constructed  from the corresponding eigenvectors.
Now, the key observation  is that a non-symmetric RDE  like  Eq. (\ref{EqRDE1}) has at most one fixed point $\hat L^*$  which is asymptotically stable  as $t\to \infty$~\cite{M1982,FJ1995,F2002}.  This is the so-called {\it dichotomic} solution of Eq. (\ref{EqARE1}) which corresponds to the set ${\cal S}^*=\{\nu_1,\nu_2,...\nu_{n+2}\}$ such that
\begin{align} 
\label{EqSeq}
\mbox{Re}(\nu_1)\ge \mbox{Re}(\nu_2) > \mbox{Re}(\nu_j) \ , \: \: \: j=3...n+2 \ .
\end{align}
If $L^*$ exists, $L(t)$  converges at an exponential rate to  this fixed point and then, from Eq. (\ref{Eqmu1b}),  
 \begin{align} 
\label{Eqmusigma1}
\mu_{\sigma}(1) =-\frac{2g}{Q_0^2}\hat L_{n1}^*\,.
\end{align}
On the other hand, if  $F$ is not dichotomically separable (i.e., there are no eigenvalues $\nu_1$ and $\nu_2$ of $F$ whose real parts are {\it strictly smaller} than the real parts of the other eigenvalues),  the RDE   (\ref{EqRDE1}) has no asymptotically stable fixed point.

In fact, the eigenvalues $\nu_i$ of  $F$ are  closely related to the roots  $s_i^*$ of the polynomial $q^*(s)$ introduced in Sec. \ref{SubsecVA}. Indeed, by comparing  Eq.  (\ref{EqpAtilde}) to the characteristic polynomial of $F$ 
\begin{align} 
p_F(s) =(\frac{n}{\tau})^n\left[(1+s^2-\frac{1}{4Q_0^2})[1+\frac{\tau}{n}(s+\frac{1}{2Q_0})]^n-\frac{g}{Q_0}\right] \ ,
\end{align}
we find that 
\begin{align} 
\label{Eqnus}
\nu_i=-s_i^*-\frac{1}{2Q_0}\ . 
\end{align}
It follows that the condition (\ref{EqSeq}) for the existence of $\hat L^*$ can be rewritten as 
 \begin{align} 
\label{EqSeq1}
\mbox{Re}(s^*_1)\le \mbox{Re}(s^*_2) < \mbox{Re}(s^*_j) \ , \: \: \: j=3...n+2 \ .
\end{align}
 The explicit expression of $\hat L^*$ in terms of the eigenvectors of $F$ associated with $\nu_1$ and $\nu_2$ is derived in Appendix \ref{AppendI}. This eventually leads to 
\begin{align} 
\label{Eqmusigma}
\mu_{\sigma}(1)=\frac{1}{Q_0}+ (s_1^*+s_2^*)\ ,
\end{align}
which is  identical to the expression (\ref{EqSdotJ1}) of  $\dot {\cal S}_{\cal J}$. Since two and only two roots of $q^*(s)$ must be located on the left side of  the Bromwich contour $\mbox{Re}(s)=c$ in the integral formula (\ref{EqSdotJ}), we see that the condition  (\ref{EqSeq1}) for the existence of a stable fixed point of  the RDE (\ref{EqRDE1}) [and thus for the  existence of the SCGF $\mu_{\sigma}(1)$] is also the condition for the existence of the quantity $\dot {\cal S}_{\cal J}$. It is clear that this condition cannot be realized  if $s^*_1$ is real while $s^*_{2}$ and $s^*_{3}$ are complex conjugates.

To conclude, we note that according to Eqs. (\ref{EqSJ:subeqns}) the equality $\mu_{\sigma}(1)= \dot {\cal S}_{\cal J}$ implies that $\mu_{\sigma}(1)\ne f^+(1)$ when all roots of the polynomial $q^*(s)$ have a positive real part or more that two roots have a negative real part. In these two cases, the SCGF $\mu_{\sigma}(\lambda)$  displays a positive jump discontinuity at $\lambda=1$ (provided $\mu_{\sigma}(1)$ exists) and the rate function $I_{\sigma}(a)$ has a linear tail for $a\le a^-=-df^+(\lambda)/d\lambda\vert_{\lambda=1}$.

\section{Conclusion} 
\label{SecVI}

We  have implemented a theoretical and numerical scheme to study  fluctuations of dynamical observables such as work, heat, or entropy production  in stochastic systems governed by linear Langevin equations with time delay.  We have then been able to derive the complete asymptotic form of the generating functions and probability distributions, and  we have characterized (for the first time for a non-Markovian Langevin system)  the effective process that  describes how fluctuations are created dynamically in the long-time limit. In this way, we have extended the current large-deviation description of statistical fluctuations  to the harder problem of a non-Markovian diffusion dynamics. 

Central to our analysis  are differential  Riccati equations, and we have put a lot of emphasis on the properties of these equations as they differ from those typically encountered  in (linear-quadratic) optimal control problems. This makes the behavior of the solutions more complicated, and we have shown that it is fruitful, both analytically and numerically,  to express these solutions  in terms of the eigenvalues and eigenvectors of the associated Hamiltonian matrices.  This procedure allows us to study the statistics of observables at arbitrary finite time and to  build explicitly the generic fixed point reached asymptotically, from which the SCGF and the sub-dominant factors are easily computed. We have also clarified the conditions under which the probability distributions exhibit exponential tails, the role of the symmetry of the observables, and we have unveiled  the nontrivial  behavior occurring at the limits of the domain of existence of the SCGF, something that cannot be predicted by only solving the spectral problem for the dominant eigenvalue of the tilted generators. 

Although we have mainly focused on a specific non-Markovian model, it is clear that our methods can be used for any linear multidimensional diffusions (as an illustration, we present in Appendix \ref{AppendJ} an application of  the Riccati formalism to an active particle model). We thus expect that many of the results presented here have a more general validity. We therefore hope that this study will motivate additional fruitful work in the field.

 %This will be the subject of future studies. 

%We expect our theoretical and numerical methods to find applications in a broad class of observables in chaotic maps, so we anticipate that our results will motivate additional fruitful work in the field. For instance, it would be interesting to use these ideas to study rare events in natural systems such as climate models [7, 8, 31], which are of great significance in a changing climate. Finally, 

\newpage

\begin{appendices}

\numberwithin{equation}{section}

\setcounter{equation}{0}

\renewcommand{\theequation}{A\arabic{equation}} 

\section{Tilted generators}
\label{AppendA}

Inserting the expression of the drift matrix $A$ [Eq. (\ref{EqDrift})] and of the vector function ${\bf g}_{\rm o}$ [Eqs. (\ref{EqgaU}) and (\ref{EqBa})] into the general definition of the tilted generator  [Eq. (\ref{Eqtiltedgen})],  we obtain
\begin{align} 
\label{EqLa}
{\cal L}_{\rm o,\lambda}&={\cal L}_0-\frac{2\lambda}{Q_0}\Delta {\cal L}_{\rm o,\lambda}\ ,
\end{align}
with 
\begin{align} 
\label{EqL0}
{\cal L}_0&=(-\frac{1}{Q_0}u_1-u_2+\frac{g}{Q_0}u_{n+2})\frac{\partial}{\partial u_1}+u_1\frac{\partial}{\partial u_2}\nonumber\\
&+\frac{n}{\tau}\sum_{j=3}^{n+2}(u_{j-1}-u_j)\frac{\partial}{\partial u_j}+\frac{1}{2}\frac{\partial^2}{\partial u_1^2}
\end{align}
and
\begin{align} 
\label{EqLw}
\Delta {\cal L}_{w,\lambda}&=\frac{g}{Q_0}u_{n+2}u_1\ ,
\end{align}
\begin{align} 
\label{EqLq}
\Delta {\cal L}_{q,\lambda}&=\frac{1}{Q_0}(1-\lambda)u_1^2-\frac{1}{2}-u_1\frac{\partial}{\partial u_1}\ ,
\end{align}
\begin{align} 
\label{EqLsigma}
\Delta {\cal L}_{\sigma,\lambda}&=(\frac{T}{T_x}-\frac{T}{T_v})u_1u_2+\frac{g}{Q_0}\frac{T}{T_v}u_{n+2}u_1\nonumber\\
&-\frac{1}{Q_0}\frac{T-T_v}{T_v} (1+\frac{T-T_v}{T_v}\lambda)u_1^2\nonumber\\
&+\frac{T-T_v}{T_v}(\frac{1}{2}+u_1\frac{\partial}{\partial u_1}) \ .
\end{align}

\renewcommand{\theequation}{B\arabic{equation}} 

\section{Derivation of the Riccati differential equations (RDEs)}
\label{AppendB}

\subsubsection{Solutions of the tilted Fokker-Planck equations}

\label{SecB1}

Here, we  carry out the calculations that lead to the RDE (\ref{EqRic:subeq1}), assuming that the solution of the partial differential equation (\ref{EqFP1}) has the form
\begin{align}
G^r_{\rm o,\lambda}({\bf u}_0,t)&=c^r_{\rm o}(\lambda,t)e^{-\frac{1}{2}{\bf u}_0^TC^r_{\rm o}(\lambda,t){\bf u}_0}\ ,
\end{align}
where $C^r_{\rm o}(\lambda,t)$ is a  symmetric matrix and $c^r_{\rm }(\lambda,t)$ is a scalar function.
The initial condition (\ref{Eqinit1}) imposes that $C^r_{\rm o}(\lambda,0)=0$ and $c^r_{\rm o}(\lambda,0)=1$. 

To simplify the  notation, we henceforth replace ${\bf u}_0$ by ${\bf u}$ and drop the dependence  of the functions on $t$ and ${\bf u}$. On the left-hand side of  Eq. (\ref{EqFP1}), we obtain
 \begin{align}
\partial_t G^r_{\rm o,\lambda}=\left(\frac{\dot c^r_{\rm o}(\lambda)}{ c^r_{\rm o}(\lambda)}-\frac{1}{2}{\bf u}^T\dot C^r_{\rm o}(\lambda){\bf u}\right)G^r_{\rm o,\lambda}\ .
\end{align}
 On the right-hand side, using ${\bf F}=A{\bf u}$, ${\bf g}_{\rm o}=B_{\rm o} {\bf u}$, we find
 \begin{align}
\label{EqF}
  {\bf F}\cdot(\nabla -\lambda {\bf g}_{\rm o})G^r_{\rm o,\lambda}=&-\frac{1}{2}{\bf u}^T\big[A^T(C^r_{\rm o}(\lambda)+\lambda B_{\rm o})+(C^r_{\rm o}(\lambda)+\lambda B_{\rm o}^T)A)\big]{\bf u}G^r_{\rm o,\lambda} \ ,
 \end{align}
 where we have symmetrized the scalar product and used the fact  that $C^r_{\rm o}(\lambda)$ is symmetric. Likewise, 
\begin{align}
\frac{1}{2}\nabla\cdot [D \nabla G^r_{\rm o,\lambda}]=&-\frac{1}{2}\nabla \cdot[D C^r_{\rm o}(\lambda){\bf u}G^r_{\rm o,\lambda}]=\frac{1}{2}{\bf u}^T[C^r_{\rm o}(\lambda)DC^r_{\rm o}(\lambda)]{\bf u}G^r_{\rm o,\lambda} -\frac{1}{2}\mbox{Tr}[DC^r_{\rm o}(\lambda)]G^r_{\rm o,\lambda} \ ,
\end{align}
 \begin{align}
 -\frac{\lambda}{2}\nabla \cdot (D {\bf g}_{\rm o}G^r_{\rm o,\lambda} )&=-\frac{\lambda}{2} \nabla \cdot  (D B_{\rm o}{\bf u}G^r_{\rm o,\lambda} )=\frac{\lambda}{2}\big[{\bf u}^T[B_{\rm o}^TDC^r_{\rm o}(\lambda)]{\bf u}-\mbox{Tr}( DB_{\rm o})\big]G^r_{\rm o,\lambda} \ ,
\end{align}
and 
\begin{align}
-\frac{\lambda}{2} {\bf g}_{\rm o}\cdot D (\nabla -\lambda {\bf g}_{\rm o})G^r_{\rm o,\lambda} =\frac{\lambda}{2}{\bf u}^T\big[(C^r_{\rm o}(\lambda)+\lambda B_{\rm o}^T)DB_{\rm o}\big]{\bf u}G^r_{\rm o,\lambda} 
\end{align}
Collecting all these results, we obtain from Eq. (\ref{EqFP1}) the matrix differential equation
\begin{align}
\label{Eqmatrix}
{\bf u}^T\dot C^r_{\rm o}(\lambda){\bf u}&={\bf u}^T\Big[(A^T-\lambda B_{\rm o}^TD)C^r_{\rm o}(\lambda)+C^r_{\rm o}(\lambda)(A-\lambda DB_{\rm o})-C^r_{\rm o}(\lambda)DC^r_{\rm o}(\lambda)+\lambda(A^TB_{\rm o}+B_{\rm o}^TA-\lambda B_{\rm o}^TDB_{\rm o})\Big] {\bf u}
\end{align}
together with the scalar equation
\begin{align}
\label{Eqscalar}
\frac{\dot c^r_{\rm o}(\lambda)}{ c^r_{\rm o}(\lambda)}=-\frac{1}{2}\mbox{Tr}\left[D(C^r_{\rm o}(\lambda)+\lambda B_{\rm o})\right]\ .
\end{align}
Eq. (\ref{Eqmatrix}) then leads to the RDE (\ref{EqRic:subeq1}) with the matrices $A_{\rm o}(\lambda)$ and $K_{\rm o}(\lambda)$ given by Eqs. (\ref{EqtildeA}) and (\ref{EqKa}) respectively.  Furthermore, by integrating  Eq. (\ref{Eqscalar}) with initial condition $c^r_{\rm o,\lambda}(0)=1$, we obtain
\begin{align}
c^r_{\rm o}(\lambda,t)&=e^{\int_0^t dt'\: f^r_{\rm o}(\lambda,t')}
\end{align}
with  $f^r_{\rm o}(\lambda,t)$ defined by Eq. (\ref{Eqmua1:subeq1}). 

Equivalent results are derived for  the forward partial differential equation (\ref{EqFP2}) using the definition of the dual generator
\begin{align} 
{\cal L}^\dag_{\rm o,\lambda}=-\nabla {\bf F}-\lambda {\bf F}{\bf g}_{\rm o}+\frac{1}{2}(\nabla +\lambda{\bf  g}_{\rm o})\cdot D(\nabla +\lambda {\bf g}_{\rm o})
\end{align}
and the ansatz
\begin{align}
G^l_{\rm o,\lambda}({\bf u},t)&=c^l_{\rm o}(\lambda,t)e^{-\frac{1}{2}{\bf u}^TC^l_{\rm o}(\lambda)(t){\bf u}}
\end{align}
with initial conditions $C^l_{\rm o}(\lambda,0)=C$ and $c^l_{\rm o}(\lambda,0)=\sqrt{\det(C)/(2\pi)^{n+2}}$.  For instance, Eq. (\ref{EqF}) is  replaced by
 \begin{align}
\label{EqFdual}
 & -\nabla \cdot ({\bf F}G^l_{\rm o,\lambda})-\lambda {\bf F}\cdot{\bf g}_{\rm o}G^l_{\rm o,\lambda} =-\mbox{Tr}(A)+\frac{1}{2}{\bf u}^T[A^T(C^l_{\rm o}(\lambda)-\lambda B_{\rm o})+(C^l_{\rm o}(\lambda)-\lambda B_{\rm o}^T)A)]{\bf u}G^l_{\rm o,\lambda} \ .
 \end{align}
 This eventually leads to the RDE  (\ref{EqRic:subeq2}) and  Eq. (\ref{Eqmua1:subeq2}) for $f^l_{\rm o}(\lambda,t)$.

\subsubsection{Explicit expressions of the matrices  $K_{\rm o}(\lambda)$}
\label{SecB2}

We now give the expressions of  the  matrices  $K_{w,\lambda},K_{q,\lambda}$, and $K_{\sigma,\lambda}$.  From the definition of the matrices $B_{\rm o}$ [Eq. (\ref{EqBa})], we obtain
\begin{align}
\label{EqKw}
K_w(\lambda)=\frac{2\lambda g}{Q_0^2}
\begin{bmatrix}
 0&0&0 &...&1\\
 0&0&0&...&0 \\
  0&0&0 &...& 0\\
  \vdots & \vdots  & \vdots & \vdots& \vdots \\
1&0&0 &...& 0&
\end{bmatrix}\ ,
\end{align}

\begin{align}
\label{EqKq}
K_q(\lambda)= \frac{4\lambda(1-\lambda)}{Q_0^2}
\begin{bmatrix}
1&0&0 &...&0\\
 0&0&0&...&0 \\
  0&0&0 &...& 0\\
  \vdots & \vdots  & \vdots & \vdots& \vdots \\
0&0&0 &...& 0&
\end{bmatrix}\ ,
\end{align}
and
\begin{align}
\label{EqKsigma}
K_{\sigma}(\lambda)=\frac{2\lambda}{Q_0}
\begin{bmatrix}
\frac{2}{Q_0}(1- \frac{T}{T_v})[1-\lambda(1- \frac{T}{T_v})]&\frac{T}{T_x}-\frac{T}{T_v}&0 &...&\frac{g}{Q_0}\frac{T}{T_v}\\
 \frac{T}{T_x}-\frac{T}{T_v}&0&0&...&0 \\
  0&0&0 &...& 0\\
  \vdots & \vdots  & \vdots & \vdots& \vdots \\
\frac{g}{Q_0}\frac{T}{T_v}&0&0 &...& 0&
\end{bmatrix}\ .
\end{align}

\subsubsection{Derivation of Eq.  (\ref{EqZAnew})}
\label{SecB3}

We  now turn to the derivation of Eq. (\ref{EqZAnew}).  We first multiply the RDE (\ref{EqRic:subeq1})  by $\Sigma^l_{\rm o}(\lambda,t)\equiv \big[C^l_{\rm o}(\lambda,t)\big]^{-1}$ from the right to obtain
\begin{align}
\frac{\partial C^l_{\rm o}(\lambda,t)}{\partial t}\Sigma^l_{\rm o}(\lambda,t)=- A_{\rm o}^T(\lambda)-C^l_{\rm o}(\lambda,t)A_{\rm o}(\lambda)\Sigma^l_{\rm o}(\lambda,t)-C^l_{\rm o}(\lambda,t)D+K_{\rm o}(\lambda)\Sigma^l_{\rm o}(\lambda,t)\ .
 \end{align}
Taking the trace and using  the general identity $d (\det M(t))/dt=\det M(t)\mbox{Tr}[(M^{-1}(t)d (M(t))/dt]$, we find
\begin{align}
\frac{\partial}{\partial t} \ln \det C^l_{\rm o}(\lambda,t)&=-2\mbox{Tr}(A-\lambda DB_{\rm o})-\mbox{Tr}(DC^l_{\rm o}(\lambda,t))+\mbox{Tr}\big(K_{\rm o}(\lambda)\Sigma^l_{\rm o}(\lambda,t)\big)\ .
 \end{align}
This allows us to rewrite  Eq. (\ref{Eqmua1:subeq2}) as 
\begin{align}
f^l_{\rm o}(\lambda,t)=-\frac{1}{2} \mbox{Tr}\big((K_{\rm o}(\lambda)\Sigma^l_{\rm o}(\lambda,t)+\lambda DB_{\rm o}\big)+\frac{1}{2}\frac{\partial}{\partial t}\ln \det C^l_{\rm o}(\lambda,t)\ .
 \end{align}
As a result,
\begin{align}
e^{\int_0^t f^l_{\rm o}(\lambda,t')dt'}=\Big[\frac{\det C^l_{\rm o}(\lambda,t)}{\det C}\Big]^{1/2}e^{-\frac{1}{2} \int_0^t dt \:\mbox{Tr}\big(K_{\rm o}(\lambda)\Sigma^l_{\rm o}(\lambda,t)+\lambda DB_{\rm o}\big ) } \ ,
 \end{align}
and Eq. (\ref{EqZAnew}) is then obtained from Eq. (\ref{EqDynZ:subeq2}). 

\renewcommand{\theequation}{C\arabic{equation}} 

\section{Global existence of the generating functions $G_{w,\lambda}(t)$ and $G_{\sigma,\lambda}(t)$ for $0\le \lambda\le 1$}
\label{AppendC}

In this appendix, we present some results about the  existence of the generating functions $G_{w,\lambda}(t)$ and $G_{\sigma,\lambda}(t)$ for $0\le \lambda\le 1$. To this aim we  discuss the existence and positiveness of the matrices $C+C^r_w(\lambda,t)$  and $C+C^r_{\sigma}(\lambda,t)$. 

We first note  from Eqs.  (\ref{EqSq}) and (\ref{EqSsigma}) in the main text that $S_q\le 0$ and $S_{\sigma} -S_q\ge 0$.  We then exploit  the  order-preserving property of Riccati differential equations which states that  the solutions  depend monotonically on the initial values~\cite{note8}. Accordingly,  the solutions  $X_w(\lambda,t)$ and $X_{\sigma}(\lambda,t)$ of the RDE  $\dot X ={\cal R}_{q,\lambda}[X]$, which correspond respectively  to the initial conditions $X_w(\lambda,0)=-\lambda S_q\ge 0$  and $X_{\sigma}(\lambda,0)=\lambda (S_{\sigma}-S_q)\ge 0$, satisfy $X_w(\lambda,t)\ge C^r_q(\lambda,t)$  and $X_{\sigma}(\lambda,t)\ge C^r_q(\lambda,t)$.  From the ``invariance" relations (\ref{EqRq1}) and (\ref{EqRq2}) (with $S_w=0$) we then obtain 
\begin{align}
\label{Eqineq1}
C_w^r(\lambda,t)=X_w(\lambda,t)+\lambda S_q\ge  C^r_q(\lambda,t)+\lambda S_q
 \end{align}
 and
 \begin{align}
\label{Eqineq2}
C_{\sigma}^r(\lambda,t)=X_{\sigma}(\lambda,t)-\lambda (S_{\sigma}-S_q)\ge  C^r_q(\lambda,t)-\lambda (S_{\sigma}-S_q)\ .
 \end{align}
 Furthermore, it can be shown that
\begin{align}
\label{EqDet1}
\det [C+\lambda S_q]&=\det [I_{n+2}+\lambda S_q\Sigma]\:\det C=(1-\lambda \frac{T_v}{T})(1-\lambda \frac{T_x}{T})\:\det C
 \end{align}
 and 
\begin{align}
\label{EqDet2}
 \det [C+\lambda(S_q-S_{\sigma})]=(1-\lambda)^2\:\det C\ . 
 \end{align} 
 As a result, $C+\lambda S_q>0$ for $\lambda<\min(T/T_x,T/T_v)$ and $C+\lambda(S_q-S_{\sigma})>0$ for $\lambda<1$ (since all eigenvalues of these matrices  are positive for $\lambda=0$ and  remain positive as long as the respective determinants do not vanish). From this set of inequalities we  conclude  that 
\begin{align}
\label{Eqineq3}
 C^r_w(\lambda,t)+C\ge C^r_q(\lambda,t)+C+\lambda S_q>0\  , \: \: 0\le \lambda<\min(1, T/T_x,T/T_v)
 \end{align} 
 and
 \begin{align}
 \label{Eqineq4}
 C^r_{\sigma}(\lambda,t)+C\ge C^r_q(\lambda,t)+C-\lambda (S_{\sigma}-S_q)>0\ , \: \:  0\le \lambda<1\ .
 \end{align}
In consequence, the corresponding generating functions $G_{w,\lambda}(t)$ and $G_{\sigma,\lambda}(t)$ are always finite for values of $\lambda$ within the above ranges. We stress that these conditions are sufficient but not necessary.  Note also that the inequality (\ref{Eqineq3})  still holds for $\lambda=1$  if both $T_x$ and $T_v$ are smaller than $T$. On the other hand, the strict inequality  (\ref{Eqineq4}) is replaced by  $C^r_{\sigma}(1,t)+C\ge 0$. This special but important case $\lambda=1$ is treated in detail in Sec. \ref{SecV}.

\renewcommand{\theequation}{D\arabic{equation}} 

\section{Spectral properties of the Hamiltonian matrices}
\label{AppendD}

\subsubsection{Characteristic polynomial}
\label{AppendD1}

Here, we show that  the Hamiltonian matrices $H_{\rm o}^r(\lambda)$ and $H_{\rm o}^l(\lambda)$ have the same eigenvalue spectrum which  is also independent of the observable. 
We then derive the explicit expression of the characteristic polynomial $p_H(\lambda,s)$ [Eq. (\ref{EqpH}) in the main text].

Consider the ``right" Hamiltonian matrix $H_{\rm o}^r(\lambda)$ defined by Eq. (\ref{EqHr}) with $A_{\rm o}(\lambda)$ and $K_{\rm o}(\lambda)$ given by Eqs. (\ref{EqtildeA}) and (\ref{EqKa}), respectively. It is easy to see that $H_{\rm o}^r(\lambda)$  can be rewritten as 
\begin{align} 
H_{\rm o}^r(\lambda)=\hat H_{\rm o}^r(\lambda)E_{\rm o}(\lambda)\ ,
\end{align}
where 
\begin{align}
\label{EqhatHr}
\hat H_{\rm o}^r(\lambda)=
\begin{bmatrix}
A&D\\
-\lambda B_{\rm o}^TA& A^T-\lambda B_{\rm o}^TD&
\end{bmatrix}
\end{align}
and 
\begin{align}
E_{\rm o}(\lambda)=
\begin{bmatrix}
-I_{n+2}&0\\
\lambda B_{\rm o}& I_{n+2}&
\end{bmatrix}\ .
\end{align}
Since $E_{\rm o}^2(\lambda)=I_{2(n+2)}$ and $\det E_{\rm o}(\lambda)=(-1)^n$, we have
\begin{align} 
 \det[sI_{2(n+2)}-H^r_{\rm o}(\lambda)]=\det \Big ([sE_{\rm o}(\lambda)-\hat H_{\rm o}^r(\lambda)]E_{\rm o}(\lambda)\Big )=(-1)^n\det[sE_{\rm o}(\lambda)-\hat H_{\rm o}^r(\lambda)]
\end{align}
with 
\begin{align}
sE_{\rm o}(\lambda)-\hat H_{\rm o}^r(\lambda)=
\begin{bmatrix}
-sI_{n+2}-A&-D\\
\lambda (sB_{\rm o}+B_{\rm o}^TA)& sI_{n+2}-A^T+\lambda B_{\rm o}^TD&
\end{bmatrix}\ .
\end{align}
Let $s$ be an eigenvalue of $H_{\rm o}^r(\lambda)$. If $\lambda\ne 0$, neither $s$ nor $-s$ are  eigenvalues of $A$ and  the two matrices $sI_{n+2}+A$ and $sI_{n+2}-A^T$ are  thus invertible [if $\lambda=0$, one simply has $ \det[sI_{2(n+2)}-H^r_{\rm o}(0)]=(-1)^np_A(s)p_A(-s)$ where $p_A(s)=\det(sI_{n+2}-A)$ is the characteristic polynomial of $A$ given by Eq. (\ref{Eqchar})].  The standard formula for the determinant of block matrices then yields 
\begin{align}
\label{Eqrel1}
 \det[sI_{2(n+2)}-H^r_{\rm o}(\lambda)]&=(-1)^np_A(-s)\det \left[sI_{n+2}-A^T+\lambda B_{\rm o}^TD-\lambda(sB_{\rm o}+B_{\rm o}^TA)(sI_{n+2}+A)^{-1}D\right]\ .
 \end{align}
By using the  identity  $(sB_{\rm o}+B_{\rm o}^TA)(sI_{n+2}+A)^{-1}=s(B_{\rm o}-B^T_{\rm o})(sI_{n+2}+A)^{-1}+B_{\rm o}^T$, this can be rewritten as
\begin{align}
\label{Eqrel2}
 \det[sI_{2(n+2)}-H^r_{\rm o}(\lambda)]=(-1)^np_A(s)p_A(-s)\det\left [I_{n+2}+2\lambda sP^T(-s)B_{\rm o,antisym}P(s)D\right]\ ,
\end{align}
where
\begin{align}
\label{EqPs}
P(s)=(sI_{n+2}+A)^{-1}\ ,
\end{align}
and  
\begin{align}
B_{\rm o,antisym}=\frac{B_{\rm o}-B^T_{\rm o}}{2}
\end{align}
 is the antisymmetric part of the matrix $B_{\rm o}$. Moreover, since  $-s$ is also an eigenvalue of $H^r_{\rm o}(\lambda)$,  the characteristic polynomial  is an even function of $s$, and changing $s$ into $-s$ in Eq. (\ref{Eqrel1}) readily shows that $\hat H_{\rm o}^r(\lambda)$ and $\hat H_{\rm o}^l(\lambda)$, the ``left" Hamiltonian matrix, have the same eigenvalue spectrum\footnote{The fact that  $ \det[sI_{2(n+2)}-H^r_{\rm o}(\lambda)]$  is an even function  of $s$ is also immediately recovered  from Eq. (\ref{Eqrel1}).  Define $f(s)=\det[I_{n+2}+2\lambda sP(-s)^TB_{\rm o}^{AS}P(s)D]$. Then $f(-s)=\det[I_{n+2}-2\lambda sP(s)^TB_{\rm o}^{AS}P(-s)D]=\det[I_{n+2}+2\lambda s[P(-s)^TB_{\rm o}^{AS}P(s)]^TD]$. By Sylvester's identity, $\det(I+MN)=\det(I+NM)$, one obtains $f(-s)=f(s)$.}. 

Note that Eq. (\ref{Eqrel2}) holds for the Hamiltonian matrix associated with {\it any} linear current observable. In the case at hand, the three matrices  $B_w,B_q$ and $B_{\sigma}$ have the same antisymmetric part (see Eq. (\ref{EqBa}))
\begin{align}
\label{EqBwAS}
B_{\rm o,antisym}=\frac{B_w-B_w^T}{2}=
 \frac{g}{Q_0^2} 
\begin{bmatrix}
 0&0&0 &...&0\\
 0&0&0&...& 1 \\
  0&0&0 &...& 0\\
  \vdots & \vdots  & \vdots & \vdots& \vdots \\
0&-1&0 &...& 0&
\end{bmatrix}\ ,
\end{align}
and therefore the eigenvalue spectrum of the matrices $H^r_{\rm o}(\lambda)$ and $H^l_{\rm o}(\lambda)$ does not depend on the observable. Moreover, the fact that $D_{ij}=\delta_{i1}\delta_{j1}$ implies that the matrix $I_{n+2}+2\lambda sP(-s)^TB_{\rm o,antisym}P(s)D$ has nonzero elements only on the first column and along the diagonal. This yields 
\begin{align}
\det\left [I_{n+2}+2\lambda sM(s)D\right]=1+2\lambda sM_{11}(s)\ ,
\end{align}
 where
\begin{align}
M(s)=P(-s)^TB_{\rm o,antisym}P(s)\ .
\end{align}
Therefore,  the characteristic polynomial is linear in $\lambda$. (For a general diffusion matrix $D$, even simply diagonal, $p_H(\lambda,s)$ is generically a polynomial in $\lambda$.)

It remains to compute $M_{11}(s)$. From Eq. (\ref{EqBwAS}) we obtain
\begin{align}
M_{11}(s)=\frac{2g}{Q_0^2}[P_{11}(-s)P_{(n+2)1}(s)+P_{11}(s)P_{(n+2)1}(-s)]\ ,
\end{align}
and it is easily found from the definition of the matrix $P(s)$ [Eq. (\ref{EqPs})] that  
\begin{align} 
&p_A(s)P_{11}(-s)=-(\frac{n}{\tau})^n s(1+\frac{s\tau}{n})^n\nonumber\\
&p_A(s)P_{(n+2)1}(-s)=-(\frac{n}{\tau})^n\ .
\end{align}
Collecting all these results, we finally obtain  Eq. (\ref{EqpH}) of the main text.

\subsubsection{Eigenvectors}
\label{AppendD2}

We now give the expression of the eigenvectors of the Hamiltonian matrices $H_w^r(\lambda)$ and $H_w^l(\lambda)$ associated with the Riccati operator ${\cal R}_{w,\lambda}$.  We recall that the matrices are  assumed to be diagonalizable for simplicity. In addition, we  suppose that all eigenvalues are distinct  so that we can use the  Faddeev-Leverrier procedure~\cite{FS1952}. (Note that this method can also be used to find the generalized eigenvectors when the matrices are defective~\cite{HWV1993}.)

Consider first $H_w^r(\lambda)$  and let  ${\bf e}_w^r(s_i)$ be the eigenvector  associated with the eigenvalue $s_i(\lambda)$.  ${\bf e}_w^r(s_i)$ is decomposed  as 
\begin{align}
{\bf e}_w^r(s_i)=
\begin{bmatrix}
{\bf y}_w^r(s_i)\\
{\bf z}_w^r(s_i)\\
\end{bmatrix}\ ,
\end{align}
where ${\bf y}_w^r(s_i)$ and  ${\bf z}_w^r(s_i)$ correspond to the first and last $n+2$ components, respectively. By definition, the four blocks of the matrix $W_w^r(\lambda)$ defined by Eq. (\ref{EqWr}) are given by 
\begin{align}
W_w^{r,11}=
\begin{bmatrix}
{\bf y}_w^r(s_1^+)&{\bf y}_w^r(s_2^+)&...&{\bf y}_w^r(s_{n+2}^+)\\
\end{bmatrix}\ ,
\end{align}
\begin{align}
W_w^{r,21}=
\begin{bmatrix}
{\bf z}_w^r(s_1^+)&{\bf z}_w^r(s_2^+)&...&{\bf z}_w^r(s_{n+2}^+)\\
\end{bmatrix}\ ,
\end{align}
\begin{align}
W_w^{r,12}=
\begin{bmatrix}
{\bf y}_w^r(s_1^-)&{\bf y}_w^r(s_2^-)&...&{\bf y}_w^r(s_{n+2}^-)\\
\end{bmatrix}\ ,
\end{align}
\begin{align}
W_w^{r,22}=
\begin{bmatrix}
{\bf z}_w^r(s_1^-)&{\bf z}_w^r(s_2^-)&...&{\bf z}_w^r(s_{n+2}^-)\\
\end{bmatrix}\ .
\end{align}
 After  some straightforward but tedious algebra we find  
\begin{align}
\label{Eqyr}
{\bf y}_w^r(s)=-p_A(s)
\begin{bmatrix}
-s(1-\frac{s\tau}{n})^n\\
(1-\frac{s\tau}{n})^n\\
(1-\frac{s\tau}{n})^{n-1}\\
(1-\frac{s\tau}{n})^{n-2}\\
.\\
.\\
1\\
\end{bmatrix}
\end{align}
and
\begin{align}
\label{Eqzr}
{\bf z}_w^r(s)&=(-1)^n\frac{2g \lambda}{Q_0^2} \begin{bmatrix}
 (-1)^ns[(1 - \frac{s\tau}{n})^n-(1 + \frac{s\tau}{n})^n]\\
q^*(s)+(-1)^n[(1 +\frac{s\tau}{n})^n-(1 - \frac{s\tau}{n})^n]\\
 \frac{s\tau}{n}q^*(s)\\
 \frac{s\tau}{n}(1+\frac{s\tau}{n})q^*(s)\\
 \frac{s\tau}{n}(1+\frac{s\tau}{n})^2q^*(s)\\
.\\
.\\
 \frac{s\tau}{n}(1+\frac{s\tau}{n})^{n-1}q^*(s)\\
\end{bmatrix}\ ,
\end{align}
where $q^*(s)=p^{\tau \to -\tau}_A(s)=(\frac{-n}{\tau})^n[(s^2+\frac{s}{Q_0}+1)(1-\frac{s\tau}{n})^n-\frac{g}{Q_0}]$.

A similar calculation for the eigenvectors of the Hamiltonian matrix $H_w^l(\lambda)$  yields
\begin{align}
\label{Eqyl}
{\bf y}_w^l(s)=p_A(-s)
\begin{bmatrix}
s(1+\frac{s\tau}{n})^n\\
(1+\frac{s\tau}{n})^n\\
(1+\frac{s\tau}{n})^{n-1}\\
(1+\frac{s\tau}{n})^{n-2}\\
.\\
.\\
1\\
\end{bmatrix}
\end{align}
and
\begin{align}
\label{Eqzl}
{\bf z}_w^l(s)=(-1)^n\frac{2g \lambda}{Q_0^2}
\begin{bmatrix}
(-1)^ns[(1 - \frac{s\tau}{n})^n-(1+\frac{s\tau}{n})^n\\
q^*(-s)+(-1)^n[(1 - \frac{s\tau}{n})^n-(1 + \frac{s\tau}{n})^n]\\
-\frac{s\tau}{n}q^*(-s)\\
- \frac{s\tau}{n}(1-\frac{s\tau}{n})q^*(-s)\\
- \frac{s\tau}{n}(1-\frac{s\tau}{n})^2q^*(-s)\\
.\\
.\\
 -\frac{s\tau}{n}(1-\frac{s\tau}{n})^{n-1}q^*(-s)\\
\end{bmatrix}\ .
\end{align}
It is readily seen that the relation between the matrices $W_w^r(\lambda)$ and $W_w^l(\lambda)$ (Eq. (\ref{EqWrWl}) in the main text) is recovered via the change $s\to -s$.

\renewcommand{\theequation}{E\arabic{equation}} 

\section{Derivation of Eq. (\ref{Eqsol4})}
\label{AppendE}

Let us partition the inverse of the matrix $W^r_{\rm o}(\lambda)$  into four submatrix blocks as 
\begin{align}
\label{EqQr1}
[W^r_{\rm o}(\lambda)]^{-1}\equiv Q^r_{\rm o}(\lambda)=
\begin{bmatrix}
 Q^{r,11}_{\rm o}(\lambda)& Q^{r,12}_{\rm o}(\lambda)\\
 Q^{r,21}_{\rm o}(\lambda)& Q^{r,22}_{\rm o}(\lambda)&
\end{bmatrix}\ .
\end{align}
For notational simplicity, we drop the dependence on $\lambda$ hereafter. Then, 
\begin{align}
\label{EqQr2}
[W^r_{\rm o}]^{-1}
\begin{bmatrix}
I_{n+2}\\
0
\end{bmatrix}
=
\begin{bmatrix}
Q^{r,11}_{\rm o}\\
Q^{r,21}_{\rm o}
\end{bmatrix}
\end{align}
and Eq. (\ref{Eqsol2})  in the main text can be rewritten as
\begin{align}
\label{EqQr3}
\begin{bmatrix}
 U^r_{\rm o}(t)\\
V^r_{\rm o}(t)
\end{bmatrix}
=W^r_{\rm o}\begin{bmatrix}
e^{ J t}&0\\
0&e^{-J t}&
\end{bmatrix}
\begin{bmatrix}
Q^{r,11}_{\rm o}\\
Q^{r,21}_{\rm o}
\end{bmatrix}\ .
\end{align}
Assuming that the submatrix $Q^{r,11}_{\rm o}$ is invertible, we  introduce the matrix $T^r_{\rm o}\equiv Q^{r,21}_{\rm o}[Q^{r,11}_{\rm o}]^{-1}$. Then, from the inversion of Eq. (\ref{EqQr2}),  
\begin{align}
\label{EqTr1}
W_{\rm o}^{r,11}+W_{\rm o}^{r,12}T^r_{\rm o}&=[Q^{r,11}_{\rm o}]^{-1}\nonumber\\
W_{\rm o}^{r,21}+W_{\rm o}^{r,22}T^r_{\rm o}&=0\ ,
\end{align}
and
\begin{align}
\begin{bmatrix}
Q^{r,11}_{\rm o}\\
Q^{r,21}_{\rm o}
\end{bmatrix}
=\begin{bmatrix}
I_{n+2}\\
T^r_{\rm o}
\end{bmatrix}Q^{r,11}_{\rm o}
=\begin{bmatrix}
I_{n+2}\\
T^r_{\rm o}
\end{bmatrix}
[W_{\rm o}^{r,11}+W_{\rm o}^{r,12}T^r_{\rm o}]^{-1}\ .
\end{align}
 Eq. (\ref{EqQr3}) becomes
\begin{align}
\label{EqTr2}
\begin{bmatrix}
 U^r_{\rm o}(t)\\
 V^r_{\rm o}(t)
\end{bmatrix}
&=\begin{bmatrix}
 W_{\rm o}^{r,11}e^{J t}+ W_{\rm o}^{r,12}e^{-J t}T^r_{\rm o}\\
 W_{\rm o}^{r,21}e^{J t}+ W_{\rm o}^{r,22}e^{-J t}T^r_{\rm o}
\end{bmatrix}
[W_{\rm o}^{r,11}+W_{\rm o}^{r,12}T^r_{\rm o}]^{-1}\ ,
\end{align}
and inserting into Eq. (\ref{Eqsol}) in the main text we easily obtain  Eq. (\ref{Eqsol4}).

It remains to show that the invertibility of the submatrix $Q^{r,11}_{\rm o}$ is equivalent to that of $W_{\rm o}^{r,22}$, so that  $T^r_{\rm o}=-[W_{\rm o}^{r,22}]^{-1}W_{\rm o}^{r,21}$ from Eq. (\ref{EqTr1}) [Eq. (\ref{EqTr}) in the main text]. 

Let us first assume that $W_{\rm o}^{r,22}$ is invertible. Then, its Schur complement $W_{\rm o}^r/W_{\rm o}^{r,22}\equiv W_{\rm o}^{r,11}-W_{\rm o}^{r,12}[W_{\rm o}^{r,22}]^{-1}W_{\rm o}^{r,21}$ is also invertible  and  the general formula for the inversion of a block matrix yields $Q^{r,11}_{\rm o}=[W_{\rm o}^r/W_{\rm o}^{r,22}]^{-1}$. Consequently, $Q^{r,11}_{\rm o}$ is invertible with $[Q^{r,11}_{\rm o}]^{-1}=W_{\rm o}^r/W_{\rm o}^{r,22}$.

We now assume that $Q^{r,11}_{\rm o}$ is invertible. Then, its Schur complement $Q_{\rm o}^r/Q_{\rm o}^{r,11}\equiv Q^{r,22}_{\rm o}-Q_{\rm o}^{r,21}[Q_{\rm o}^{r,11}]^{-1}Q_{\rm o}^{r,12}$ is also invertible and the same general formula applied to $W_{\rm o}^r=[Q_{\rm o}^r]^{-1}$ yields $W_{\rm o}^{r,22}=[Q_{\rm o}^r/Q_{\rm o}^{r,11}]^{-1}$. Hence, $W_{\rm o}^{r,22}$ is invertible, with $[W_{\rm o}^{r,22}]^{-1}=Q_{\rm o}^r/Q_{\rm o}^{r,11}$.

\renewcommand{\theequation}{F\arabic{equation}} 

\section{Solutions of the CAREs  for $\lambda=0$}
\label{AppendF}

\renewcommand{\theequation}{F\arabic{equation}} 

In this appendix  we show that $\hat C^{r,+}_{\rm o}(\lambda)$ and $\hat C^{l,+}_{\rm o}(\lambda)$ are the only solutions of the CAREs (\ref{EqCARE:subeqns})  that satisfy $\hat C^{r,+}_{\rm o}(0)=0$ and $\hat C^{l+}_{\rm o}(0)=C$. To this end, we need to consider the behavior of the solutions as $\lambda\to 0$. Since there is no dependence on the observable for $\lambda=0$, we choose $\rm o=w$  in order to take advantage of  the explicit expressions of the eigenvectors given in Appendix \ref{AppendD2}. 

Let  $\hat C_{w}^{r,(\alpha)}(\lambda)$ and $\hat C_{w}^{l,(\alpha)}(\lambda)$ be the solutions of the CAREs built from a set ${\cal S}^{(\alpha)}_{\lambda}$  with $m^{(\alpha)}$ eigenvalues with a positive real part and $n+2-m^{(\alpha)}$ with a negative real part. 
 Since $(-1)^n p_H(\lambda,s) \to p_A(s)p_A(-s)$ as $\lambda \to 0$, $n+2-m^{(\alpha)}$ eigenvalues are roots  of  $p_A(s)$ and $m^{(\alpha)}$ eigenvalues are roots of $p_A(-s)$ in this limit.  Eqs. (\ref{Eqyr}) and (\ref{Eqyl}) then tell us that $n+2-m^{(\alpha)}$ columns of the matrix  $Y_{w}^{r,(\alpha)}(\lambda)$ and $m^{(\alpha)}$ columns of the matrix $Y_{w}^{l,(\alpha)}(\lambda)$  are ${\cal O}(\lambda)$. On the other hand, from Eqs. (\ref{Eqzr}) and (\ref{Eqzl}), all columns of the matrices $Z_{w}^{r,(\alpha)}(\lambda)$ and $Z_{w}^{l,(\alpha)}(\lambda)$ are ${\cal O}(\lambda)$ regardless the value of $m^{(\alpha)}$.

We now successively consider the two cases ${\cal S}^{(\alpha)}_{\lambda}={\cal S}^+_{\lambda}$, i.e., $m^{(\alpha)}=n+2$, and ${\cal S}^{(\alpha)}_{\lambda}\ne {\cal S}^+_{\lambda}$, i.e., $m^{(\alpha)}<n+2$. 

In the first case,  the matrix $Y_{w}^{r,+}(\lambda)=W_{w}^{r,11}(\lambda)$ is ${\cal O}(1)$ in the limit $\lambda \to 0$ whereas all columns of $Y_{w}^{l,+}(\lambda)=W_{w}^{l,11}(\lambda)$ and all columns of $Z_{w}^{l,+}(\lambda)=W_{w}^{l,21}(\lambda)$ are ${\cal O}(\lambda)$. Hence, $\hat C_{w}^{r,+}(0)=\lim_{\lambda \to 0}Z_{w}^{r,+}(\lambda)[Y_{w}^{r,+}(\lambda)]^{-1}=0$ whereas  $\hat C_{w}^{l,+}(0)=\lim_{\lambda \to 0}Z_{w}^{l,+}(\lambda)[(Y_{w}^{l,+}(\lambda)]^{-1}={\cal O}(1)$. Moreover, both $\det W_{w}^{l,21}(\lambda)$ and $\det W_{w}^{l,11}(\lambda)$ are ${\cal O}(\lambda^{n+2})$ so that $\lim_{\lambda \to 0}\det \hat C^{l,+}_{\lambda}={\cal O}(1)$. The matrix $\hat C_{w}^{l,+}(0)$ is thus invertible and its inverse is the unique solution of  the linear equation $(\hat C_{w}^{l,+}(0)^{-1}A^T+A[\hat C_{w}^{l,+}(0)]^{-1} =-D$ which is nothing but the Lyapunov equation (\ref{EqLya}). As a result,  $\hat C_{w}^{l,+}(0)=C$.

In the second case, $\hat C_{w}^{r,(\alpha)}(0)=\lim_{\lambda \to 0} Z_{w}^{r,(\alpha)}(\lambda)[(Y_{w}^{r,(\alpha)}(\lambda)]^{-1}={\cal O}(1)$. Likewise, $\hat C_{w}^{l,(\alpha)}(0)=\lim_{\lambda \to 0} Z_{w}^{l,(\alpha)}(\lambda)(Y_{w}^{l,(\alpha)}(\lambda)]^{-1}={\cal O}(1)$ with  $\det(\hat C_{w}^{l,(\alpha)}(\lambda)={\cal O}(\lambda^{n+2-m^{(\alpha)}})$. In consequence, $\hat C_{w}^{l,(\alpha)}(0)$ is not invertible and thus differs from $C$.

\setcounter{figure}{0}                      
\renewcommand\thefigure{G.\arabic{figure}} 
\renewcommand{\theequation}{G\arabic{equation}} 

\section{Long-time behavior of $\det C^l_w(\lambda,t)$ and $\det [C^r_w(\lambda,t)+C]$ at the boundaries of the interval $\hat {\cal D}_w$}
\label{AppendG}

In this Appendix, we show that there is no contradiction between the fact that  $\det C^l_w(\lambda,t)$ and $\det [C^r_w(\lambda,t)+C]$ vanish for the same values of $\lambda$ (i.e., $\lambda_w^-(t)$ or $\lambda_w^+(t)$), as shown in Figs. \ref{Fig2} and \ref{Fig3}), and the observation in Fig. \ref{Fig7}  that  $\det \hat C^{l,+}_w(\lambda_{w1})=0$ and $\det (\hat C^{r,+}_w(\lambda_{w 2})+C)=0$. To this aim, we must focus on the long-time behavior of the determinants in the vicinity of boundaries of the interval $\hat {\cal D}_w$.
For conciseness we only consider the vicinity of $\lambda_{w 1}\simeq -0.963$ for $\tau=1$, as shown in Fig. \ref{Fig8}. 
 \begin{figure}[hbt]
\begin{center}
\includegraphics[trim={0cm 0cm 0cm 0cm},clip,width=10cm]{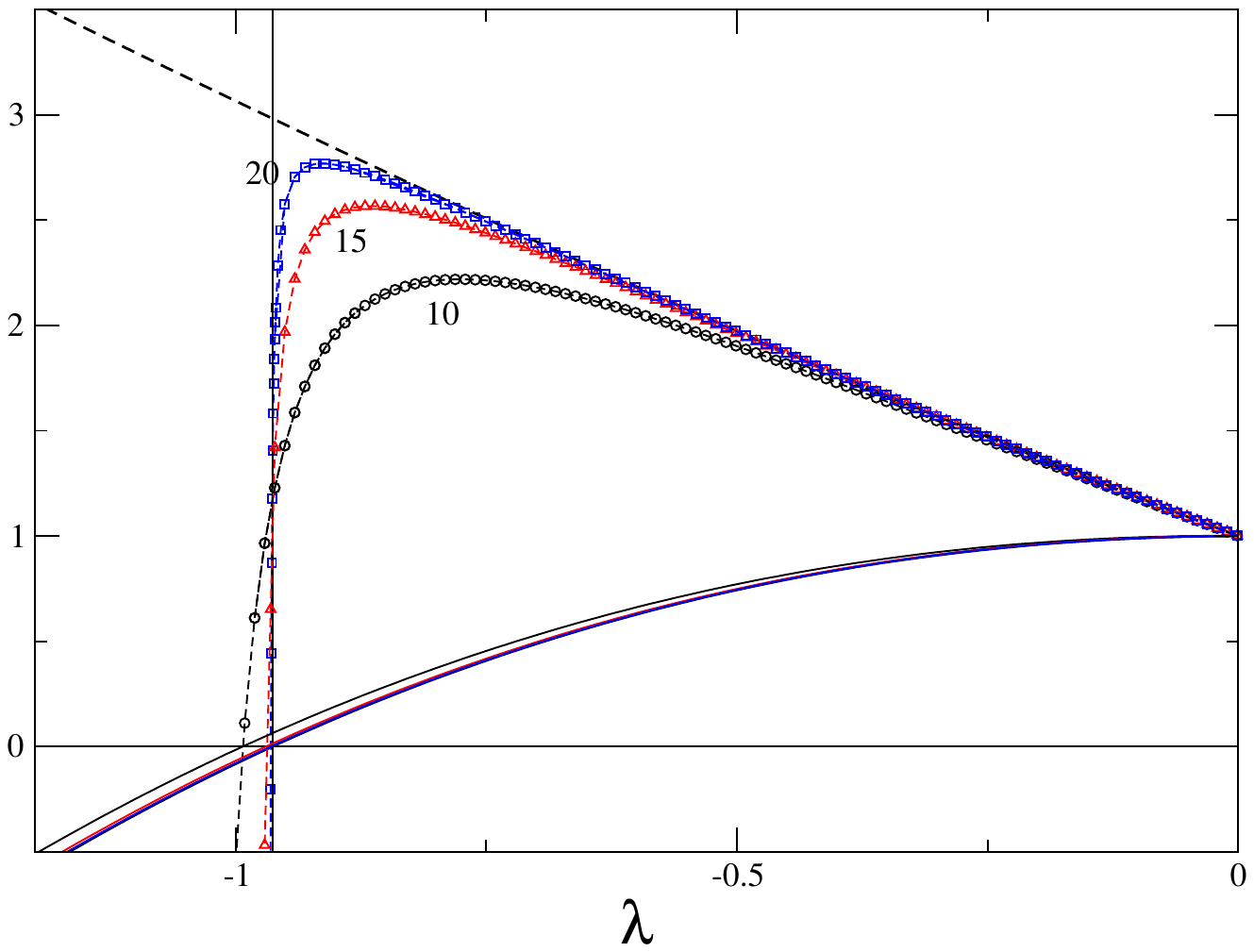}
\caption{ (Color on line) $\det (C^r_w(\lambda,t)+C)/\det C$ (symbols) and  $\det (C^l_w(\lambda,t)/\det C$ (solid lines) as a function of $\lambda$ for $\tau=1$ and $t=10,15,20$. The solid and dashed black lines represent $\det \hat C_w^{l,+}(\lambda)/\det C$ and  $\det (\hat C_w^{r,+}(\lambda)+C)/\det C$, respectively, like in Fig. \ref{Fig7}.}
\label{Fig8}
\end{center}
\end{figure}

 What this figure reveals is that  $\det (C^r_w(\lambda,t)+C)$ strongly deviates from $ \det (\hat C^{r,+}_w(\lambda)+C)$ as $\lambda$ decreases before vanishing at  $\lambda=\lambda_w^-(t)$. This contrasts with the behavior of $\det C^l_w(\lambda,t)$ that smoothly approaches $ \det \hat C^{l,+}_w(\lambda)$ as $t$ increases. In addition, $\det (C^r_w(\lambda,t)+C)$  diverges to $-\infty$ for $\lambda$  very close to (but smaller than) $\lambda_w^-(t)$, signaling that the solution of the RDE  (\ref{EqRic:subeq1}) blows up. Another striking feature is that the  value of the determinant for $\lambda=\lambda_{w 1}$ (indicated by the vertical line in the figure) is approximately independent of time (as the three representative curves cross each other at almost the same point). This suggests that $\lim_{t\to \infty}\det(C^r_w(\lambda_{w1},t)+C)$ is finite but differs from $ \det (\hat C^{r,+}_w(\lambda_{w1})+C)$. This is indeed confirmed by  Fig. \ref{Fig9}  which  shows the time evolution of $\det C_w^l(\lambda,t)$ and $\det (C^r_w(\lambda,t)+C)$ for  values  of $\lambda$ close to $\lambda_{w 1}$. One can  see that $\det(C^r_w(\lambda_{w1},t)+C)/\det C\approx 1.17$ at large times whereas $\det (\hat C^{r,+}_w(\lambda_{w1})+C)/\det C\approx 2.98$. 
% Values of $\lambda$ from right to left: $-0.965,-0.97,-0.975,-0.98,-1,-1.1$ ($-1.2$ est disponible mais non imprimÂ)
\begin{figure}[hbt]
\begin{center}
\includegraphics[trim={0cm 0cm 0cm 0cm},clip,width=10cm]{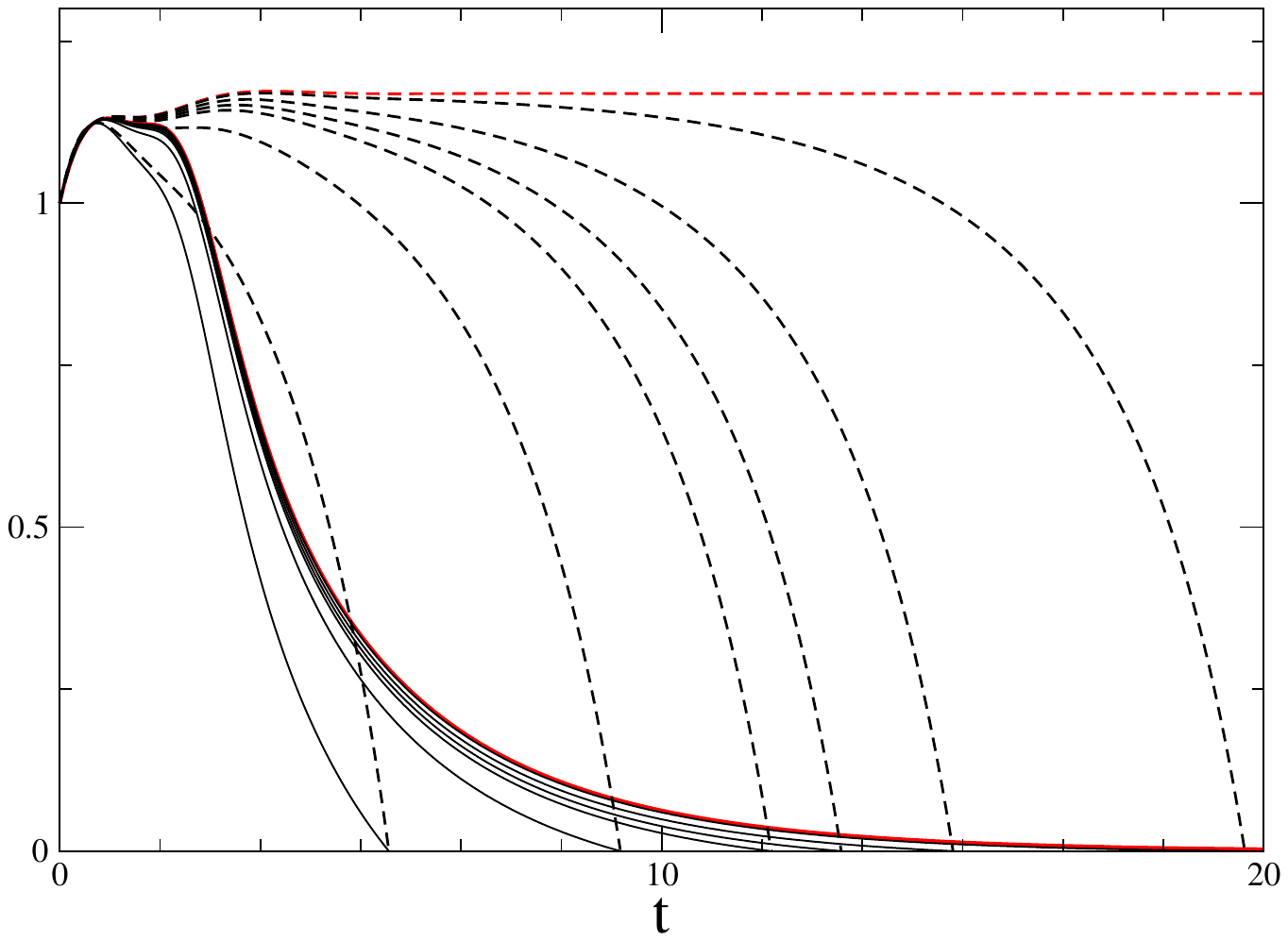}
\caption{ (Color on line) Time evolution of $\det C_w^l(\lambda,t)/ \det C$ (solid lines) and $\det (C^r_w(\lambda,t)+C)/\det C$ (dashed lines)  for $\tau=1$ and different values  of $\lambda$ close to $\lambda_{w 1}\simeq -0.963$. The red lines correspond to $\lambda=\lambda_{w 1}$.}
\label{Fig9}
\end{center}
\end{figure}

This phenomenon is rather subtle but should come as no surprise. Indeed, as predicted by Eq. (\ref{EqEquiv:subeq1}), the condition $\det C_{\rm o}^{l,+}(\lambda_{w 1})=0$ implies that the solution of the RDE (\ref{EqRic:subeq1}) {\it does not} go asymptotically to $\hat C_{\rm o}^{r,+}(\lambda_{w 1})$ and, as a result, $\det(C^r_w(\lambda_{w1},t\to \infty)+C)\neq \det(C^{r,+}_w(\lambda_{w1})+C)$.
As a matter of fact, a similar behavior takes place in the vicinity of $\lambda_{w 2}$, with the roles of $C^r_w(\lambda,t)+C$  and $C^l_w(\lambda,t)$ inverted\footnote{This explains the abrupt decrease of $\det C^l(\lambda,t)$  in the vicinity of $\lambda_w^+(t)\approx 0.65$  for $t=10$  in Fig. \ref{Fig2}(a).}, and it is also observed for  $\tau=3$.  More generally, this is one of the two scenarios that may occur  at the limits of the domain of definition of the SCGF when $\hat {\cal D}_{\rm o}\subset {\cal D}_H$ (i.e., $\lambda_{\rm o 1}>\lambda_{\min}$ and $\lambda_{\rm o 2}<\lambda_{\max}$).  This is a nontrivial issue which is  discussed in Sec. \ref{SubsecIVC}. 
An important  case is $\lambda_{\rm o 2}=1$, as it concerns the fluctuation theorem (\ref{EqIFTheat}) for the heat and the conjecture (\ref{EqIFTentropy}) for the apparent entropy production.

\renewcommand{\theequation}{H\arabic{equation}} 

\section{Expression of the SCGF as an integral over frequency}
\label{AppendH}

In this Appendix, we derive the expression of the SCGF $f^+(\lambda)$ as an integral over frequency [Eq. (\ref{Eqmunlambda}) in the main text]. To this end, we use the property that each symmetric  solution, $\hat C_{\rm o}^{r,(\alpha)}(\lambda)$ or  $\hat C_{\rm o}^{l,(\alpha)}(\lambda)$, of the CAREs gives rise to a factorization of the characteristic polynomial $p_H(\lambda,s)$ of $H^r_{\rm o}(\lambda)$ and $H^l_{\rm o}(\lambda)$ in the form~\cite{K2010} 
\begin{align} 
\label{Eqfactorisation}
p_H(\lambda,s)=(-1)^{n}q^{(\alpha)}(\lambda,s)q^{(\alpha)}(\lambda, -s)\ ,
\end{align}
where $q^{(\alpha)}(\lambda,s)$ is the characteristic polynomial  of the matrices $-A_{\rm o}(\lambda)+D\hat C_{\rm o}^{r,(\alpha)}(\lambda)$ and $A_{\rm o}(\lambda)+D\hat C_{\rm o}^{l,(\alpha)}(\lambda)$. (Owing to the invariance relations (\ref{Eqrel3}), $q^{(\alpha)}(\lambda,s)$ does not depend on the observable.)  Eq. (\ref{Eqfactorisation}) follows from 
\begin{align}
\label{Eqgraph1}
\begin{bmatrix}
 I_{n+2}&0\\
  \hat C^{r,(\alpha)}_{\rm o}(\lambda)&I_{n+2}&
\end{bmatrix}^{-1}H^r_{\rm o}(\lambda)\begin{bmatrix}
 I_{n+2}&0\\
  \hat C^{r,(\alpha)}_{\rm o}(\lambda)&I_{n+2}&
\end{bmatrix}=\begin{bmatrix}
 -A_{\rm o}(\lambda)+D \hat C^{r,(\alpha)}_{\rm o}(\lambda)&D\\
 0&[A_{\rm o}(\lambda)-D \hat C^{r,(\alpha)}_{\rm o}(\lambda)]^T&
\end{bmatrix}
\end{align}
and a similar relation involving $C^{l,(\alpha)}_{\rm o}(\lambda)$ and $H^l_{\rm o}(\lambda)$.
Focusing on the maximal solutions   $\hat C^{r,+}_{\rm o}(\lambda)$ and $\hat C^{l,+}_{\rm o}(\lambda)$, we then define the rational function 
\begin{align} 
\label{EqGs}
G^+(\lambda,s)=\frac{q^+(\lambda, s)}{p_A(s)}\ ,
\end{align}
where  $p_A(s)$ is the characteristic polynomial  of  the drift matrix $A$ given by Eq. (\ref{Eqchar}). As a result, $G^+(\lambda,s)$ has $n+2$ zeros  but no poles in the closed right half plane (since all eigenvalues of $A$ have a negative real part). In addition, 
\begin{align} 
\label{Eqqplus}
q^+(\lambda,s)=\prod_{i=1}^{n+2}(s-s_i^+)= s^{n+2}-(\sum_i^{n+2} s_i^+ )s^{n+1}+...
\end{align}
and
\begin{align} 
p_A(s)=s^{n+2}-\mbox{Tr}(A)s^{n+1}+...\ ,
\end{align}
so that the  function $L^+(\lambda,s)$ defined through 
\begin{align} 
G^+(\lambda,s)=[1+L^+(\lambda,s)]^{-1}
\end{align}
has a relative degree equal to $1$. (The relative degree is the difference between the degrees of the polynomials in the denominator and  in the numerator. In control theory,  $L^+_{\lambda}(s)$ would be interpreted  as a proper, scalar rational  loop transfer function of a feedback system~\cite{B2021,AM2008}.)  A standard application of Cauchy's residue theorem, known  as Bode's sensitivity integral in control theory~\cite{B2021,AM2008}, then yields
\begin{align} 
\label{EqBode}
\frac{1}{2\pi i}\int_{-i\infty}^{i\infty}ds\: \ln \vert G^+(\lambda,s)\vert=\sum_{i=1}^{n+2}s_i^+(\lambda)-\frac{1}{2}\kappa^+(\lambda)\ ,
\end{align}
where the integration is performed along the imaginary axis in the complex $s$-plane,  and 
\begin{align} 
\label{Eqkappa}
\kappa^+(\lambda)&\equiv \lim_{s\to \infty} sL^+(\lambda,s)=\lim_{s\to \infty}s\frac{p_A(s)-q^+(\lambda,s)}{p_A(s)}\nonumber\\
&=\sum_{i=1}^{n+2}s_i^+(\lambda)-\mbox{Tr}(A)\ .
\end{align}
Inserting Eqs. (\ref{EqBode}) and (\ref{Eqkappa}) into  Eq. (\ref{Eqmustar}), we obtain that
\begin{align} 
f^+(\lambda)=-\frac{1}{2\pi i}\int_{-i\infty}^{i\infty} ds\:\ln \vert G^+(\lambda,s)\vert,\nonumber\\
\end{align}
which thanks to Eqs. (\ref{Eqfactorisation}), (\ref{EqGs}), and Eq. (\ref{EqpH}) in the main text can  be rewritten as
\begin{align} 
\label{Eqmu2}
f^+(\lambda)&=-\frac{1}{4\pi i}\int_{-i\infty}^{i\infty} ds\: \ln \frac{(-1)^np_H(\lambda,s)}{\vert p_{A}(s)\vert^2} \nonumber\\
&=-\frac{1}{4\pi i}\int_{-i\infty}^{i\infty} ds\: \ln\Big[1-\frac{2\lambda g}{Q_0^2}(\frac{n}{\tau})^{2n}s\frac{(1-\frac{s\tau}{n})^n-(1+\frac{s\tau}{n})^n}{\vert p_A(s)\vert^2}\Big]\ .
\end{align}
Note that the integral is finite as $p_H(\lambda,s)$ has no purely imaginary roots for  $\lambda\in{\cal D}_H$. 

We can further transform Eq. (\ref{Eqmu2})  by  replacing the Laplace variable $s$ by the frequency $\omega=is$ and  introducing the response function  $ \chi_n(\omega)$. This eventually leads  Eq. (\ref{Eqmunlambda}) in the main text.

\section{Derivation of the expression of the SCGF  $\mu_{\sigma}(1)$ for the entropy production}

\label{AppendI}

\renewcommand{\theequation}{I\arabic{equation}} 

In this appendix, we derive the expression of the dichotomic solution $\hat L^*$ of the CARE   
\begin{align} 
 LF_{11}-F_{22}L+ LF_{12} L-F_{21}=0\ ,
\end{align}
with 
\begin{align} 
F_{11}=\begin{bmatrix}
\frac{1}{2Q_0}&-1\\
1&-\frac{1}{2Q_0}\\
\end{bmatrix}\ , 
\end{align}
\begin{align} 
F_{12}=\frac{2g}{Q_0^2} \begin{bmatrix}
0&0&.&.&1\\
0&0&.&.&0\\
\end{bmatrix}\ , 
\end{align}
\begin{align} 
F_{21}=\frac{nQ_0}{2\tau} \begin{bmatrix}
0&1\\
0&0\\
.&.\\
.&.\\
0&0\\
\end{bmatrix}\ , 
\end{align}
and 
\begin{align} 
F_{22}= \begin{bmatrix}
-\frac{1}{2Q_0}-\frac{n}{\tau}&0&.&.&0\\
\frac{n}{\tau}&-\frac{1}{2Q_0}-\frac{n}{\tau}&.&.&0\\
0&\frac{n}{\tau}&-\frac{1}{2Q_0}-\frac{n}{\tau}&.&0\\
.&.&.&.&.\\
.&.&.&.&.\\
0&0&.&\frac{n}{\tau}&-\frac{1}{2Q_0}-\frac{n}{\tau}\\
\end{bmatrix}\ .
\end{align}
 Assuming that the eigenvalues $\nu_i$ of the matrix $F=
\begin{bmatrix}
F_{11}&F_{12}\\
  F_{21}&F_{22}&
\end{bmatrix}$ satisfy the condition (\ref{EqSeq}) in the main text,  we introduce the matrix  $W^*$ of dimension $(n+2)\times 2$ formed by the eigenvectors ${\bf e}_{1}$ and ${\bf e}_{2}$ associated with  $\nu_{1}$ and $\nu_{2}$. The dichotomic solution $\hat L^*$ is then obtained  as  
\begin{align} 
\label{EqhatL}
\hat L^*=Z^*(Y^*)^{-1}\ ,
\end{align}
where $Y^*$ is the sub-matrix of $W^*$ of dimension $2\times 2$ formed by the first two components of the vectors ${\bf e}_{1}$ and ${\bf e}_{2}$, and $Z^*$ is the sub-matrix of dimension $n\times 2$ formed by the other $n$ components (Eq. (\ref{EqdetY}) below shows that the matrix $Y^*$ is  invertible).

Using the Faddeev-Leverrier's recursive method~\cite{FS1952} to compute the eigenvector ${\bf e}$ associated with the eigenvalue $\nu$, we find\footnote{For simplicity, we here assume  that $F$ is diagonalizable, as we did for $H_{\rm o}^r(\lambda)$ and $H_{\rm o}^l(\lambda)$. Otherwise, one must consider the generalized eigenvectors~\cite{M1982,FJ1995,F2002}.}
\begin{align}
{\bf e}(\sigma)=\frac{Q_0}{2}
\begin{bmatrix}
\frac{2}{Q_0}(\sigma-\frac{n}{\tau})\sigma^n\\
\frac{2}{Q_0} \sigma^n\\
\frac{n}{\tau} \sigma^{n-1}\\
(\frac{n}{\tau})^2 \sigma^{n-2}\\
.\\
.\\
(\frac{n}{\tau})^n\\
\end{bmatrix}\ ,
\end{align}
where $\sigma=\nu+1/(2Q_0)+n/\tau$. This yields
\begin{align}
Y^*(\sigma_{1},\sigma_{2})=
\begin{bmatrix}
(\sigma_{1}-\frac{n}{\tau})\sigma_{1}^n&(\sigma_{2}-\frac{n}{\tau})\sigma_{2}^n\\
\sigma_{1}^n&\sigma_{2}^n\\
\end{bmatrix}\ ,
\end{align}
and 
\begin{align}
Z^*(\sigma_{1},\sigma_{2})=\frac{Q_0}{2}
\begin{bmatrix}
\frac{n}{\tau} \sigma_{1}^{n-1}&\frac{n}{\tau} \sigma_{2}^{n-1}\\
(\frac{n}{\tau})^2 \sigma_{1}^{n-2}&(\frac{n}{\tau})^2 \sigma_{2}^{n-2}\\
.&.\\
.&.\\
(\frac{n}{\tau})^n&(\frac{n}{\tau})^n\\
\end{bmatrix}\ .
\end{align}
Note in passing that 
\begin{align} 
\label{EqdetY}
\det(Y^*)&=(\sigma_{1}\sigma_{2})^n(\sigma_{1}-\sigma_{2})=(\nu_{1}+\frac{1}{2Q_0}+\frac{n}{\tau})^n(\nu_{2}+\frac{1}{2Q_0}+\frac{n}{\tau})^n(\nu_{1}-\nu_{2})\ne 0
\end{align}
since $-1/(2Q_0)-n/\tau$ is not an eigenvalue of $F$ (one has $p_F(-1/(2Q_0)-n/\tau)=-(n/\tau)^ng/Q_0\ne 0$). 

We thus have from Eq. (\ref{EqhatL}) an explicit expression of $\hat L^*$ in terms of  $\sigma_{1}$ and $\sigma_{2}$.  In particular, 
\begin{align} 
\hat L_{n1}^* =(\frac{n}{\tau})^n\frac{Q_0}{2(\sigma_{1}-\sigma_{2})}(\frac{1}{\sigma_{1}^n}-\frac{1}{\sigma_{2}^n})\ .
\end{align}
This can be further simplified by using the fact that $\sigma_i=-s^*_i+n/\tau$ from Eq. (\ref{Eqnus}). Hence, $\sigma_i$ is a root of $q^*(-s+n/\tau)$, which yields
\begin{align} 
\sigma_i^{-n}=(\frac{\tau}{n})^n\frac{Q_0}{g}[1+\sigma_i^2-\sigma_i (\frac{1}{Q_0}+\frac{2n}{\tau})+\frac{n}{\tau}(\frac{1}{Q_0}+\frac{n}{\tau})]
\end{align}
and then
\begin{align} 
L^*_{n,1}&=\frac{Q_0^2}{2g}[\sigma_{1}+\sigma_{2}-(\frac{1}{Q_0}+\frac{2n}{\tau})]\nonumber\\
&=\frac{Q_0^2}{2g}(-s_1^*-s_2^*-\frac{1}{Q_0})\ .
\end{align}
The expression  of the SCGF $\mu_{\sigma}(1)$ [Eq. (\ref{Eqmusigma}) in the main text] is finally obtained from Eq. (\ref{Eqmusigma1}) .

\section{Work fluctuations of an active Ornstein-Uhlenbeck particle}

\label{AppendJ}

\renewcommand{\theequation}{J\arabic{equation}} 

In this Appendix, we apply the formalism developed in the main text to a model of an overdamped active Ornstein-Uhlenbeck particle (AOUP) confined in a harmonic potential. The fluctuations of the ``active work" have been  investigated in a recent paper~\cite{SGSZ2023} where it was shown that  the harmonic confinement may induce  linear tails in the rate function  for large values of the work.  We  here show that these  results, which were obtained after long and cumbersome calculations,  are readily recovered by using the Riccati formalism, 
 
The model  is defined by the two equations
 \begin{align} 
 \label{EqSem1}
\gamma \dot r(t)&=a(t)-kr(t)+\sqrt{2\gamma k_BT}\: \xi(t)\nonumber\\
\dot a(t)&=-\nu a(t)+F\sqrt{2\nu}\: \eta(t)\ ,
\end{align}
 where $r(t)$ is the position of the particle and $a(t)$ represents a self-propulsion force with an amplitude $F$ and a decay rate $\nu=k_BT/(\gamma d^2)$ where $d$ is a length proportional to the particle diameter~\cite{SGSZ2023}.  $\xi(t)$ and $\eta(t)$ are two independent standard white noises. Eqs. (\ref{EqSem1}) can be made dimensionless by introducing the elastic constant $\kappa=kd^2/(k_BT)$  and the P\'eclet number $\mbox{Pe}=(Fd)/(k_BT)$. 

One is thus dealing with a bivariate  Ornstein-Uhlenbeck process for the dynamical variable ${\bf u}_t=[r_t,a_t]^T$ with a drift matrix $A=\begin{bmatrix}
-\kappa &1\\
0&-1
\end{bmatrix}$ and a diffusion matrix $D=2\begin{bmatrix}
1&0\\
0&\mbox{Pe}^2
\end{bmatrix}$. The time-integrated  ``active work" is defined by
\begin{align} 
{\cal W}_t=\int_0^t a(t')\circ dr(t')=\int_0^t (B{\bf u}_{t'})\circ d{\bf u}_{t'} \ ,
\end{align}
where $B=\begin{bmatrix}
0&1\\
0&0
\end{bmatrix}$. Accordingly, the  matrices $C^r_w(\lambda,t)$ and $C^l_w(\lambda,t)$ defined by Eqs. (\ref{EqSol1:subeqns}) in the main text  are   $2\times 2$  matrices which  satisfy the RDEs  (\ref{EqRic:subeqns}) with $A_{\lambda}=A-\lambda DB=\begin{bmatrix}
-\kappa&1-2\lambda \\
0 &-1
\end{bmatrix}$ and $K_{\lambda}=\lambda(A^T  B + B^T A) - \lambda^2 B^T D B=\lambda \begin{bmatrix}
0&-\kappa\\
 -\kappa &2(1-\lambda)
\end{bmatrix}$ (note that $\lambda$ in Ref. \cite{SGSZ2023} has an opposite sign).

The low dimensionality of the system allows us to easily obtain the analytical expressions of the  eigenvalues and eigenvectors of the Hamiltonian matrices $H^r_{w,\lambda}$ and $H^l_{w,\lambda}$. The characteristic polynomial reads
 \begin{align} 
 p_H(\lambda,s)=s^4+(4Pe^2\lambda^2 - 4Pe^2\lambda - \kappa^2 - 1)s^2+ \kappa^2\ ,
\end{align}
from which we get the eigenvalues $s_1=(1/2)(\alpha_++\alpha_-),s_2=(1/2)(\alpha_+-\alpha_-), s_3=-s_1,s_4=-s_2$, with 
$\alpha_{\pm}=\sqrt{(1 \pm \kappa)^2 + 4\lambda(1-\lambda)Pe^2}$.
% \begin{align}
%\label{Eqalp} 
% \alpha_{\pm}&=\sqrt{(1 \pm \kappa)^2 + 4\lambda(1-\lambda)Pe^2)}\ . 
% \end{align}
The eigenvectors ${\bf e}^r(s_i)$ and ${\bf e}^l(s_i)$ associated with the eigenvalue $s_i$ are then obtained by using the Faddeev-Leverrier's  method~\cite{FS1952}, which yields
\begin{align}
{\bf e}^r(s_i)=
\begin{bmatrix}
s_i^3 + \kappa s_i^2+[4Pe^2\lambda(\lambda-1) -1]s_i - \kappa(1+2Pe^2\lambda)\\
-2Pe^2\lambda\kappa(\kappa + s_i)\\
2Pe^2\lambda^2\kappa^2\\
\lambda\kappa(1-s_i)(\kappa + s_i)\\
\end{bmatrix}\ ,
\end{align}
and 
\begin{align}
\label{Eqel}
{\bf e}^l(s_i)=
\begin{bmatrix}
s_i^3 - \kappa s_i^2+[4Pe^2\lambda(\lambda-1) -1]s_i + \kappa(1+2Pe^2\lambda)\\
2Pe^2\lambda\kappa(\kappa - s_i)\\
2Pe^2\lambda^2\kappa^2\\
\lambda\kappa(1+s_i)(\kappa -s_i)\\
\end{bmatrix}\ .
\end{align}

The expression of the SCGF $\mu_w(\lambda)$ is  obtained at once from Eq. (\ref{Eqmustar}) in the main text:
 \begin{align}
\label{Eqmu} 
\mu_w(\lambda)=f^+(\lambda)=-\frac{1}{2}[\mbox{Tr}A+(s_1+s_2)]=\frac{1}{2}(1+\kappa+\alpha_+)\ ,
\end{align}
and the two boundaries of the interval ${\cal D}_H$ in which the  eigenvalues $s_i$ are not purely imaginary  (and accordingly the corresponding algebraic Riccati equation (\ref{EqCARE:subeqns}) have real symmetric solutions) are  given  by 
 \begin{align}
\label{Eqlim}
 \lambda_{\min}&=\frac{1}{2}[1-\frac{\sqrt{Pe^2 +(1+ \kappa)^2 }}{Pe}]\nonumber\\
 \lambda_{\max}&=\frac{1}{2}[1+\frac{\sqrt{Pe^2 + (1+ \kappa)^2}}{Pe}]\ .
 \end{align}
Eqs. (\ref{Eqmu})  and  (\ref{Eqlim}) are identical to Eqs. (5) and (6) in Ref. \cite{ SGSZ2023} (with $\lambda$ changed into $-\lambda$).
% which are obtained after cumbersome calculations. 

The maximal solutions of the CAREs (\ref{EqCARE:subeqns}) are computed  from the formulas $\hat C^{r,+}_w(\lambda)=W_w^{r,21}(\lambda)[W_w^{r,11}(\lambda)]^{-1}$ and $\hat C^{l,+}_w(\lambda)=W_w^{l,21}(\lambda)[W_w^{l,11}(\lambda)]^{-1}$, with the matrices $W_w^r(\lambda)$ and $W_w^l(\lambda)$  built from the eigenvectors ${\bf e}^r(s_i)$ and ${\bf e}^l(s_i)$ (see Sec. \ref{SubsecRic2b})). For the sake of conciseness, the explicit expressions of $\hat C^{r,+}_w(\lambda)$ and $\hat C^{l,+}_w(\lambda)$ are not reproduced here and can be found in the Supplemental material of Ref. \cite{SGSZ2023}. Indeed, the matrices  ${\cal L}_{\lambda}$ and ${\cal R}_{\lambda}$ whose  expressions are derived in Ref. \cite{SGSZ2023} after long  calculations correspond exactly to $C+\hat C^{r,+}_w(\lambda)$ and $\hat C^{l,+}_w(\lambda)$, respectively. 

As discussed in Sec. (\ref{SubsecIVB1}), the  actual domain of existence of the SCGF may be smaller that ${\cal D}_H$ and corresponds to the interval $\hat {\cal D}_w=[\lambda_{w,1},\lambda_{w,2}]$ in which the matrices $C+\hat C^{r,+}_w(\lambda)$ and $\hat C^{l,+}_w(\lambda)$ are both positive definite. 
The  boundaries $\lambda_{w,1}$ and $\lambda_{w,2}$ are  then obtained by solving the equations $\det[C+\hat C^{r,+}_w(\lambda)]=0$ and $\det \hat C^{l,+}_w(\lambda)=0$. 
Consider for instance the second equation. Since $\hat C^{l,+}_w(\lambda)=W_w^{l,21}(\lambda)[W_w^{l,11}(\lambda)]^{-1}$, one only needs to compute the values of $\lambda$ for which $\det W_w^{l,21}(\lambda)=0$. From Eq. (\ref{Eqel}) and the expressions of the eigenvalues $s_i$, we find
\begin{align} 
W_w^{l,21}(\lambda)=\frac{\kappa \lambda}{2}\begin{bmatrix}
4 Pe^2 \kappa \lambda&(1-\kappa +\alpha_+)(1-\kappa+\alpha_-)+4\lambda (1-\lambda)Pe^2\\
4 Pe^2\kappa \lambda&(1-\kappa +\alpha_+)(1-\kappa+\alpha_-)-4\lambda (1-\lambda)Pe^2
\end{bmatrix}\ . 
\end{align}
This yields
\begin{align} 
\det W_w^{l,21}(\lambda)=2\alpha_- Pe^2\kappa^3 \lambda^3 (1-\kappa +\alpha_+)\ ,
\end{align}
and since $\alpha_+>0$  we conclude that $\det \hat C^{l,+}_w(\lambda)=0$ for  $\kappa >1$ only and 
\begin{align} 
\lambda=\frac{1}{2}[1\pm \sqrt{1+\frac{4\kappa}{Pe^2}}\: ] \ .
\end{align} 
The complete description of the  domain of definition of the SCGF as a function of $\kappa$ and $Pe$ is given in Ref. \cite{ SGSZ2023}. As  discussed in Sec. (\ref{SubsecIVB2}), the finite slopes of the SCGF at the two cutoff $\lambda_{w,1}$ and $\lambda_{w,2}$ imply that the rate function  exhibits linear branches for large values of the work.

Finally, we recall that the Riccati  formalism  also allows us to study the statistics of the active work at arbitrary finite time.  Note in particular that finite-time divergences may occur in the present model since the matrix $K_{\lambda}$ is not positive definite.

\end{appendices}

\vspace{1cm}

\hspace{7cm}  {\bf Notation}

\vspace{0.5cm}

%We first list the notations used in the paper.

Matrices are denoted by capital latin letters while boldface, lower-case latin letters denote vectors. $X^T$ denotes the transpose of the matrix $X$ and $I_{n+2}$ denotes the $(n+2)\times (n+2)$  identity matrix. If the real symmetric matrix $X$ is positive definite (resp. semidefinite), we write $X >0$ (resp. $X\ge 0$) and $X_1>X_2$ (resp. $X_1\ge X_2$) means that $X_1-X_2>0$ (resp. $X_1-X_2\ge 0$).

$G_{\rm o,\lambda}(t)$, $G^r_{\rm o,\lambda}({\bf u}_0, t),G^l_{\rm o,\lambda}({\bf u}, t)$ -  moment generating functions ($\rm o=w,q,\sigma$ denotes the observable)

$\mu_{\rm o}(\lambda)$ - scaled cumulant generating function  

$I_{\rm o}(a)$ - large deviation function 

$A$ - drift matrix 

$p_A(s)$ - characteristic polynomial of $A$

$p({\bf u})$ stationary probability density

$\Sigma$ - stationary covariance matrix 
%$\dot X(\lambda,t)={\cal R}_{\rm o,\lambda}[X(\lambda,t)]$   Riccati differential equation (RDE)

${\cal R}_{\rm o,\lambda}$ -  Riccati differential operator

$C^r_{\rm o}(\lambda,t), C^l_{\rm o}(\lambda,t)$ - solutions of the Riccati differential equations (RDEs)

$H^r_{\rm o}(\lambda),H^l_{\rm o}(\lambda)$ - Hamiltonian matrices associated with the RDEs

$p_H(\lambda,s)$ - characteristic polynomial of $H^r_{\rm o}(\lambda)$ and $H^l_{\rm o}(\lambda)$

$\hat C^{r,(\alpha)}_{\rm o}(\lambda), \hat C^{l,(\alpha)}_{\rm o}(\lambda)$ -  solutions of the continuous-time Riccati algebraic equations (CAREs)

$\hat C^{r,+}_{\rm o}(\lambda),\hat C^{l,+}_{\rm o}(\lambda)$ - maximal real symmetric solutions of the CAREs

$f^{r,(\alpha)}(\lambda),f^{l,(\alpha)}(\lambda)$ - scalar functions associated with $\hat C^{r,(\alpha)}_{\rm o}(\lambda)$ and $\hat C^{l,(\alpha)}_{\rm o}(\lambda)$

$f^+(\lambda) $ -  scalar function associated with the maximal solutions

${\cal D}_H=(\lambda_{\min},\lambda_{\max})$ - interval for which $H^r_{\rm o}(\lambda),H^l_{\rm o}(\lambda)$ have no purely imaginary eigenvalues

$ {\cal D}_{\rm o}(t)=(\lambda^-_{\rm o}(t),\lambda^+_{\rm o}(t))$ - domain of existence of $G_{\rm o,\lambda}(t)$

$\hat {\cal D}_{\rm o}=(\lambda_{\rm o 1},\lambda_{\rm o,2})={\cal D}_{\rm o}(\infty)$ - domain of existence of  $\mu_{\rm o}(\lambda)$

$\hat A(\lambda)$ - drift matrix of the effective process

$\hat p_{\lambda}({\bf u})$ stationary probability density of the effective process

$\hat \Sigma(\lambda)$ - covariance matrix of the effective process

\end{document}